\documentclass[1p]{elsarticle}
\usepackage{etex}
\usepackage{theorem,mathtools,xspace}
\usepackage{amsmath}
\usepackage{amsfonts}
\usepackage{amssymb}

\usepackage{ascmac}
\usepackage{hyperref}

\usepackage[pdftex,all]{xy}
\SelectTips{cm}{}  
\xyoption{v2}
\xyoption{curve}
\xyoption{2cell}
\UseAllTwocells
\SilentMatrices

\newif\ifignore 
\ignorefalse
\newcommand{\auxproof}[1]{
\ifignore\mbox{}\newline
\textbf{BEGIN: AUX-PROOF} \dotfill\newline
{#1}\mbox{}\newline
\textbf{END: AUX-PROOF}\dotfill\newline
\fi}

\usepackage{wrapfig}
\usepackage{tikz}

\usepackage{cancel}

\usepackage{makeidx}  

\newcommand{\pow}{\mathcal{P}}
\newcommand{\dist}{\mathcal{D}}
\newcommand{\place}{\underline{\phantom{n}}\,} 
\newcommand{\Sets}{\mathbf{Sets}}

\newcommand{\lang}{L}

\newcommand{\kto}[0]{\mathrel{\mspace{-4mu} \to \mspace{-16mu} \raisebox{0.1ex}[0ex][0ex]{$\shortmid$} \mspace{6mu}}}

\newcommand{\longkto}{\mathrel{\ooalign{\hfil\raisebox{.3pt}{$\shortmid$}\hfil\crcr$\longrightarrow$}}}
\newcommand{\Kl}[0]{\mathcal{K}\mspace{-1mu}\ell}
\newcommand{\mcr}[2]{#1_{\mathcal{#2}}}

\newcommand{\bvec}[1]{\mbox{\boldmath $#1$}}
\newcommand{\mss}[0]{\mathcal{M}_{\mathcal{S}}}

\newcommand{\id}[0]{\mathrm{id}}
\newcommand{\tr}[0]{{\sf tr}}

\newcommand{\APEX}{{\sc apex}}

\newcommand{\Max}[0]{\mathrm{Max}}
\newcommand{\Min}[0]{\mathrm{Min}}

\newcommand{\spt}[0]{\mathcal{S}_{+,\times}}

\newcommand{\smp}[0]{\mathcal{S}_{\max,+}}

\newcommand{\ptlangfwdalgo}[0]{{\sf pt\_langincl\_fwd}}
\newcommand{\ptlangbwdalgo}[0]{{\sf pt\_langincl\_bwd}}
\newcommand{\ptsimfwdalgo}[0]{{\sf pt\_simulation\_fwd}}
\newcommand{\ptsimbwdalgo}[0]{{\sf pt\_simulation\_bwd}}
\newcommand{\ptfpealgo}[0]{{\sf pt\_FPE}}
\newcommand{\ptbpealgo}[0]{{\sf pt\_BPE}}

\newcommand{\mplangfwdalgo}[0]{{\sf mp\_langincl\_fwd}}
\newcommand{\mplangbwdalgo}[0]{{\sf mp\_langincl\_bwd}}
\newcommand{\mpsimfwdalgo}[0]{{\sf mp\_simulation\_fwd}}
\newcommand{\mpsimbwdalgo}[0]{{\sf mp\_simulation\_bwd}}
\newcommand{\mpfpealgo}[0]{{\sf mp\_FPE}}
\newcommand{\mpbpealgo}[0]{{\sf mp\_BPE}}

\newcommand{\pttreesimfwdalgo}[0]{{\sf pttree\_simulation\_fwd}}
\newcommand{\pttreesimbwdalgo}[0]{{\sf pttree\_simulation\_bwd}}

\newcommand{\ptsim}[0]{{\sf $+\times$-sim}}
\newcommand{\ptPE}[0]{{\sf $+\times$-PE}}

\newcommand{\mpsim}[0]{{\sf $\max$$+$-sim}}
\newcommand{\mpPE}[0]{{\sf $\max$$+$-PE}}
\newcommand{\pttsim}[0]{{\sf $+\times$-treesim}}

\newcommand{\Tree}[0]{\text{Tree}}

\newcommand{\FPE}[0]{{\sf FPE}}
\newcommand{\BPE}[0]{{\sf BPE}}
\newcommand{\trans}{{}^t\!}

\newcommand{\kar}{\ar|-*\dir{|}}

\newcommand{\fwd}[0]{\sqsubseteq_{\bf F}}
\newcommand{\bwd}[0]{\sqsubseteq_{\bf B}}

\makeatletter
  \def\@thmcountersep{.}
\makeatother

\theorembodyfont{\itshape}
\newtheorem{mytheorem}{Theorem}[section]
\newtheorem{mylemma}[mytheorem]{Lemma}
\newtheorem{myproposition}[mytheorem]{Proposition}

\newtheorem{mycorollary}[mytheorem]{Corollary}
\theorembodyfont{\rmfamily}

\newtheorem{myremark}[mytheorem]{Remark}
\newtheorem{myexample}[mytheorem]{Example}
\newtheorem{mydefinition}[mytheorem]{Definition}

\newtheorem{myalgorithm}[mytheorem]{Algorithm}

\newproof{myproof}{Proof}

\begin{document}
\begin{frontmatter} 
\title{Quantitative Simulations by Matrices${}^\star$}
 \tnotetext[concur]{An earlier version of this paper~\cite{UrabeH14} has been
 presented at: \emph{Concurrency Theory---25th International Conference
 (CONCUR 2014)}, 
September 2--5 2014, Rome, Italy.}
%
%
\author[ut]{Natsuki Urabe}

\author[ut]{Ichiro Hasuo}


\address[ut]{Department of Computer Science, Graduate School of Information Science and Technology, the University of Tokyo, 
7-3-1, Hongo, Bunkyo-ku, Tokyo, Japan}

%


\begin{abstract}
We introduce notions of simulation between semiring-weighted automata as
models of quantitative systems. Our simulations are instances of the
categorical/coalgebraic notions previously studied by Hasuo---hence 
soundness against language inclusion comes for free---but are concretely presented as
matrices that are subject to linear inequality constraints.
Pervasiveness of these formalisms allows us to exploit  existing
 algorithms in: searching for a simulation, and hence
 verifying quantitative correctness 
that is formulated as language inclusion.
  Transformations of automata that aid search
 for simulations are introduced, too. 
This verification workflow is implemented for the plus-times
 and max-plus semirings. Furthermore, an extension to weighted \emph{tree}
 automata is presented and implemented.
\end{abstract}


\end{frontmatter}

\section{Introduction}
 \emph{Quantitative} aspects of various systems are more and
 more emphasized in recent verification scenarios. Probabilities in
 randomized or fuzzy systems are a classic example; utility in economics
 and game theory is another. Furthermore, now that many computer systems are
 integrated into physical ambience---realizing so-called
 \emph{cyber-physical systems}---physical
 quantities like energy consumption are necessarily taken into account.


\subsection{Semiring-Weighted Automata}
It is standard in the concurrency community to model 
 such quantitative systems by state-transition systems in which
 \emph{weights} are assigned to their states and/or transitions. The
 semantics of such systems  varies, however,
 depending on the interpretation
 of weights. If they are probabilities, they are accumulated by
 $\times$ along a path and summed across different paths; if
 weights are  (worst-case) costs, they are summed up along
 a path and we would take $\max$ across different paths. 

The algebraic structure of \emph{semirings}  then arises as a 
 uniform mathematical language for  different notions of
 ``weight,'' as is widely acknowledged in the community. The subject of
 the current study is state-based systems with labeled transitions, in
 which each transition is assigned a weight from a prescribed
 semiring $\mathcal{S}$. We shall call them \emph{$\mathcal{S}$-weighted
 automata}; and we are
more specifically interested in the (weighted, finite) \emph{language
inclusion} problem  and 
a \emph{simulation-based} approach to it.


\subsection{Language Inclusion}
Let $\mathcal{A}$ be an $\mathcal{S}$-weighted automaton with labels 
from an alphabet $\Sigma$. It assigns  to each word $w\in
\Sigma^{*}$ a weight taken from $\mathcal{S}$---this is much like a
(purely) probabilistic automaton assigns a probability to each word. Let
us denote this function by $L(\mathcal{A})\colon \Sigma^{*}\to
\mathcal{S}$ and call it the \emph{(weighted) language} of
$\mathcal{A}$ by analogy with classic automata theory. The \emph{language
inclusion} problem  $L(\mathcal{A})\sqsubseteq L(\mathcal{B})$ asks if:
 $L(\mathcal{A})(w)\sqsubseteq L(\mathcal{B})(w)$ for each word $w\in\Sigma^{*}$,
where  $\sqsubseteq$ is a natural order on the semiring $\mathcal{S}$.

It is not hard to see that language inclusion 
$L(\mathcal{A})\sqsubseteq L(\mathcal{B})$ has numerous applications in
verification. In a typical scenario, one of $\mathcal{A}$ and $\mathcal{B}$ is a
model of a \emph{system} and the other expresses \emph{specification};
and $L(\mathcal{A})\sqsubseteq L(\mathcal{B})$ gives the definition of
``the system meeting the specification.'' More concrete examples are as follows.
\begin{itemize}
 \item $\mathcal{S}$ represents probabilities;  $\mathcal{A}$
models a system; and $\mathcal{B}$ expresses the specification that
certain bad behaviors---identified with words---occur with a certain
probability. Then $L(\mathcal{A})\sqsubseteq L(\mathcal{B})$ is
 a \emph{safety} statement:  each  bad behavior occurs in $\mathcal{A}$
 at most as likely as in $\mathcal{B}$.
 \item $\mathcal{S}$ represents profit,  $\mathcal{A}$  is a
       specification and $\mathcal{B}$ is  a
system. Then $L(\mathcal{A})\sqsubseteq L(\mathcal{B})$
 guarantees the minimal profit yielded by the system
 $\mathcal{B}$. 
 \item  There are other properties   reduced to  language
	inclusion in a less trivial manner. An example is \emph{probable
	innocence}~\cite{reiter98crowdsanonymity}, a quantitative notion
	of anonymity. See~\cite{hasuo10probabilisticanonymity}.
\end{itemize}


\subsection{Simulation}
Direct check of language inclusion is
simply infeasible because there are infinitely many words
$w\in\Sigma^{*}$. One finitary proof method---well-known for nondeterministic (i.e.\ possibilistic)
systems---is by \emph{(forward or backward) simulations}, whose
systematic study is initiated in~\cite{lynch95forwardand}.
       In the nondeterministic setting, a simulation $R$ is a relation
       between states of $\mathcal{A}$ and $\mathcal{B}$ that witnesses
       ``local language inclusion''; moreover, from the coinductive way
       in which it is defined, a simulation persistently witnesses local language
       inclusion---ultimately yielding (global) language inclusion. This property---existence of a simulation implies language inclusion---is called \emph{soundness}.

\subsection{Contribution: Weighted Forward/Backward Simulations by
  Matrices}
In this paper we extend this simulation approach to language inclusion~\cite{lynch95forwardand} 
to the quantitative setting of semiring-weighted automata. Our notions
of (forward and backward) weighted simulation  are not given by
relations, but by \emph{matrices} with entries from a
semiring $\mathcal{S}$. 

Use of matrices in automata theory is  classic---in fact our
framework instantiates to that in~\cite{lynch95forwardand} when we take
as $\mathcal{S}$ the Boolean semiring. This is not how we arrived here; conversely, the current results are obtained as
instances of a more general theory of \emph{coalgebraic
simulations}~\cite{hasuo06genericforward,hasuo07generictrace,hasuo10genericforward}. 
There various systems are identified with a categorical construct of
\emph{coalgebras} in a Kleisli
category; and forward and backward simulations
are characterized as lax/oplax morphisms between coalgebras. A generic
soundness result (with respect to language/trace inclusion) is also
proved in the general categorical terms. 

 This paper is devoted to concrete presentations of these categorical
notions  by matrices, and to their application to
actual verification of quantitative systems.  Presentation by matrices
turns out to be an advantage: a simulation is now a matrix $X$ that
satisfies certain \emph{linear inequalities}; and existence of such
$X$---i.e.\ feasibility of linear inequalities---is so common a problem
in many fields that there is a large body of existing work that is waiting to be
applied. For example \emph{linear programming (LP)} can be exploited for the
plus-times semiring for probabilities; and there are  algorithms  proposed for other
semirings such as the max-plus (tropical) one.

Our (mostly semiring-independent) workflow is as follows. A verification
goal is formulated as language inclusion $\lang(\mathcal{A})\sqsubseteq
\lang(\mathcal{B})$, which we aim to establish by finding a forward or
backward simulation from $\mathcal{A}$ to $\mathcal{B}$. Soundness of
simulations follows from the general result
in~\cite{hasuo06genericforward}. A simulation we seek for is a matrix
subject to certain linear inequalities, existence of which is checked
by  various algorithms that
exist for different semirings. We implemented this workflow for the
plus-times and max-plus semirings.

This simulation-based method is sound but not necessarily complete
with respect to language inclusion.  Therefore we introduce
transformations of weighted automata---called \emph{(forward/backward)
partial execution}---that potentially create matrix simulations.  Via our
equivalence results between our matrix simulation and some
known ones (including the one
in~\cite{chatterjee10quantitativelanguages}), 
 the partial
execution transformations potentially create those  simulations, too.

Compared to the earlier version~\cite{UrabeH14} of this paper, 
the current version additionally contains the following materials.
\begin{itemize}
 \item Section~\ref{sec:matSimPolyF} is added, where we exploit the
       coalgebraic theory behind and generalize matrix simulation 
       from
       weighted (word) automata to 
       weighted \emph{tree} automata. We describe the definition 
       of
       forward partial execution in categorical terms, too, so that it transfers
       to weighted tree automata. We also have a preliminary implementation. 
 \item We now have more extensive discussions of  related work,
       including~\cite{beal05equivalenceZautomata,esik11categorysimulations,ciric12bisimulationfuzzy}
       of which we were not aware before.
 \item We conducted  experiments again with a faster machine,
       enlarging the size of problem instances that can be handled.
 \item We have some examples that were absent in the previous version~\cite{UrabeH14}.
 \item Concrete description of the procedure (forward/backward)
partial execution is included.
 \item Some proofs were omitted in~\cite{UrabeH14} for space reasons;
       they are present here.
\end{itemize}


\subsection{Organization of the paper}
In~Section~\ref{sec:prelim} that is devoted to preliminaries, we define semiring-weighted automata,
characterize them in coalgebraic terms and recap the coalgebraic
theory in~\cite{hasuo06genericforward}. These are combined to yield
the notion of simulation matrix
in~Section~\ref{sec:simulationMatrices}. In~Section~\ref{sec:partExec} partial
execution transformations of automata are described and proved
correct. The framework  obtained so far is applied to the plus-times and max-plus
semirings, in~Section~\ref{sec:plustimesWeightedAutomata}
and~Section~\ref{sec:maxplusWeightedAutomata}, respectively. There our
proof-of-concept implementations~\cite{implPtMpTree}
 and relationship to other known
simulation notions are discussed, too. 
In Section~\ref{sec:matSimPolyF} we generalize the framework so far from 
words to trees: the generalization is straightforward---thanks to the 
coalgebraic backend---although linearity of constraints, as well as backward
partial execution, is lost.
In Section~\ref{sec:relatedWork} (and in earlier sections) we discuss related work; in~Section~\ref{sec:concl} we
conclude.

\section{Preliminaries}\label{sec:prelim}
We review
the generic theory of traces and simulations
in~\cite{hasuo07generictrace,hasuo06genericforward} that is based on
$(T,F)$-systems, which
will eventually lead to the notion of simulation matrix in~Section~\ref{sec:simulationMatrices}.

\subsection{Semiring-Weighted Automata}\label{subsec:semiringWA}
The notion of semiring-weighted automaton is parametrized by 
a semiring $\mathcal{S}$. For our purpose of applying coalgebraic theory
in~\cite{hasuo06genericforward,hasuo07generictrace}, we impose the
following properties.
The notion seems to be new, though hardly original.
Indeed, a similar notion called \emph{complete ordered semiring} is introduced and used in~\cite[Chapter 1]{droste09handbookof}.
It is different from ours in that only  countable additions are allowed in our definition while arbitrary additions
are allowed in the notion in~\cite[Chapter 1]{droste09handbookof}.



\begin{mydefinition}\label{def:commCppoSemiring}
A \emph{commutative cppo-semiring} is a tuple $\mathcal{S}=(S,
+_{\mathcal{S}},0_{\mathcal{S}},\times_{\mathcal{S}},1_{\mathcal{S}},\sqsubseteq)$ that satisfies the following conditions.
\begin{itemize}
\item $(S,+_{\mathcal{S}},0_{\mathcal{S}},\times_{\mathcal{S}},1_{\mathcal{S}})$ is a semiring in which $\times_{\mathcal{S}}$, in addition to $+_{\mathcal{S}}$, is commutative.
\item A relation $\sqsubseteq$ is a partial order on $S$ and $(S,\sqsubseteq)$ is
      \emph{$\omega$-complete}, i.e.\ an increasing chain $s_{0}\sqsubseteq
      s_{1}\sqsubseteq\cdots$ has a supremum.
\item Any element $s\in S$ is \emph{positive} in the sense that $0_{\mathcal{S}}\sqsubseteq
      s$.
\item Addition $+_{\mathcal{S}}$ and multiplication $\times_{\mathcal{S}}$ are monotone with
      respect to $\sqsubseteq$.
\end{itemize}
\end{mydefinition}
It follows from positivity and $\omega$-completeness that countable sum
can be straightforwardly defined in a commutative cppo-semiring
$\mathcal{S}$. We will use this fact throughout the paper.

\begin{myexample}[semirings $\spt,\smp,\mathcal{B}$]\label{example:semirings}
\sloppy
 The \emph{plus-times semiring} $\spt=([0,\infty],+,0,\times,1,\leq)$
 is a commutative cppo-semiring, where
 $+$ and $\times$ are usual addition and multiplication of 
 real numbers. This is
 the semiring that we will use for modeling probabilistic
 branching. Specifically, probabilities of successive transitions are
 accumulated using $\times$, and those of different branches are
 combined with $+$.

\fussy

The \emph{max-plus semiring}
 $\smp=([-\infty,\infty],\max,-\infty,+,0,\leq)$---also sometimes
 called the \emph{tropical semiring}~\cite{pin98tropicalsemirings}---is
 also a commutative cppo-semiring.
 Here a number $r\in [-\infty,\infty]$ can be understood as
 (best-case) \emph{profit}: they are summed up along a path, and an optimal one
 ($\max$) is chosen among different branches. Another possible
 understanding of $r$ is as (worst-case) \emph{cost}. The unit for the semiring addition $\max$ is given by $-\infty$; since
 it must also be a zero element of the semiring multiplication $+$, we
 define $(-\infty) + \infty=-\infty$.
In the two examples $\spt$ and $\smp$ we added $\infty$ so that they become $\omega$-complete.

Finally, the \emph{Boolean semiring} $\mathcal{B}=(\{0,1\}, \lor, 0, \land, 1,\le)$ is  an
example that is qualitative rather than quantitative.
\end{myexample}





\begin{mydefinition}[$\mathcal{S}$-weighted
 automaton, weighted language]\label{def:SWeightedAutom}
Let
 $\mathcal{S}=(S,+_{\mathcal{S}},0_{\mathcal{S}},\times_{\mathcal{S}},
\\
1_{\mathcal{S}},\sqsubseteq)$ be a commutative
 cppo-semiring. An \emph{$\mathcal{S}$-weighted automaton} $\mathcal{A}
 = (Q, \Sigma, M, \alpha, \beta)$ consists of a countable state space
 $Q$, a
 countable
 alphabet $\Sigma$, transition matrices
$M(a)\in S^{Q\times Q}$ for all $a\in \Sigma$, 
the initial row vector $\alpha \in S^{Q}$ and the final column vector $\beta \in S^{Q}$. 

Let $x,y\in Q$ and $a\in\Sigma$. We write $\alpha_x$ and $\beta_x$ for the $x$-th entry
 of $\alpha$ and $\beta$, respectively, and $M(a)_{x,y}$ for the
 $(x,y)$-entry of the matrix $M(a)$. Note that these entries are
 all elements of the semiring $\mathcal{S}$.

\sloppy
An $\mathcal{S}$-weighted  automaton $\mathcal{A}=(Q, \Sigma, M, \alpha,
 \beta)$ yields a \emph{weighted language} 
$\lang (\mathcal{A})\colon
 \Sigma^{*}\to \mathcal{S}$. It is  given by the following  multiplication of
 matrices and vectors.


\begin{equation}\label{eq:weightForWords}
 \lang (\mathcal{A})(w) \;:=\; \alpha\cdot M(a_{1}) \cdot\;
\cdots\;\cdot M(a_{k})\cdot \beta\qquad 
 \text{for each $w = a_{1}\cdots a_{k} \in \Sigma^{*}$.}
\end{equation}


\fussy
\end{mydefinition}
\begin{myremark}[size of state space $Q$]
In the above definition, differently from the usual definitions of weighted automata (e.g.~\cite{droste07weightedautomata}),
we allow $Q$ to be infinite. This is because in Theorem~\ref{thm:finalCoalg},
such an infinite-state automaton arises as a final coalgebra that captures the trace semantics of weighted automata.

However, at the same time, we require $Q$ to be at most countable.
This is so that matrix multiplications in~(\ref{eq:weightForWords})
are well-defined.
In fact, it is possible to remove this restriction 
by  instead requiring the support of each transition to be countable.
We  note that in this case, some definitions in the later sections become slightly complicated:
for example, a forward or backward simulation matrix $X$ in
 Definition~\ref{def:simMat} will be required to have 
only a countable number of non-zero elements in each its row.
Therefore, mostly for simplicity of presentation,
we stick to state spaces $Q$ that are at most countable.
\end{myremark}

Our  interest is in establishing language inclusion between
two weighted automata.

\begin{mydefinition}[language inclusion]\label{def:languageIncl}
 We write $\lang(\mathcal{A})\sqsubseteq \lang(\mathcal{B})$ if, for
 each $w\in \Sigma^{*}$, $\lang(\mathcal{A})(w)\sqsubseteq
 \lang(\mathcal{B})(w)$. The last $\sqsubseteq$ is the order of $\mathcal{S}$.
\end{mydefinition}

\subsection{Coalgebraic Modeling of Semiring-Weighted Automata}\label{subsec:coalgModelingWA}
Here we characterize semiring-weighted automata as   instances of a
generic coalgebraic model of branching systems---so-called
\emph{$(T,F)$-systems} with parameters
$T,F$~\cite{hasuo07generictrace,hasuo06genericforward}. 




\begin{mydefinition}[$(T,F)$-system]\label{def:TFSys}
 Let  $T$ be a monad and  $F$ be a  functor, both on the category $\mathbf{Sets}$ of
 sets and functions. A \emph{$(T,F)$-system} is a triple

\begin{displaymath}
\mathcal{X}
 \;=\; \bigl(\;X,\;\; s\colon \{\bullet\}\to TX,\;\;c\colon X\to
TFX\;\bigr)
\end{displaymath}
of a set $X$ (the \emph{state space}), and functions $s$ (the \emph{initial
 states}) and $c$ (the \emph{dynamics}). 
\end{mydefinition}
This modeling is coalgebraic~\cite{jacobs12introductionto} in the sense that $c$ is
so-called a $TF$-coalgebra.
More precisely,
a $(T,F)$-system $(X,s,c)$ is a \emph{pointed coalgebra} in a Kleisli category $\Kl(T)$ though 
what is ``pointed'' by $s:\{\bullet\}\to TX$ is not necessarily an element in $X$ but a distribution over $X$.
In the definition we have two parameters $T$ and $F$. Let us forget
about their categorical
structures (a \emph{monad} or a \emph{functor}) for a moment
and  think of them simply as
constructions on sets. Intuitively speaking, $T$  specifies what kind 
of \emph{branching} the systems in question exhibit; and $F$ specifies
a type of \emph{linear-time behaviors}. Here are some examples; in the
example $F=1+\Sigma\times(\place)$ the only element of $1$ is denoted by
$\checkmark$ (i.e.\ $1=\{\checkmark\}$).
\begin{center}
 \begin{tabular}{c||c}
  $T$ & ``branching''
 \\\hline
  $\pow$ & \; nondeterministic
 \\
  $\dist$ & probabilistic 
   \\
  $\mss$ & $\mathcal{S}$-weighted 
 \end{tabular}
 \;
 \begin{tabular}{c||c}
  $F$ & ``linear-time behavior''
 \\\hline
  $1+\Sigma\times(\place)$ & \; $\to\checkmark$ \; or \;
      $\stackrel{a}{\rightarrow}$ (where $a\in\Sigma$) 
 \\
  $(\Sigma+(\place))^*$  &
\multicolumn{1}{l}{\quad words over terminals ($a\in\Sigma$)}
   \\
   & \quad \& nonterminals, suited for CFG~\cite{hasuoJ05contextfree}
 \end{tabular}
\end{center}
 The above examples of a monad $T$---the \emph{powerset monad} $\mathcal{
 P}$, the \emph{subdistribution
 monad} $\mathcal{D}$, and the \emph{$\mathcal{S}$-multiset monad}
 $\mss$ for  $\mathcal{S}$---are described as follows.
\begin{equation}\label{eq:monads}
\begin{aligned}
 \pow X &=\{X'\mid X' \subseteq X\} \qquad
 \dist X = \{f:X\to [0,1]\mid \textstyle\sum_{x\in X} f(x)\leq 1\} \\
 \mss X &=\{f :X \to S \mid \mathrm{supp}(f) \text{ is countable}\}
\end{aligned}
\end{equation}
Here $\mathrm{supp}(f)=\{x\in X\mid f(x)\neq 0_{\mathcal{S}}\}$. 
Countable support in $\mss$ is a technical requirement so that
the multiplication of $\mss$ is well-defined and therefore
we can define composition $\odot$ of Kleisli arrows (Definition~\ref{def:KlArrow}).

\begin{myremark}\label{rem:omegaCompletenessToKappa}
In Definition~\ref{def:commCppoSemiring}, we required the semiring
$\mathcal{S}$ to be $\omega$-complete.  In fact, it is possible to
strengthen this restriction by requiring $\kappa$-completeness where
$\kappa$ is an arbitrary ordinal, or even directed completeness (that
 essentially amounts to $\kappa$-completeness for any $\kappa$; see e.g.~\cite[\S{}2.2.4]{AbramskyJ94domaintheory}).  This allows us to define $\mss$ so
that $f\in\mss X$ (and transitions of weighted automata) can have an
uncountable support.
However, weighted automata have many applications even if we require their transitions to have countable supports.
Hence in many cases it suffices to assume $\omega$-completeness; we
 shall do so also for the sake of simplicity of presentation.
\end{myremark}

It should not be hard to see that a $(T,F)$-system models a
state-based
system with $T$-branching and $F$-linear-time behaviors.
 For example, when $T=\pow$  and $F=1+\Sigma\times(\place)$, $s\colon
 \{\bullet\}\to \pow X$ represents the set of initial states and 
 $c\colon X\to \pow (1+\Sigma\times X)$ represents one-step
 transitions---that $\checkmark\in c(x)$ means $x$ is accepting ($x\to\checkmark$), and
 $(a,x')\in c(x)$ means there is a transition $x\stackrel{a}{\to}x'$. Overall, a
 $(\pow, 1+\Sigma\times(\place))$-system is nothing but a
 nondeterministic automaton.

Analogously we obtain the following, by the definition
of $\mss$ in~(\ref{eq:monads}). 

\begin{myproposition}[weighted automata as $(T,F)$-systems]\label{prop:SemiringWeightedAutomAsCoalg}
Let $\mathcal{S}$ be a commutative cppo-semiring. 
There is a bijective
 correspondence between: 1)   $\mathcal{S}$-weighted automata (Definition~\ref{def:SWeightedAutom}); and 2)
 $\bigl(\mss,1+\Sigma\times(\place)\bigr)$-systems whose state spaces
 are at most countably infinite. 

Concretely, 
an $\mathcal{S}$-weighted automaton $\mathcal{A}=(Q, \Sigma, M, \alpha,
 \beta)$
gives rise to an $\bigl(\mss,1+\Sigma\times(\place)\bigr)$-system
 $\mathcal{X_{A}}=(Q,s_{\mathcal{A}},c_{\mathcal{A}})$ defined as follows.
$s_{\mathcal{A}}\colon \{\bullet\}\to\mss Q$ is given by $s_{\mathcal{A}}(\bullet)(x)=\alpha_x$; and $c_{\mathcal{A}}\colon Q\to\mss (1+\Sigma\times Q)$ is given by
 $c_{\mathcal{A}}(x)(\checkmark)=\beta_{x}$ and 
 $c_{\mathcal{A}}(x)(a,y)=M(a)_{x,y}$. \qed
\end{myproposition}



\subsection{Coalgebraic Theory of Traces and Simulations}\label{subsec:coalgTheoryTraceSim}
\label{subsec:coalgebraicTraceAndSim}
We review the  theory of traces and simulations
in~\cite{hasuo07generictrace,hasuo06genericforward} that is based on
$(T,F)$-systems. In presentation we
 restrict to $T=\mss$ and $F=1+\Sigma\times (\place)$ for simplicity.



\subsubsection{Kleisli Arrows}
One notable success of coalgebra was a uniform characterization, in
terms of the same categorical diagram, of \emph{bisimulations} for various kinds
of systems (nondeterministic, probabilistic, etc.)~\cite{jacobs12introductionto}. This works quite well for
branching-time process semantics. For linear-time semantics---i.e.\ trace
semantics---it is noticed in~\cite{power99categorytheory} that so-called
a \emph{Kleisli category}, in place of the category $\Sets$, gives a
suitable base category for coalgebraic treatment. This idea---replacing functions $X\to Y$ with
\emph{Kleisli arrows} $X\kto Y$ and drawing the same diagrams---led to the
development
in~\cite{hasuo07generictrace,hasuo06genericforward,hasuo10genericforward}
of an extensive theory of traces and simulations. 
The notion of  Kleisli arrow is parametrized by a monad $T$: a
$T$-Kleisli arrow $X\kto_{\,T} Y$ (or simply $X\kto Y$) is defined to be a function $X\to TY$, 
hence represents a ``$T$-branching function from $X$ to $Y$.''


We restrict to $T=\mss$ for simplicity of presentation. 
 An $\mss$-Kleisli arrow $f\colon X\kto Y$ below is ``an
$\mathcal{S}$-weighted function from $X$ to $Y$.'' In particular, for
each $x\in X$ and $y\in Y$ it assigns a \emph{weight} $f(x)(y)\in S$.

\begin{mydefinition}[Kleisli arrow]\label{def:KlArrow}
Let $X,Y$ be sets. An \emph{$\mss$-Kleisli arrow} (or simply a
 \emph{Kleisli arrow}) from $X$ to
 $Y$,  denoted by $X\kto Y$, is a function from $X$ to $\mss Y$. 

We list some special Kleisli arrows: $\eta_{X}$, $g\odot f$ and $Jf$.
\begin{itemize}
\item For each set $X$, the \emph{unit arrow} $\eta_X\colon X\kto X$ is
			  given by:
$\eta(x)(x)=1_{\mathcal{S}}$; and $\eta(x)(x')=0_{\mathcal{S}}$ for $x'\neq x$. Here $0_{\mathcal{S}}$ and $1_{\mathcal{S}}$
       are  units in the semiring
      $\mathcal{S}$.
\item For consecutive Kleisli arrows $f:X\kto Y$ and $g:Y\kto Z$, their
      \emph{composition} $g\odot f:X\kto Z$ is given as follows:
\begin{displaymath}
 (g\odot f)(x)(z)
\;:=\;
\textstyle\sum_{y\in \mathrm{supp}(f(x))} f(x)(y)\times_{\mathcal{S}} g(y)(z)\enspace.
\end{displaymath}
Since
      $\mathrm{supp}(f(x))$ is countable, the above sum in a
      cppo-semiring $\mathcal{S}$ is well-defined. 
\item For a (usual) function $f:X\to Y$, its \emph{lifting} to a Kleisli
      arrow $Jf:X\kto Y$ is given by $Jf=\eta_Y\circ f$. Here we
      identified $\eta_{Y}\colon Y\kto Y$ with a function
      $\eta_{Y}\colon Y\to \mss Y$.
\end{itemize}
\end{mydefinition}
Categorically speaking: the first two ($\eta$ and $\odot$) organize
Kleisli arrows as a category (the \emph{Kleisli category} $\Kl(\mss)$);
and the third gives a functor $J\colon \Sets\to \Kl(\mss)$ that is
identity on objects.

In Proposition~\ref{prop:SemiringWeightedAutomAsCoalg} we characterized 
an $\mathcal{S}$-weighted automaton $\mathcal{A}$ in coalgebraic terms. Using Kleisli
arrows it is presented as a triple 

\begin{equation}\label{eq:weightedAutomInKlArrows}
 \mathcal{X_{A}} \;=\;
 \bigl(\;Q,\;\; s_{\mathcal{A}}\colon \{\bullet\} \longkto Q,\;\; c_{\mathcal{A}}\colon Q \longkto 1+\Sigma\times Q\;\bigr)\enspace.
\end{equation}


\subsubsection{Generic Trace Semantics}\label{subsec:GenericTraceSemantics}
In~\cite{hasuo07generictrace}, for  monads $T$ with a suitable order, a final coalgebra in $\Kl(T)$ is
identified. It (somehow interestingly) coincides with an initial algebra in
$\Sets$. Moreover, the universality of this final coalgebra is shown
to capture natural notions of (finite) \emph{trace semantics} for a variety of
branching systems---i.e.\ for different $T$ and $F$. What is important
for the current work is the fact that the weighted language $\lang(\mathcal{A})$
in~(\ref{eq:weightForWords}) is an instance of this generic trace
semantics, as we will show in Theorem~\ref{thm:traceSem}.

\def\LS{0.9}
 \begin{equation} \label{eq:traceSem}  
\vcenter{   \begin{xy}
  \xymatrix@R=1.5em@C+4em{
  {1+\Sigma\times X} \ar@{}[dr]|{\scalebox{\LS}{$=$}} \kar@{-->}[r]^{\scalebox{\LS}{$1+\Sigma\times(\tr(c))$}} & {1+\Sigma\times\Sigma^*} \\
  {X} \kar[u]_{\scalebox{\LS}{$c$}} \kar@{-->}[r]_{\scalebox{\LS}{$\tr(c)$}} & {\Sigma^*}
   \kar[u]^{\scalebox{\LS}{final}}_{\scalebox{\LS}{$J([\mathsf{nil},\mathsf{cons}]^{-1})$}} 
  \\
  {\{\bullet\}}
    \kar[u]_{\scalebox{\LS}{$s$}}
    \kar@/_.7pc/[ru]_(.6){\scalebox{\LS}{$\tr(\mathcal{X})$}}
 }
  \end{xy}
}
 \end{equation}
We shall state the   results in~\cite{hasuo07generictrace} on
coalgebraic traces, restricting again to 
 $T=\mss$ and $F=1+\Sigma\times(\place)$ for simplicity.
In the diagram~(\ref{eq:traceSem}) above, composition of Kleisli arrows are given by $\odot$ in
Definition~\ref{def:KlArrow}; $J$ on the right is the lifting in Definition~\ref{def:KlArrow}; and $\mathsf{nil}$ and $\mathsf{cons}$ are the obvious constructors of
words in $\Sigma^{*}$. 
The top arrow $1+\Sigma\times(\tr(c))$ is the
functor $1+\Sigma\times (\place)$ on $\Sets$, lifted to the Kleisli
category $\Kl(\mss)$, and applied to the Kleisli arrow $\tr(c)$; its
concrete description is as follows. See~\cite{hasuo07generictrace} for more details.

%
%
\begin{mydefinition}\label{def:functorActionOnArrows}
  For a Kleisli arrow $f\colon X\kto Y$, its lifting $1+\Sigma\times f \colon 1+\Sigma\times X\kto 1+\Sigma\times Y$ is defined as follows:
\[
  (1+\Sigma\times f)(*)(t) = \begin{cases} 1_{\mathcal{S}} & (t=*) \\ 0_{\mathcal{S}} & (\text{otherwise}) \end{cases} \;\;\;
  (1+\Sigma\times f)(a,x)(t) = \begin{cases} f(x)(y) & (t=(a,y)) \\ 0_{\mathcal{S}} & (\text{otherwise}) \end{cases}
\] 
\end{mydefinition}

\begin{mytheorem}[final coalgebra in $\Kl(\mss)$]\label{thm:finalCoalg}
 Given any set $X$ and any Kleisli arrow $c\colon X\kto
	1+\Sigma\times X$,  there exists a unique Kleisli arrow ${\sf
	tr}(c)$ that makes the top square in the  diagram~(\ref{eq:traceSem}) commute. 
 \qed
\end{mytheorem}


\begin{mydefinition}[$\tr(\mathcal{X})$]\label{def:traceSemForTFSys}
 Given an
 $\bigl(\mss,1+\Sigma\times(\place)\bigr)$-system $\mathcal{X}=(X,s,c)$
 (this is on the left in the diagram~(\ref{eq:traceSem})), 
 its component $c$ induces an arrow $\tr(c)\colon X\kto \Sigma^{*}$
 by Theorem~\ref{thm:finalCoalg}. We define $\tr(\mathcal{X})$ to be the
 composite $\tr(c)\odot s$  (the bottom triangle in the
 diagram~(\ref{eq:traceSem})), and call it the \emph{trace semantics} of $\mathcal{X}$.
\end{mydefinition}


\begin{mytheorem}[weighted language as trace semantics]\label{thm:traceSem}
 Let $\mathcal{A}$ be an $\mathcal{S}$-weighted automaton.
For $\mathcal{X_{A}}=(Q,s_{\mathcal{A}},
 c_{\mathcal{A}})$ induced by $\mathcal{A}$
 in~(\ref{eq:weightedAutomInKlArrows}), its trace semantics 
 $\tr(\mathcal{X_{A}})\colon\{\bullet\}\kto \Sigma^{*}$---identified with a function $\{\bullet\}\to \mss
 \Sigma^{*}$, hence with a function $\Sigma^{*}\to \mathcal{S}$---coincides with
the weighted language $\lang(\mathcal{A})\colon \Sigma^{*}\to \mathcal{S}$
 in~(\ref{eq:weightForWords}).
\qed
\end{mytheorem}
In the last theorem we need that $\Sigma^{*}$ is countable; this is why
we assumed that $\Sigma$ is countable in
Definition~\ref{def:SWeightedAutom}. Henceforth we do not distinguish
$\lang(\mathcal{A})$ and $\tr(\mathcal{X_A})\colon
\{\bullet\}\kto \Sigma^{*}$.

\subsubsection{Forward and Backward \emph{Kleisli} simulations}

In~\cite{hasuo06genericforward}, the classic results in~\cite{lynch95forwardand}  on forward and
backward simulations---for (nondeterministic) labeled transition
systems---are generalized to $(T,F)$-systems. Specifically,  forward and backward simulations are characterized
as \emph{lax/oplax coalgebra homomorphisms} in a Kleisli category; and
\emph{soundness}---their existence witnesses trace inclusion---is proved
once for all in a general categorical setting.

As before,  we present those notions and results
in~\cite{hasuo06genericforward} restricting to $T=\mss$ and
$F=1+\Sigma\times (\place)$. If $T=\pow$ and $F=1+\Sigma\times(\place)$ they instantiate to the results
in~\cite{lynch95forwardand}.

\begin{mydefinition}[Kleisli simulation]\label{def:ksim}
Let $\mathcal{X} = (X,s,c)$ and $\mathcal{Y} = (Y,t,d)$ be
 $(\mss,1+\Sigma\times(\place))$-systems (cf.\ Definition~\ref{def:TFSys},
 Proposition~\ref{prop:SemiringWeightedAutomAsCoalg}
 and~(\ref{eq:weightedAutomInKlArrows})).
\begin{enumerate}
\item A \emph{forward (Kleisli) simulation} from $\mathcal{X}$ to $\mathcal{Y}$ is a Kleisli arrow $f : Y \kto X$ such that $s \sqsubseteq f \odot t$ and $c \odot f \sqsubseteq (1+\Sigma\times f)  \odot d$. 
See Figure~\ref{fig:coalgSim}. \label{def:ksimfwd}
\item A \emph{backward simulation} from $\mathcal{X}$ to $\mathcal{Y}$ is a Kleisli arrow $b : X \kto Y$ such that $s \odot b \sqsubseteq t$ and $(1+\Sigma\times b)  \odot c \sqsubseteq d \odot b$. 
\item A \emph{forward-backward simulation} from $\mathcal{X}$ to
      $\mathcal{Y}$ consists of: a $(T,F)$-system $\mathcal{Z}$;
      a forward simulation $f$  from $\mathcal{X}$ to $\mathcal{Z}$; and
      a  backward simulation $b$ from $\mathcal{Z}$ to $\mathcal{Y}$. 

      
\item A \emph{backward-forward simulation} from $\mathcal{X}$ to
      $\mathcal{Y}$ consists of: a $(T,F)$-system $\mathcal{Z}$;
      a backward simulation $b$  from $\mathcal{X}$ to $\mathcal{Z}$; and
      a  forward simulation $f$ from $\mathcal{Z}$ to $\mathcal{Y}$. 

      
\end{enumerate}

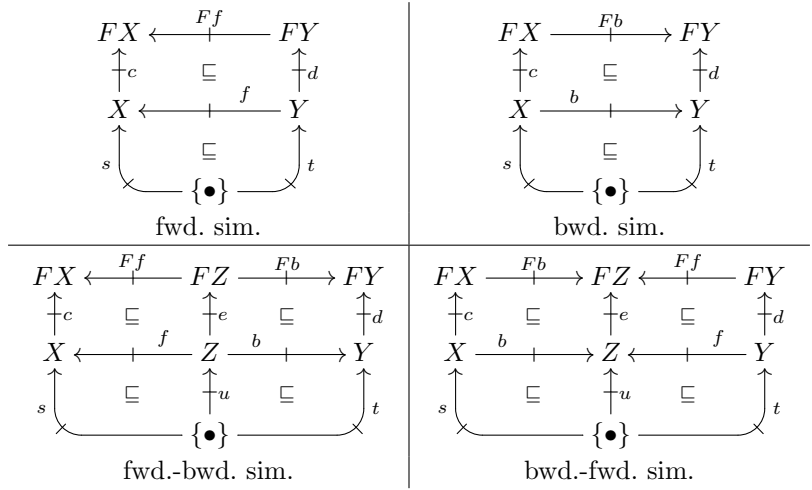
\begin{figure}[htbp]
\begin{center}
 \begin{tabular}{c|c}
 \begin{xy}
 \xymatrix@R=1.6em@C=1.2em{
 {F X} \ar@{}[drr]|{\sqsubseteq}  & & {F Y} \kar[ll]_{F f } \\
 {X} \kar[u]_{c} \ar@{}[drr]|{\sqsubseteq} & & {Y} \kar[u]_{d} \ar_(.3){f}|-*\dir{|}[ll]  \\
 {} \ar@{}[ru]|(.10)*\dir{-} & {\{\bullet\}} \ar`l[lu][lu]^(0){s}  \ar`r[ru][ru]_(0){t}  & \ar@{}[lu]|(.10)*\dir{-} }
 \end{xy}
& 
\begin{xy}
 \xymatrix@R=1.6em@C=1.2em{
 {F X} \ar@{}[drr]|{\sqsubseteq} \kar[rr]^{F b } & & {F Y}  \\
 {X} \kar[u]_{c} \ar@{}[drr]|{\sqsubseteq} \ar^(.3){b}|-*\dir{|}[rr]  & & {Y} \kar[u]_{d}  \\
 {} \ar@{}[ru]|(.10)*\dir{-} & {\{\bullet\}} \ar`l[lu][lu]^(0)s \ar`r[ru][ru]_(0)t  & \ar@{}[lu]|(.10)*\dir{-} }
 \end{xy}
 \\ 
  fwd.\ sim.
  & 
 bwd.\ sim. 
  \\ \hline
 \begin{xy}
 \xymatrix@R=1.6em@C=3.6em{
 {F X} \ar@{}[dr]|{\sqsubseteq}  & {F Z} \ar@{}[dr]|{\sqsubseteq} \kar[l]_{F f } \kar[r]^{F b } & {F Y}  \\
 {X} \kar[u]_{c} \ar@{}[dr]|{\sqsubseteq} & {Z} \kar[u]_{e} \ar@{}[dr]|{\sqsubseteq} \ar_(.3){f}|-*\dir{|}[l] \ar^(.3){b}|-*\dir{|}[r]  & {Y} \kar[u]_{d}  \\
 {} \ar@{}[ruu]|(.04)*\dir{-} & {\{\bullet\}} \ar`l[lu][lu]^(0)s \kar[u]_{u} \ar`r[ru][ru]_(0)t  & \ar@{}[luu]|(.04)*\dir{-} }
 \end{xy}
 &
  \begin{xy}
 \xymatrix@R=1.6em@C=3.6em{
 {F X} \ar@{}[dr]|{\sqsubseteq} \kar[r]^{F b } & {F Z} \ar@{}[dr]|{\sqsubseteq}   & {F Y} \kar[l]_{F f } \\
 {X} \kar[u]_{c} \ar@{}[dr]|{\sqsubseteq} \ar^(.3){b}|-*\dir{|}[r] & {Z} \kar[u]_{e} \ar@{}[dr]|{\sqsubseteq}   & {Y} \kar[u]_{d} \ar_(.3){f}|-*\dir{|}[l]  \\
 {} \ar@{}[ruu]|(.04)*\dir{-} & {\{\bullet\}} \ar`l[lu][lu]^(0)s \kar[u]_{u} \ar`r[ru][ru]_(0)t  & \ar@{}[luu]|(.04)*\dir{-} }
 \end{xy}
 \\ 
 fwd.-bwd.\ sim. 
  & 
 bwd.-fwd.\ sim. 
 \\ 
 \end{tabular}
 \end{center}
\caption{{Kleisli simulations (here 
$F=1+\Sigma\times(\_)$)}}
\label{fig:coalgSim}
\end{figure}

We write $\mathcal{X}\fwd\mathcal{Y}$, $\mathcal{X}\bwd\mathcal{Y}$,
 $\mathcal{X}\sqsubseteq_{\bf FB}\mathcal{Y}$ or
 $\mathcal{X}\sqsubseteq_{\bf BF}\mathcal{Y}$ if there exists a forward, backward, forward-backward, or backward-forward simulation, respectively.
\end{mydefinition}

(Generic) soundness is proved using the maximality of $\tr(c)$ in~(\ref{eq:traceSem}) among (op)lax
coalgebra homomorphisms, arguing in the language of enriched category
theory~\cite{hasuo06genericforward}.

\begin{mytheorem}[soundness]
\label{thm:soundnessOfKlesliSim}
Let $\mathcal{X}$ and $\mathcal{Y}$ be $(\mss,1+\Sigma\times(\place))$-systems. 
Each of the following yields $\tr(\mathcal{X})
\sqsubseteq
\tr(\mathcal{Y})
\colon \{\bullet\}\kto \Sigma^{*}
$ (cf.\ Definition~\ref{def:traceSemForTFSys}).  \\
\begin{tabular}{l@{\hspace{1cm}}l@{\hspace{1cm}}l@{\hspace{1cm}}l}
1. $\mathcal{X} \sqsubseteq_{\bf F} \mathcal{Y}$ &  2. $\mathcal{X} \sqsubseteq_{\bf B} \mathcal{Y}$ &
3. $\mathcal{X} \sqsubseteq_{\bf FB} \mathcal{Y}$ & 4. $\mathcal{X} \sqsubseteq_{\bf BF} \mathcal{Y}$ 
\end{tabular} \qed
\end{mytheorem}


\begin{mytheorem}[completeness]\label{thm:completenessOfBF}
The converse of soundness holds for backward-forward
 simulations. That is:
$\tr(\mathcal{X}) \sqsubseteq \tr(\mathcal{Y})$ implies
 $\mathcal{X}\sqsubseteq_{\bf BF}\mathcal{Y}$\;.
\qed
\end{mytheorem}

\section{Simulation Matrices for Semiring-Weighted Automata}
\label{sec:simulationMatrices}
In this section we fix parameters $T=\mss$ and $F=1+\Sigma\times(\place)$
in the generic theory in~Section~\ref{subsec:coalgebraicTraceAndSim} and
rephrase the coalgebraic framework in terms of matrices (whose entries are taken from
$\mathcal{S}$). Specifically: Kleisli arrows become matrices; and 
Kleisli simulations become  matrices subject to certain
linear inequalities. Such matrix representations ease implementation, a
feature we will exploit in later sections.



Recall that  a Kleisli arrow $A\kto B$ is  a function $A\to
\mss B$ (Definition~\ref{def:KlArrow}).

\begin{mydefinition}[matrix representation $M_{f}$]\label{def:matrixReprOfKlArr}
 Given a Kleisli arrow $f\colon A\kto B$, its \emph{matrix
 representation} $M_f\in S^{A\times B}$  is given by
 $(M_f)_{x,y}=f(x)(y)$.
\end{mydefinition}
In what follows we shall use the
notations $f$ and $M_{f}$ interchangeably. 



%

\begin{mylemma}\label{lem:kleisliArrAsMatrices}
Let $f,f' \colon  A \kto B$ and $g \colon  B \kto C$ be Kleisli arrows.
\begin{enumerate}
\item $f \sqsubseteq f'$ if and only if $M_{f} \sqsubseteq M_{f'} $. Here the
      former $\sqsubseteq$ is between $\mss$-Kleisli arrows, and the latter
      order $\sqsubseteq$ is between matrices, 
      both defined entrywise: namely,
      $f\sqsubseteq f' \;\overset{\text{def.}}{\Leftrightarrow}\; \forall x\in A.\,\forall y\in B.\, f(x)(y)\sqsubseteq f'(x)(y)$, and
      $M \sqsubseteq M' \;\overset{\text{def.}}{\Leftrightarrow}\; \forall x\in A.\,\forall y\in B.\, M_{x,y}\sqsubseteq M'_{x,y}$ .
\item $M_{g \odot f} = M_{f}M_{g}$, computed by matrix multiplication.
\qed
\end{enumerate}
\end{mylemma}
The correspondence from $ A\stackrel{f}{\longkto} B$
to 
$1+\Sigma\times A \stackrel{ 1+\Sigma\times f}{\longkto} 1+\Sigma\times B$---
used
in~(\ref{eq:traceSem}) and in Figure~\ref{fig:coalgSim}---can be described using matrices, too. 


\begin{mylemma}\label{lem:actionOfFOnArrowsByMatrices}
Let $f : A \kto B$ be a Kleisli arrow and $M_f$ be its matrix representation. 
Then the matrix representation   $M_{1+\Sigma\times f}$ is given by

\begin{displaymath}
 I_{1}\oplus (I_{\Sigma}\otimes
 M_f)\;\in\; \mathcal{S}^{(1+\Sigma\times A)\times
 (1+\Sigma\times B)}\enspace,
\end{displaymath}
where $\oplus$ and $\otimes$ denote \emph{coproduct} and \emph{the
 Kronecker product} of matrices:

\begin{equation}
 \scriptsize
 \begin{aligned}
 &
 X \oplus Y \;=\; {\renewcommand\arraystretch{1.3}\begin{pmatrix}
 \begin{xy}(0,0)*+[F]{X}\end{xy}\, & O \\
 O & \begin{xy}(0,0)*+[F]{Y}\end{xy} 
 \end{pmatrix}}\enspace,
 &&
 \begin{pmatrix}
 & \vdots &  \\
 \cdots & x_{i,j} & \cdots \\
 & \vdots & 
 \end{pmatrix}
 \otimes \begin{xy}(0,0)*+[F]{Y}\end{xy} \;=\; \begin{pmatrix}
  & \vdots &  \\
 \cdots & \begin{xy}(0,0)*+[F]{x_{i,j}Y}\end{xy} & \cdots \\
 & \vdots & 
 \end{pmatrix}\enspace.
 \end{aligned}
 \tag*{\raisebox{-7mm}{\qed}}
\end{equation}


\end{mylemma}
 This description of $M_{Ff}$ generalizes from $F=1+\Sigma\times(\place)$ to
any polynomial functor $F$, inductively on the construction of $F$. In
this paper  the generality is not needed.

Using
 Lemma~\ref{lem:kleisliArrAsMatrices}--\ref{lem:actionOfFOnArrowsByMatrices}, 
we can present Kleisli simulations (Definition~\ref{def:ksim}) as
 matrices. Recall that a state space of a weighted automaton is assumed
 to be
 countable (Definition~\ref{def:SWeightedAutom}); hence all the matrix
 multiplications in the definition below make sense.


\begin{mydefinition}[forward/backward simulation matrix]\label{def:simMat}
Let $\mathcal{A}=(\mcr{Q}{A},\Sigma,\mcr{M}{A},\mcr{\alpha}{A},\mcr{\beta}{A})$ and $\mathcal{B}=(\mcr{Q}{B},\Sigma,\mcr{M}{B},\mcr{\alpha}{B},\mcr{\beta}{B})$ be
$\mathcal{S}$-weighted automata.
 \begin{itemize}
  \item A matrix $X \in S^{\mcr{Q}{B}\times\mcr{Q}{A}}$ is a
	\emph{forward simulation matrix} from $\mathcal{A}$  to
	$\mathcal{B}$ if
	\begin{equation} 
	          \mcr{\alpha}{A} \sqsubseteq \mcr{\alpha}{B} X\enspace,
	 \quad
            X\cdot\mcr{M}{A}(a) \sqsubseteq \mcr{M}{B}(a)\cdot X
	 \quad(\forall a \in \Sigma)\enspace,\quad \text{and}\quad
           X\mcr{\beta}{A} \sqsubseteq \mcr{\beta}{B}\enspace.
           \label{eq:simMatFwd}
	\end{equation} 
  \item A matrix $X \in S^{\mcr{Q}{A}\times\mcr{Q}{B}}$ is  a
	\emph{backward simulation matrix} from $\mathcal{A}$  to
	$\mathcal{B}$ if
	\begin{equation} 
	 \alpha_{\mathcal{A}}X \sqsubseteq \alpha_{\mathcal{B}}
	 \enspace,
	 \quad
	 M_{\mathcal{A}}(a)\cdot X \sqsubseteq X\cdot M_{\mathcal{B}}(a)
	 \quad(\forall a \in \Sigma)\enspace,\quad \text{and}\quad
\beta_{\mathcal{A}} \sqsubseteq X\beta_{\mathcal{B}}           
	 \enspace.
	            \label{eq:simMatBwd}
	\end{equation} 
 \end{itemize}
\end{mydefinition}
The requirements on $X$ are obtained by first translating
Figure~\ref{fig:coalgSim} into matrices, and then breaking them up into
smaller matrices using Lemma~\ref{lem:actionOfFOnArrowsByMatrices}.
It is notable that the requirements are given in the form of \emph{linear
inequalities}, a format often used in constraint solvers. 
Solving them is a topic of  extensive research efforts that
include~\cite{akian12tropicalpolyhedra,butkovic06astrongly}. This fact becomes 
an advantage in implementing search algorithms, as we see later.

We also note that \emph{forward} and \emph{backward} simulation matrices
have different   dimensions. This difference comes from
the different directions of arrows in Figure~\ref{fig:coalgSim}.

\begin{myremark}\label{rem:oppositeAutom}
 The \emph{opposite} of an $\mathcal{S}$-weighted automaton 
 $\mathcal{A}=(Q, \Sigma, M, \alpha, \beta)$---obtained by reversing
 transitions and swapping initial/final states---can be naturally
 defined
 by matrix transpose, that is,
 \begin{math}
  \trans\mathcal{A}:= (Q,\Sigma,\trans M, \trans\beta,\trans\alpha).
 \end{math} 
It is easy to see that: if $X$ is a forward simulation matrix from
$\mathcal{A}$
 to $\mathcal{B}$, then $\trans X$ is a backward simulation matrix from
$\trans\mathcal{A}$ to $\trans\mathcal{B}$.

 This characterization of backward simulation matrices---by matrix
 transposes---does not seem to 
 generalize easily, however,
to weighted \emph{tree} automata (that are discussed in
 Section~\ref{sec:matSimPolyF}). The reason is simply that: a word
 reversed is another word; but a tree, with its parent-child
 relationship reversed, is not a tree any more.

 Recently a (co)algebraic reincarnation of \emph{Brzozowski's
 minimization algorithm}~\cite{brzozowski62canonicalregular} is
 presented
in~\cite{bonchiBHPRS14algebracoalgebra}. 
 The algorithm essentially relies on reversing word automata; its
 relationship to the current work is yet to be seen.
\end{myremark}

\begin{mytheorem}\label{thm:matsimIsKlSim}
Let
 $\mathcal{A}
$ and $\mathcal{B}
$ be
$\mathcal{S}$-weighted automata.
 There is a bijective correspondence between: 1) forward simulation
 matrices from $\mathcal{A}$ to $\mathcal{B}$; and 2) forward Kleisli
 simulations from $\mathcal{X_A}$ to $\mathcal{X_B}$. The same holds for the
 backward variants.
\qed
\end{mytheorem}
In what follows we write $\fwd,\bwd$ also between $\mathcal{S}$-weighted
automata. Theorem~\ref{thm:matsimIsKlSim} yields:
$\mathcal{A}\fwd\mathcal{B}$ if and only if there is a forward 
simulation matrix.

Here is our core result; the rest of the paper is devoted to its application.
\begin{mycorollary}[soundness of simulation
 matrices]\label{cor:soundnessOfMatSim}
\sloppy
Let
 $\mathcal{A}
$ and $\mathcal{B}
$ be
$\mathcal{S}$-weighted automata.
Existence of a forward (or backward) simulation matrix from
 $\mathcal{A}$ to $\mathcal{B}$---i.e.\ 
 $\mathcal{A}\fwd\mathcal{B}$ or  $\mathcal{A}\bwd\mathcal{B}$---witnesses 
language inclusion $
 \lang(\mathcal{A})
\sqsubseteq\lang(\mathcal{B})
$.
\fussy
\end{mycorollary}
\begin{myproof}
\begin{math}
 \begin{array}[t]{l}
  \exists\;(\text{fwd./bwd.\ simulation matrix 
      from $\mathcal{A}$ to $\mathcal{B}$})
      \\
  \stackrel{\text{Thm.~\ref{thm:matsimIsKlSim}}}{\Longleftrightarrow}
  \quad
  \exists\;(\text{fwd./bwd.\ Kleisli simulation 
      from $\mathcal{X_A}$ to $\mathcal{X_B}$})
      \\
  \stackrel{\text{Thm.~\ref{thm:soundnessOfKlesliSim}}}{\Longrightarrow}
  \quad
  \tr(\mathcal{X_{A}})\sqsubseteq \tr(\mathcal{X_{B}})
  \quad
  \stackrel{\text{Thm.~\ref{thm:traceSem}}}{\Longleftrightarrow}
  \quad
  \lang(\mathcal{A})\sqsubseteq   \lang(\mathcal{B})\enspace.
  \qquad  \qquad  \qquad\; \text{\qed}
 \end{array}
\end{math}
\end{myproof}
%
%
%
It is classic to represent  nondeterministic automata by
 Boolean matrices. This corresponds to the special case
 $\mathcal{S}=\mathcal{B}$ (the Boolean semiring) of the current framework; and a simulation
 matrix becomes the same thing as a (relational) simulation in~\cite{lynch95forwardand}.
\section{Forward and Backward Partial Execution}\label{sec:partExec}
In this section we  introduce for   semiring-weighted automata
their transformations---called \emph{forward} and \emph{backward partial
execution}---that increase the number of forward or backward simulation matrices. We also
prove some correctness results. 

\subsection{Incompleteness of Matrix Simulations}
We have  four different notions of simulation (Definition~\ref{def:ksim}):
forward, backward, forward-backward, and backward-forward. Our view on these is as (possibly
finitary) witnesses of language inclusion.

The combined ones (forward-backward and backward-forward) subsume the one-direction
ones (forward and backward)---simply take the identity arrow as one of the two
simulations required.
Moreover, backward-forward is complete
(Theorem~\ref{thm:completenessOfBF}). Despite these theoretical advantages, the combined simulations
are generally harder to find: in addition to two simulations, we 
have to find an intermediate system too ($\mathcal{Z}$ in
Definition~\ref{def:ksim}). Furthermore,  since language inclusion for finite
$\spt$-weighted automata---models of probabilistic
systems---is known to be
undecidable~\cite{blondel03undecidableproblems}, existence of a
backward-forward simulation is undecidable
too.

Therefore in what follows we focus on the one-directional (i.e.\ forward 
or backward)  simulations as proof methods for language inclusion. They have
convenient matrix presentations, too, as we saw
in~Section~\ref{sec:simulationMatrices}.  
However, it is easy to see that the classic one-directional simulations---introduced in~\cite{lynch95forwardand} for \emph{nondeterministic} automata---are
not necessarily complete.
This incompleteness is carried over to the current quantitative setting.
We can see it from the following counterexample.

\begin{myexample}[$\fwd$ and $\bwd$ are not complete]\label{example:traceInclButNoFwdOrBwd}
The following $\spt$-weighted automata exhibit
 $\lang(\mathcal{A})\sqsubseteq \lang(\mathcal{B})$ (in fact
 $\lang(\mathcal{A})=\lang(\mathcal{B})$). 
\begin{displaymath}
\begin{xy}
(0,15)*{\mathcal{A}} ="",
(0,0)*+[Fo]{x_0}   ="x0",
(15,15)*+[Fo]{x_1} ="x1",
(15,-15)*+[Fo]{x_2}="x2",
(30,0)*+[Fo]{x_3}  ="x3",
(40,0)*{\checkmark}  ="xc",
(60,15)*{\mathcal{B}} ="",
(60,0)*+[Fo]{y_0}   ="y0",
(75,15)*+[Fo]{y_1} ="y1",
(75,-15)*+[Fo]{y_2}="y2",
(90,0)*+[Fo]{y_3}  ="y3",
(100,0)*{\checkmark}  ="yc",
\ar (-10,-10); "x0"
\ar ^{a,\frac{1}{2}} "x0"; "x1"
\ar _{b,\frac{1}{2}} "x0"; "x2"
\ar _{a,1} "x2"; "x3"
\ar @<1mm> ^{a,1} "x1"; "x3"
\ar @<1mm> ^{a,\frac{1}{2}} "x3"; "x1"
\ar ^{  \frac{1}{2}} "x3"; "xc"
\ar (50,-10); "y0"
\ar ^{a,\frac{1}{2}} "y0"; "y1"
\ar _{b,\frac{1}{2}} "y0"; "y2"
\ar @<-1mm> _{a,1} "y2"; "y3"
\ar ^{a,1} "y1"; "y3"
\ar @<-1mm> _{a,\frac{1}{2}} "y3"; "y2"
\ar ^{  \frac{1}{2}} "y3"; "yc"
\end{xy}
\end{displaymath}
Indeed,
for each word $w\in\Sigma^*$, we have
\[\lang(\mathcal{A})(w) = \lang(\mathcal{B})(w) = \begin{cases} \frac{1}{4}\bigl(\frac{1}{2}\bigr)^n & \text{($w=aaa^{2n}$ or $baa^{2n}$)} \\ 0 & \text{(otherwise)}\enspace. \end{cases}\] 
However there is no forward
 or backward simulation from $\mathcal{A}$ to $\mathcal{B}$: one
 can show by direct calculation  that there is no $X$ that satisfies the requirements in
 Definition~\ref{def:simMat}. 
 Hence this pair is a counterexample for the
  completeness of $\fwd$ and that of $\bwd$.
 
It turns out that a simulation in the sense of Jonsson and
 Larsen~\cite{jonsson91specificationand} does exist from $\mathcal{A}$
 to $\mathcal{B}$. See also
 Section~\ref{subsec:SPTcomparisonWithOtherSim} later, where we
 systematically compare our current notions of simulation with  existing ones.
%
%

%

\end{myexample}

The incompleteness of forward and backward simulations can be also
deduced from complexity arguments. See Section~\ref{subsec:SPTcomparisonWithOtherSim}.

\subsection{Forward and Backward Partial Execution for Semiring-weighted Automata}
We shall define transformations, called FPE and BPE, that increase matrix
simulations for semiring-weighted automata. 
We prove some of its properties, too.

\begin{mydefinition}[FPE, BPE]\label{def:PEFormally}
 \emph{Forward partial execution (FPE)} is a transformation of 
a weighted automaton that ``replaces some states with their one-step
 behaviors.'' Concretely, given an  $\mathcal{S}$-weighted
 automaton $\mathcal{A}=(Q, \Sigma, M, \alpha, \beta)$  and a parameter
 $P\subseteq Q$, the resulting automaton $\mathcal{A}_{\FPE,
 P}=(Q',\Sigma,M',\alpha',\beta')$ has a state space
\begin{align}\label{eq:FPEStateSpace}
 Q' &\;=\; \bigl\{\checkmark\,\bigl|\bigr.\, \exists x\in P.\,\beta_{x}\neq 0_{\mathcal{S}}  \bigr\} + \bigl\{(a,y)\,\bigl|\bigr.\,  \exists x\in P.\, M(a)_{x,y}\neq 0_{\mathcal{S}}\bigr\} + (Q \setminus P)\enspace,
\end{align}
 replacing each $p\in P$ with its one-step behaviors ($\checkmark$ or
 $(a,q)$) as new states. The other data $M',\alpha',\beta'$ are defined
 as follows. 
 For the transition matrices $M'$:
\begin{eqnarray*}
&
\begin{array}{rclrclrcl}
M'(a)_{(a,x),\checkmark}&=&\beta_x & M'(a)_{(a,x),(a',y)}&=&M(a')_{x,y} & M'(a)_{(a,x),x}&=&1_{\mathcal{S}} \\
 M'(a)_{x,\checkmark}&=&\bigl(M(a)\beta\bigr)_{x} &
  M'(a)_{x,(a',y)}&=&\bigl(M(a)M(a')\bigr)_{x,y}&
 M'(a)_{x,y}&=&M(a)_{x,y} 
\end{array}
\end{eqnarray*}
where  $a,a'\in\Sigma,x,y\in Q$. For all the other cases we define
$M'(a)_{u,v}=0_{\mathcal{S}}$, where $u,v\in Q'$.
For the initial and final vectors $\alpha'$ and $\beta'$, the definition
 is shown below.
\begin{displaymath}
 \begin{array}{rclrclrcl}
 \alpha'_{\checkmark}&=&\alpha\beta &\hspace{1cm} \alpha'_{(a,x)}&=&\bigl(\alpha M(a)\bigr)_{x} \hspace{1cm}& \alpha'_{x}&=&\alpha_{x}
 \\
 \beta'_{\checkmark}&=&1_{\mathcal{S}} & \hspace{1cm} \beta_{(a,x)}&=&0_{\mathcal{S}}\hspace{1cm} &  \beta'_{x}&=&\beta_{x} 
 \end{array}
\end{displaymath}


\emph{Backward partial execution (BPE)} in contrast ``replaces states in
 a parameter $P\subseteq Q$ with their backward one-step behaviors.''
For the same $\mathcal{A}$ as above, the resulting automaton 
$\mathcal{A}_{\BPE,
 P}=(Q',\Sigma,M',\alpha',\beta')$ has a state space
\begin{align}\label{eq:BPEStateSpace}
 Q' &\;=\; \bigl\{\bullet\,\bigl|\bigr.\, \exists x\in P.\,\alpha_{x}\neq 0_{\mathcal{S}}  \bigr\} + \bigl\{(a,y)\,\bigl|\bigr.\,  \exists x\in P.\, M(a)_{y,x}\neq 0_{\mathcal{S}}\bigr\} + (Q \setminus P)\enspace,
\end{align}
replacing each $p\in P$ with its backward one-step behaviors---$(a,q)$
 with $q\stackrel{a}{\to}p$, and $\bullet$ if $p$ is initial---as new
 states.
The other data $M',\alpha',\beta'$ are defined
 as follows. 
 For the transition matrices $M'$:
\begin{eqnarray*}
&
\begin{array}{rclrclrcl}
M'(a)_{\bullet,(a,y)}&=&\alpha_y & M'(a)_{(a',x),(a,y)}&=&M(a')_{x,y} & M'(a)_{x,(a,x)}&=&1_{\mathcal{S}} \\
 M'(a)_{\bullet,y}&=&\bigl(\alpha M(a)\bigr)_{y} & 
 M'(a)_{(a',x),y}&=&(M(a')M(a))_{x,y} & 
 M'(a)_{x,y}&=&M(a)_{x,y} 
\end{array}
\end{eqnarray*}
where  $a,a'\in\Sigma,x,y\in Q$. For all the other cases we define
$M'(a)_{u,v}=0_{\mathcal{S}}$, where $u,v\in Q'$.
For the initial and final vectors $\alpha'$ and $\beta'$, the definition
 is shown below.
\begin{displaymath}
\begin{array}{rclrclrcl}
\alpha'_{\bullet}&=&1_{\mathcal{S}} &\hspace{1cm} \alpha_{(a,x)}&=&0_{\mathcal{S}} & \hspace{1cm}\alpha'_{x}&=&\alpha_{x} \\
\beta'_{\bullet}&=&\alpha\beta &\hspace{1cm} \beta'_{(a,x)}&=&(M(a)\beta)_{x} \hspace{1cm}& \beta'_{x}&=&\beta_{x}
\end{array}
\end{displaymath}
\end{mydefinition}

\begin{figure}[htbp]
 \centering
 \vspace{2mm}
 \def\FPEheight{9}
 \begin{tabular}{ccc}
 \begin{xy}
 (-54,0)*{\bullet} = "0",
, (-60,\FPEheight)*{\circ} = "1",
 (-48,\FPEheight)*{\circ} = "2",
 (-54,-\FPEheight)*{\circ} = "3",
 (-36,0)*{\circ} = "41",
 (-24,0)*{\circ} = "42",
 (-36,\FPEheight)*{\circ} = "5",
 (-24,\FPEheight)*{\circ} = "6",
 (-30,-\FPEheight)*{\circ} = "7",
  \ar ^{b,p} "3"; "0"
 \ar ^{a_1,p_1} "0"; "1"
 \ar _(.7){a_2,p_2} "0"; "2"
 \ar ^{b,p\cdot p_1} "7"; "41"
 \ar _{b,p\cdot p_2} "7"; "42"
 \ar _{a_1,1_\mathcal{S}} "41"; "5"
 \ar _{a_2,1_\mathcal{S}} "42"; "6"
 \ar@{|->} (-48,0); (-39,0)
 \ar@{|->} (-48,0); (-39,0)
 \end{xy}
 &
 \!\!\!\!
 \begin{xy}
 (-9,0)*{\bullet} = "121",
 (3,0)*{\bullet} = "122",
 (-3,\FPEheight)*{\circ} = "13",
 (-9,-\FPEheight)*{\circ} = "14",
 (3,-\FPEheight)*{\circ} = "15",
 (21,0)*{\circ} = "8",
 (21,\FPEheight)*{\circ} = "9",
 (15,-\FPEheight)*{\circ} = "10",
 (27,-\FPEheight)*{\circ} = "11",
\ar ^{a,r_1} "121"; "13"
 \ar _{a,r_2} "122"; "13"
 \ar ^{b_1,q_1} "14"; "121"
 \ar ^{b_2,q_2} "15"; "122"
 \ar _{a,1_\mathcal{S}} "8"; "9"
 \ar ^(-0)*-{\substack{b_1,\\q_1\cdot r_1\mathstrut\,}} "10"; "8"
 \ar _(0){\substack{b_2,\\q_2\cdot r_2\mathstrut}} "11"; "8"
 \ar@{|->}  (7.5,0); (15,0)
 \end{xy}
 &
 \!\!\!\!
 \begin{xy}
 (38,-\FPEheight)*{\circ} = "21",
 (38,0)*{\bullet} = "22",
 (52,-\FPEheight)*{\circ} = "23",
 (38,5)*{} = "24",
 (38,3)*{\smallsetminus}="",
 \ar _{c,s} "21"; "22"
 \ar@{|->}  (42,0); (48,0)
 \ar "22"; "24"
 \end{xy} 
\\[+1em]
 {``split backward''} & {``merge backward''} & {``eliminate dead end'' } 
\\[+.5em]
 \multicolumn{3}{c}{\fbox{Forward Partial Execution}}
\end{tabular}

 \vspace{5mm}

 \def\BPEheight{9}
 \begin{tabular}{ccc}
 \begin{xy}
 (-54,0)*{\bullet} = "0",
 (-60,-\BPEheight)*{\circ} = "1",
 (-48,-\BPEheight)*{\circ} = "2",
 (-54,\BPEheight)*{\circ} = "3",
 (-36,0)*{\circ} = "41",
 (-24,0)*{\circ} = "42",
 (-36,-\BPEheight)*{\circ} = "5",
 (-24,-\BPEheight)*{\circ} = "6",
 (-30,\BPEheight)*{\circ} = "7",
  \ar ^{b,p} "0"; "3"
 \ar ^{a_1,p_1} "1"; "0"
 \ar _(.3){a_2,p_2} "2"; "0"
 \ar ^{b,p\cdot p_1} "41"; "7"
 \ar _{b,p\cdot p_2} "42"; "7"
 \ar _{a_1,1_\mathcal{S}} "5"; "41"
 \ar _{a_2,1_\mathcal{S}} "6"; "42"
 \ar@{|->} (-48,0); (-39,0)
 \ar@{|->} (-48,0); (-39,0)
 \end{xy}
 &
 \!\!\!\!
 \begin{xy}
 (-9,0)*{\bullet} = "121",
 (3,0)*{\bullet} = "122",
 (-3,-\BPEheight)*{\circ} = "13",
 (-9,\BPEheight)*{\circ} = "14",
 (3,\BPEheight)*{\circ} = "15",
 (21,0)*{\circ} = "8",
 (21,-\BPEheight)*{\circ} = "9",
 (15,\BPEheight)*{\circ} = "10",
 (27,\BPEheight)*{\circ} = "11",
\ar ^{a,r_1} "13"; "121"
 \ar _{a,r_2} "13"; "122"
 \ar ^{b_1,q_1} "121"; "14"
 \ar ^{b_2,q_2} "122"; "15"
 \ar _{a,1_\mathcal{S}} "9"; "8"
 \ar ^(1)*-{\substack{b_1,\\q_1\cdot r_1\mathstrut\,}} "8"; "10"
 \ar _(1){\substack{b_2,\\q_2\cdot r_2\mathstrut}} "8"; "11"
 \ar@{|->}  (7.5,0); (15,0)
 \end{xy}
 &
 \!\!\!\!
 \begin{xy}
 (38,\BPEheight)*{\circ} = "21",
 (38,0)*{\bullet} = "22",
 (52,\BPEheight)*{\circ} = "23",
 (38,-5)*{} = "24",
 (38,-3)*{\smallsetminus}="",
 \ar _{c,s} "22"; "21"
 \ar@{|->}  (42,0); (48,0)
  \ar _{} "24"; "22"
 \end{xy}   \\[+.5em]
 {``split forward''} & {``merge forward''} & {``eliminate dead end'' } 
  \\[+.5em]
  \multicolumn{3}{c}{\fbox{Backward Partial Execution}}
\end{tabular}
\caption{Fwd./bwd.\ partial execution (FPE, BPE),
 pictorially. Black nodes need to be in $P$}
\label{fig:FPEBPE}
\end{figure}
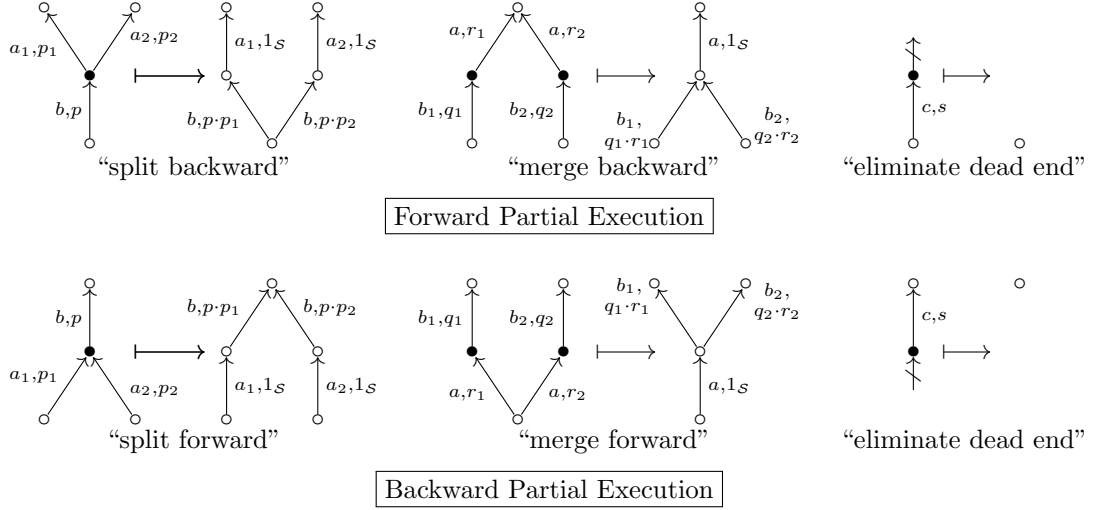

Pictorially, the actions of FPE and BPE can be illustrated as in Figure~\ref{fig:FPEBPE}.
Roughly speaking, FPE replaces a \emph{concrete} state $p\in P$ with an
\emph{abstract} state, such as $(a,q)$ in $Q'$
of~(\ref{eq:FPEStateSpace}) that is thought of as a description ``a state that makes an
$a$-transition to $q$.'' The idea comes from \emph{partial
evaluation} of a program; hence the name.







\subsection{Correctness of FPE and BPE}\label{subsec:correctnessFPEBPEWord}
The use of FPE/BPE is as follows: we aim to establish
$\lang(\mathcal{A})\sqsubseteq \lang(\mathcal{B})$; depending on whether
we search for a forward or backward simulation matrix, we apply one of
FPE and BPE to each of $\mathcal{A}$ and $\mathcal{B}$, according to the
 table~(\ref{eq:FPEBPEwhichSide}) below.
\begin{equation}\label{eq:FPEBPEwhichSide}
 \begin{array}{c||c|c}
\text{goal: } \lang(\mathcal{A})\sqsubseteq\lang(\mathcal{B}) & \mathcal{A} & \mathcal{B} \\
\hline\hline
\text{by } \fwd & \text{FPE} & \text{BPE} \\
\hline
\text{by } \bwd & \text{BPE} & \text{FPE} \\
\end{array}
\end{equation}

We shall now state correctness properties of this strategy. 
\emph{Soundness} means that discovery of a simulation after transformation indeed
witnesses the language inclusion for the \emph{original} automata. 
The second property---we call it \emph{adequacy}---states that simulations that are already
there are preserved by partial execution. 

\begin{mytheorem}[soundness of FPE/BPE]\label{thm:soundnessFPEBPE}
Let $P$ and $P'$ be  arbitrary subsets of the state spaces of
 $\mathcal{A}$ and  $\mathcal{B}$, respectively.
 Each of the following implies $\lang(\mathcal{A})\sqsubseteq
 \lang(\mathcal{B})$.
\begin{enumerate}
 \item $\mathcal{A}_{\FPE,P}\fwd\mathcal{B}_{\BPE,P'}$ 
 \item $\mathcal{A}_{\BPE,P}\bwd\mathcal{B}_{\FPE,P'}$ \qed
\end{enumerate}
\end{mytheorem}



\begin{mytheorem}[adequacy of FPE/BPE]
\label{thm:adequacyOfFPEBPE}
Let $P$ and $P'$ be  arbitrary subsets of the state spaces of
 $\mathcal{A}$ and  $\mathcal{B}$, respectively.
We have:
\begin{enumerate}
 \item
      $\mathcal{A}\fwd\mathcal{B}\;\Rightarrow\;\mathcal{A}_{\FPE,P}\fwd\mathcal{B}_{\BPE,P'}$
 \item 
$\mathcal{A}\bwd\mathcal{B}\;\Rightarrow\;\mathcal{A}_{\BPE,P}\bwd\mathcal{B}_{\FPE,P'}$ \qed
\end{enumerate}
\end{mytheorem}

To prove these two theorems, we should first prove the following lemma.

\begin{mylemma}\label{lem:peproperty}
For each subset $P$, we have the following.\\
\hspace{5mm}\begin{tabular}{l@{\hspace{7mm}}l@{\hspace{7mm}}l@{\hspace{7mm}}l}
1. $\mathcal{A} \bwd \mathcal{A}_{\FPE,P}$ &
2. $\mathcal{A}_{\FPE,P} \fwd \mathcal{A}$ &
3. $\mathcal{A} \fwd \mathcal{A}_{\BPE,P}$ &
4. $\mathcal{A}_{\BPE,P} \bwd \mathcal{A}$ 
\end{tabular}
\end{mylemma}
\begin{myproof}
1. Let $\mathcal{A}=(Q, \Sigma, M, \alpha, \beta)$ and $\mathcal{A}_{\FPE,P}=(Q',\Sigma,M',\alpha',\beta')$.
We define $X\in S^{Q\times Q'}$ as follows:
\begin{displaymath}
\begin{array}{rllcrll}
X_{x,\checkmark}&=\beta_x  &(x\in P), &\;\;\;\;\; & X_{x,(y,a)}&=M(a)_{x,y} & (x\in P), \\
X_{x,x}&=1_{\mathcal{S}}  &(x\notin P), & &X_{u,v}&=0_{\mathcal{S}} & (\text{otherwise})
\end{array}
\end{displaymath}
Then, this $X$ is a backward simulation matrix from $\mathcal{A}$ to $\mathcal{A}_{\FPE,P}$

The items 2., 3.\ and 4.\ are proved similarly. 
\qed
\end{myproof}

\newproof{proof:soundnessFPEBPE}{Proof of Theorem~\ref{thm:soundnessFPEBPE}}
\begin{proof:soundnessFPEBPE} 
1. From Proposition~\ref{lem:peproperty}.1, $\mathcal{A} \bwd \mathcal{A}_{\FPE,P}\fwd\mathcal{B}_{\BPE,P'}\bwd\mathcal{B}$.
Hence from Corollary~\ref{cor:soundnessOfMatSim}, 
$\lang(\mathcal{A}) \sqsubseteq \lang(\mathcal{A}_{\FPE,P})\sqsubseteq \lang(\mathcal{B}_{\BPE,P'})\sqsubseteq \lang(\mathcal{B})$
holds. 

The item 2.\ is proved similarly.
\qed
\end{proof:soundnessFPEBPE}


\newproof{proof:adequacyOfFPEBPE}{Proof of Theorem~\ref{thm:adequacyOfFPEBPE}}{\itshape}{\rmfamily}



%
\begin{proof:adequacyOfFPEBPE} 
1. From Proposition~\ref{lem:peproperty}.2, $\mathcal{A}_{\FPE,P} \fwd \mathcal{A}\fwd\mathcal{B}\fwd\mathcal{B}_{\BPE,P'}$.
Because $\fwd$ is transitive (see the diagram below where $F=1+\Sigma\times(\place)$), 
this implies $\mathcal{A}_{\FPE,P}\fwd\mathcal{B}_{\BPE,P'}$. 
\[
\begin{xy}
\xymatrix@R=1.8em@C=3.6em{
{F X} \ar@{}[dr]|{\sqsubseteq}  & {F X'} \ar@{}[dr]|{\sqsubseteq} \kar[l]_{F f'} & {F Y} \kar[l]_{F f} \\
{X} \kar[u]_{c} \ar@{}[dr]|{\sqsubseteq} & {X} \kar[u]_{c'} \ar@{}[dr]|{\sqsubseteq} \ar|(.3){f'}|-*\dir{|}[l] & {Y} \kar[u]_{d} \ar|(.3){f}|-*\dir{|}[l] \\
{} \ar@{}[ruu]|(.04)*\dir{-} & {\{\bullet\}} \ar`l[lu][lu]^s \kar[u]^{s'} \ar`r[ru][ru]_t  & {} \ar@{}[luu]|(.04)*\dir{-} }
\end{xy}
\]
The item 2.\ is proved similarly.
\qed
\end{proof:adequacyOfFPEBPE}




We also show that a bigger parameter $P$ yields a greater number of
simulations. In implementation, however, a bigger $P$ generally gives us a bigger
state space which slows down search for a simulation. Hence we are in a
trade-off situation.


\begin{myproposition}[monotonicity]\label{prop:monotonicityOfFPEBPE}
Assume $P_{1}\subseteq P'_{1}$ and $P_{2}\subseteq P'_{2}$. We have:\\
\hspace{5mm}\begin{tabular}[b]{l@{\hspace{1cm}}l}
1. $\mathcal{A}_{\FPE,P_{1}}\fwd\mathcal{B}_{\BPE,P_{2}}\;\Rightarrow\;\mathcal{A}_{\FPE,P'_{1}}\fwd\mathcal{B}_{\BPE,P'_{2}}$\enspace, &
\\
2. $\mathcal{A}_{\BPE,P_{1}}\bwd\mathcal{B}_{\FPE,P_{2}}\;\Rightarrow\;\mathcal{A}_{\BPE,P'_{1}}\bwd\mathcal{B}_{\FPE,P'_{2}}$\enspace.
\end{tabular}\qed
\end{myproposition}

For $\mathcal{S}=\spt$ or $\smp$, we can easily see that the complement problem of language inclusion between finite $\mathcal{S}$-weighted automata is semi-decidable. 
Since language inclusion itself is
undecidable~\cite{blondel03undecidableproblems,krob92theequality},
language inclusion is not even semidecidable.
Because existence of a simulation matrix is decidable,
it can be the case that however many times we apply FPE or BPE,
simulation matrices do not exist while language inclusion holds.
A concrete example is found in the following example.

\begin{myexample}[limitation of FPE]\label{example:traceInclButNoFwdWithFPE}
The following $\spt$-weighted automata exhibit $\lang(\mathcal{A})\sqsubseteq \lang(\mathcal{B})$, 
but a forward simulation does not exist no matter how many times FPE is applied to $\mathcal{A}$. 
\begin{displaymath}
 \begin{xy}
 (-10,16)*{\mathcal{A}}="",
 (0,0)*{\circ} = "x",
 (0,16)*{\checkmark} = "xc",
 (25,16)*{\mathcal{B}}="",
 (20,8)*{\circ} = "y0",
 (30,8)*{\circ} = "y1",
 (40,0)*{\circ} = "y2",
 (50,8)*{\circ} = "y3",
 (40,8)*{\circ} = "y4",
 (35,16)*{\checkmark} = "yc",
 (0,-8)*{} = "y100",
 (35,-8)*{} = "y101",
 \ar ^{1} "y100";"x"
 \ar @(dl,ul)^{a,\frac{1}{2}} "x";"x"
 \ar ^{\frac{1}{2}} "x";"xc"
 \ar ^{\frac{2}{3}} "y101";"y1"
 \ar _{\frac{1}{3}} "y101";"y2"
 \ar @<0.5mm>^{a,1} "y0";"y1"
 \ar @<0.5mm>^{a,\frac{1}{4}} "y1";"y0"
 \ar @<-0.5mm>_{a,\frac{1}{4}} "y4";"y3"
 \ar @<-0.5mm>_{a,1} "y3";"y4"
 \ar ^{a,1} "y2";"y4"
 \ar ^{\frac{3}{4}} "y1";"yc"
 \ar _{\frac{3}{4}} "y4";"yc"
 \end{xy}
\end{displaymath}
\end{myexample}

\begin{myremark}
We have concluded that language inclusion for $\spt$-weighted automata is not 
even semi-decidable, from the fact that its complement problem is semi-decidable.
The same type of arguments are applicable to general commutative cppo-semirings
whose addition, multiplication, and order relation are all computable.
See e.g.~\cite{esik10simulationequivalence}.
\end{myremark}

It is possible to describe FPE on the coalgebraic level of
abstraction. Besides providing an insight into the essence of the
construction, it
also allows for
the application of FPE to quantitative tree automata. 
In contrast, the definition of BPE seems to rely on the fact that we can
``reverse'' word automata, and hence is hard to generalize e.g.\ to tree
automata.
See Section~\ref{sec:matSimPolyF}.

\section{Simulation Matrices for Probabilistic Systems by $\mathcal{S}=\spt$
}\label{sec:plustimesWeightedAutomata}
In
this section
we focus on 
 $\spt$-weighted automata which we identify as
(purely) probabilistic
automata
 (cf.\ Example~\ref{example:semirings}).
 In~Section~\ref{subsec:SPTcomparisonWithOtherSim} 
our method by simulation matrices is compared with other notions of
probabilistic simulation; in~Section~\ref{subsec:sptImpl} we discuss our implementation.


\subsection{
Other Simulation Notions for Probabilistic Systems
}\label{subsec:SPTcomparisonWithOtherSim}
\[
\scalebox{0.8}{
\begin{tikzpicture}
[scale=0.8] 
\def\circletr{ (5,3) circle [x radius=4cm, y radius = 2cm]}
\def \circlefb{(5,3) circle [x radius=3.5cm, y radius = 1.7cm]}
\def\circlehj{(4,3) circle [x radius=2.2cm, y radius = 1.1cm]}
\def\circlejl{(4,3) circle [x radius=1.5cm, y radius = 0.7cm]}
\def\circlef{ (6,3) circle [x radius=2cm, y radius = 1cm]}
\fill[yellow!10] \circletr ;
\draw \circletr ;
\fill[blue!10] \circlefb ;
\draw \circlefb ;
\fill[red!20] \circlehj ;
\draw \circlehj ;
\fill[red!50] \circlejl ;
\draw \circlejl ;
\fill[blue!30] \circlef ;
\draw \circlef ;
\begin{scope}
  \clip \circlehj;
  \fill[purple!50] \circlef;
\end{scope}
\begin{scope}
  \clip \circlejl;
  \fill[purple!100] \circlef;
\end{scope}
\node at (3.3,3){$\times$};
\node at (-0.5,3){$\substack{\text{Example} \\ \ref{example:traceInclButNoFwdOrBwd}}$};
\draw (0.5,3)--(3.0,3);
\node at (7,3){$\times$};
\node at (10,3){\cite{hasuo10genericforward}};
\draw (9.5,3)--(7.2,3);
\draw (1.3,3.7)--(1,4);
\node at (0.3,4.4){lang.\ incl.};
\draw (8,3.9)--(8.9,4.3);
\node at (10,4.6){$\sqsubseteq_{\bf FB}$};
\draw (3.0,2.0)--(1.5,1.3);
\node at (1.2,0.9){$\sqsubseteq_{\bf HJ}$};
\draw (2.8,2.6)--(0.9,2.0);
\node at (0.1,1.8){$\sqsubseteq_{\bf JL}$};
 \draw (7.1,2.2)--(9.2,1.7);
\node at (10.2,1.3){ $\sqsubseteq_{\bf
 F}\text{ or } \sqsubseteq_{\bf B}$
};
\node at (10.2,.5){(i.e.\ by sim.\ matrices)
};
\end{tikzpicture}
}
\]
Various simulation notions have been introduced for probabilistic
systems, either as a behavioral order by itself or as a proof method for
language inclusion. Jonsson and Larsen's
one~\cite{jonsson91specificationand} (denoted by $\sqsubseteq_{\bf JL}$) is well-known;
it is shown in~\cite{hasuo10genericforward} to be a special case of
Hughes and Jacobs'
coalgebraic notion of simulation~\cite{hughes04simulationsin}
($\sqsubseteq_{\bf HJ}$), which in turn is a special case of forward-backward
(Kleisli) simulation ($\sqsubseteq_{\bf FB}$, Definition~\ref{def:ksim}).
Comparison of all these notions (observed in~\cite{hasuo10genericforward}) is as depicted above; it follows from
Theorem~\ref{thm:soundnessOfKlesliSim} that
all these 
simulation notions are sound with respect to language inclusion. 

We note that language inclusion between finite $\spt$-weighted automata
is known to be undecidable~\cite{blondel03undecidableproblems} while language
equivalence can be determined in polynomial
time~\cite{kiefer11languageequivalence}. 
The former undecidability result may account
for the fact that there does not seem to be many proof methods for
probabilistic/quantitative language inclusion. For example,
\emph{probabilistic simulation} in~\cite{BaierHK04} is possibilistic
simulation between systems with both probabilistic and nondeterministic
choice and not a quantitative notion like in the current study. 

We also note that given finite-state $\spt$-weighted automata $\mathcal{A}$ and $\mathcal{B}$,
if $\mathcal{A}\fwd\mathcal{B}$ or not is decidable: existence of a
solution $X$ of the linear constraints in Definition~\ref{def:simMat} can be
reduced to linear programming (LP) problems, and the latter are known to
be decidable. The same applies to $\bwd$ too.

Probabilistic systems are commonly modeled using the
 monad $\dist$ (see~(\ref{eq:monads}))---with an explicit
\emph{normalization} condition $\sum_{x}d(x)\le 1$---instead of
$\mathcal{M}_{\spt}$. 
However  there is no need to impose  normalization on
simulations: sometimes only ``non-normalized'' simulation matrices are
found 
and they are still sound. Here is such an example.

\begin{myexample}\label{example:foundOnlyInMss}
The following $\spt$-weighted automata exhibit $\lang(\mathcal{A})\sqsubseteq \lang(\mathcal{B})$.
Neither forward nor backward Kleisli simulation
 (in the categorical sense of Definition~\ref{def:ksim}) exists between them as
 long as we represent the automata $\mathcal{A}$ and $\mathcal{B}$ as
 $(\dist,1+\Sigma\times(\place))$-systems. However Kleisli
 simulations (forward and backward) are found once we represent
 $\mathcal{A}$ and $\mathcal{B}$  as $(\mss,1+\Sigma\times(\place))$-systems.
\begin{displaymath}
 \begin{xy}
 (-20,9)*{\mathcal{A}}="",
 (0,-6)*+[Fo]{x_0}="x0",
 (0,6)*+[Fo]{x_1} ="x1",
 (-15,0)*{\checkmark} ="xc",
 (20,6)*{\mathcal{B}}="",
 (40,0)*+[Fo]{y_0} ="y0",
 (25,0)*{\checkmark} ="yc",
 \ar (10,-12); "x0"
 \ar @(dr,dl)^{a,\frac{1}{8}} "x0"; "x0"
 \ar @<1mm> @(u ,d )_{a,\frac{2}{8}} "x0"; "x1"
 \ar @(ul,ur)^{a,\frac{3}{8}} "x1"; "x1"
 \ar @<1mm> @(d ,u )_{a,\frac{4}{8}} "x1"; "x0"
 \ar ^{  \frac{2}{8}} "x0"; "xc"
 \ar _{  \frac{1}{8}} "x1"; "xc"
 \ar (30,-6); "y0"
 \ar @(ur,dr)^{a,\frac{5}{8}} "y0"; "y0"
 \ar @( l, r)_{  \frac{3}{8}} "y0"; "yc"
 \end{xy}
\end{displaymath}

Indeed, the only matrix $X$ that is a forward simulation
 (Definition~\ref{def:simMat}) is  $X=\begin{pmatrix} 1 &
				      1\end{pmatrix}$, and 
the only backward simulation matrix is $X=\begin{pmatrix} 1 \\ 2
					  \end{pmatrix}$.
Neither of these satisfies the normalization condition imposed on the
 subdistribution monad $\dist$.
\end{myexample}


%
%


\begin{myremark}[language equivalence vs.\ language inclusion]\label{rem:literatureForLanguageEquivalence}
Because of the undecidability of language \emph{inclusion} between
 $\spt$-weighted automata~\cite{blondel03undecidableproblems}, a proof
 method for language inclusion that is computationally tractable---such
 as our current one via forward/backward simulation matrices whose existence is in
 PTIME---is naturally incomplete.
In contrast, several complete methods for language \emph{equivalence}
 have been introduced for various kinds of weighted automata, including
 the aforementioned one in~\cite{kiefer11languageequivalence} for
 $\spt$-weighted automata.


In~\cite{boreale09weightedbisimulation}, quantitative bisimulation that
is complete with respect to  the language equivalence is defined for
$\mathbb{R}$-weighted automata.  Differently from a classical
definition~\cite{park81concurrencyautomata} where bisimulation is
defined as a \emph{relation}, the bisimulation notion
in~\cite{boreale09weightedbisimulation} is defined as a \emph{subspace}
of a certain vector space, using that $\mathbb{R}$ is not only a
semiring but also a \emph{field}.  The definition of bisimulation
in~\cite{boreale09weightedbisimulation} is almost the same as the one
in~\cite{stark03behaviorequivalence}, except for minor relaxations.
Furthermore, the notion in~\cite{boreale09weightedbisimulation} turns
out to be an instance of the coalgebraic (Aczel-Mendler) definition of bisimulation
in~\cite{aczelM89finalcoalgebra} by spans.\footnote{
This is pointed out  by one of the
 reviewers.
Therefore the 
 seemingly surprising completeness result in~\cite{boreale09weightedbisimulation}, of bisimulations against
 language equivalence, is no wonder: here the language equivalence is
 coalgebraic bisimilarity, much like for deterministic automata as
 $2\times(\place)^{\Sigma}$-coalgebras.}
Using the linear algebraic
characterization of bisimulation in~\cite{boreale09weightedbisimulation}, a partition refinement-based algorithm
for calculating the largest bisimulation is given
in~\cite{boreale09weightedbisimulation}.

Another definition of quantitative bisimulation is found in~\cite{esik10simulationequivalence}.
The bisimulation in~\cite{esik10simulationequivalence} 
is defined as a matrix, 
in a similar manner to 
the simulation $\fwd$
in the current paper.
More concretely, 
the bisimulation in~\cite{esik10simulationequivalence} is defined as a matrix $X$ that satisfies certain equations,
 that are  obtained by replacing the inequalities in (\ref{eq:simMatFwd}) of Definition~\ref{def:simMat} with equalities.
In~\cite{esik10simulationequivalence}
the authors go on to introduce the notion of \emph{properness} of a
 semiring. A semiring  $\mathcal{S}$ is  called \emph{proper} when: 
the weighted automata exhibit language equivalence if and only if
two $\mathcal{S}$-weighted automata
are connected by a finite-length chain of bisimulations.
Properness implies decidability of language equivalence (on the
 condition that the addition, the multiplication and
the order relation of $\mathcal{S}$ are all computable): existence of 
a finite-length chain of  bisimulations is semidecidable, hence so is
 language equivalence by properness, on the one hand; on the other hand
 it is not hard to see that  the complement problem of language
 equivalence  is semidecidable too.
Some examples and counterexamples of proper semirings are given in~\cite{esik10simulationequivalence}:
for example, the semiring $\mathbb{N}$ of natural numbers and the semiring $\mathbb{Z}$ of integers are both proper, 
while $\smp$ is not.

A method other than bisimulation towards complete checking of 
quantitative language equivalence is given in~\cite{bonsangueMS13soundand}.
There, 
for automata weighted with elements in a \emph{Noetherian semiring},
sound and complete axiomatizations (in the style of equational logic) of weighted language equivalence 
are presented. 
It is known that Noetherian semirings form a strict subclass of proper semirings~\cite{esik10simulationequivalence}:
the semiring $\mathbb{Z}$ is a Noetherian semiring and hence a proper semiring,
while $\mathbb{N}$ is a proper semiring~\cite{esik10simulationequivalence} but not a Noetherian semiring~\cite{bonsangueMS13soundand}.
It is shown in~\cite{bonsangueMS13soundand} that we can translate
a state of a 
Noetherian semiring-weighted automaton to an algebraic term for such
 equational reasoning.
\end{myremark}

\subsection{Algorithms}\label{subsec:algoSpt}
The following algorithm \ptlangfwdalgo\ checks language inclusion
between $\spt$-weighted automata by searching for a forward simulation matrix.
This algorithm uses three algorithms---\ptsimfwdalgo, \ptfpealgo\ and
\ptbpealgo\ that we describe later---as sub-algorithms.

\begin{myalgorithm}[\ptlangfwdalgo]

\noindent{\bf Input}:
A pair of $\spt$-weighted automata $\mathcal{A}$ and $\mathcal{B}$.

\noindent{\bf Output}: 
``$\mathsf{Yes}$'' if the language inclusion between $\mathcal{A}$ and
 $\mathcal{B}$ is witnessed by a forward simulation matrix.

\noindent{\bf Procedure}:
This algorithm alternatively applies FPE and BPE (by the algorithms
 \ptfpealgo\ and \ptbpealgo) and transforms $\mathcal{A}$ and
 $\mathcal{B}$. After each transformation it runs the algorithm
 \ptsimfwdalgo\ to search for a forward simulation matrix. 
In each the iteration, the parameter $P$ for \ptfpealgo\ and \ptbpealgo\
 are set appropriately (see the description of our  implementation below).
\end{myalgorithm}

The behavior of the above algorithm \ptlangfwdalgo\ is as follows: when it
terminates then we can conclude (by
Corollary~\ref{cor:soundnessOfMatSim} and
Theorem~\ref{thm:soundnessFPEBPE}) that language indeed inclusion holds between
the automata given as input; in particular, its sub-algorithm
\ptsimfwdalgo\ is sound and complete with respect to existence of a
forward simulation matrix; but,
sometimes, even if language inclusion does
hold the algorithm \ptlangfwdalgo\ may not terminate
(Example~\ref{example:traceInclButNoFwdWithFPE}); and finally, in case
language inclusion does not hold the algorithm \ptlangfwdalgo\ does not
terminate. To account for the last case we can use a straightforward
semi-decision procedure for the complement of language inclusion.

The algorithm \ptlangbwdalgo\ that tries to find a backward simulation
matrix is given similarly, by replacing the sub-algorithm
\ptsimfwdalgo\ with \ptsimbwdalgo\ that searches for a backward
simulation, and switching the automata to which FPE (or BPE) is applied
(cf.\ (\ref{eq:FPEBPEwhichSide})).

Now let us turn to the sub-algorithms.
The algorithms \ptfpealgo\ and \ptbpealgo\ that apply FPE and BPE are
straightforward. We only present \ptfpealgo; the other one \ptbpealgo
for BPE is similar.
\begin{myalgorithm}[\ptfpealgo]
 \noindent{\bf Input}: 
 An $\spt$-weighted automaton $\mathcal{A}$ and a subset $P$ of the state space.

 \noindent{\bf Output}: 
 An $\spt$-weighted automaton $\mathcal{A}_{\FPE,P}$.

 \noindent{\bf Procedure}:
 We follow Definition~\ref{def:PEFormally} and construct $\mathcal{A}_{\FPE,P}$. 
\end{myalgorithm}






Finally we describe
the algorithm \ptsimfwdalgo\ that searches for a forward  simulation
matrix between $\spt$-weighted automata (\ptsimbwdalgo\ for searching
for a backward simulation matrix is similar). 

\begin{myalgorithm}[\ptsimfwdalgo]\label{algo:ptsimfwdalgo}
\noindent{\bf Input}: 
A pair of $\spt$-weighted automata $\mathcal{A}$ and $\mathcal{B}$.

\noindent{\bf Output}: 
A forward simulation matrix $X$ from $\mathcal{A}$ to $\mathcal{B}$ if such $X$ exists, and
``$\mathsf{No}$" otherwise.

\noindent{\bf Procedure}:
This algorithm first transforms the constraints (\ref{eq:simMatFwd}) in Definition~\ref{def:simMat} into an
inequality $A\bvec{x}\leq \bvec{b}$.
Then it solves a linear programming problem whose constraint and objective function 
are set to $A\bvec{x}\leq \bvec{b}$ and $0$, respectively.
If a feasible solution is found, then the algorithm outputs the solution.
Otherwise it outputs ``$\mathsf{No}$.''
\end{myalgorithm}


For solving a linear programming problem our algorithm uses the \emph{simplex method}---its worst-case
time complexity is known to be exponential~\cite{kleeM72howgood}.
It is known that there are methods that run in polynomial time (e.g.\ Karmarkar's method~\cite{karmarkar84newpolynomial}), 
though it is also known that practically, the simplex method is efficient on average~\cite{smale83averagenumber}.

We also note that the matrix $A$ in Algorithm~\ref{algo:ptsimfwdalgo} is sparse because of the way the different constraints in Definition~\ref{def:simMat} are combined.
It has $n+anm+m$ rows, $nm$ columns and at most $2nm+a(n^2m+nm^2)$ nonzero entries.





\subsection{Implementation, Experiments and Discussions}\label{subsec:sptImpl}
Our implementation consists of two components: \ptsim\ and \ptPE. Both of them  are available in~\cite{implPtMpTree}.

\begin{itemize}
 \item 
The program \ptsim\ (in C++) implements \ptsimfwdalgo\ and \ptsimbwdalgo\ in the above. 
It uses \emph{glpk}~\cite{glpk} as a linear programming solver.

 \item
The program \ptPE\ (in OCaml) 
implements \ptfpealgo\ and \ptbpealgo\ in the above. It
takes an automaton $\mathcal{A}$ and
       $d\in\mathbb{N}$ as input, and returns $\mathcal{A}_{\FPE,P}$
      (or $\mathcal{A}_{\BPE,P}$, by choice). Here  $P$ is chosen, by heuristics, to be $P=\{x\mid
      x\overbrace{\to\cdots\to}^{ d}\checkmark\}$
 (or $P=\{x\mid \bullet\overbrace{\to\cdots\to}^{ d}x\}$, respectively).




\end{itemize}
The two programs are alternately applied to the given automaton, for
$d=1,2,\dotsc$, each time incrementing the parameter $d$ for \ptPE. 
The experiments were  on 
a MacBook Pro laptop
 with a Core i5 processor (2.6 GHz, 2 cores) and 16 GB RAM. 
\subsubsection{Grades Protocol} 
 The \emph{grades
protocol} is introduced in~\cite{kiefer11languageequivalence} and is used
there as a benchmark: the protocol
and its specification are expressed as probabilistic programs
$\mathsf{P}$ and $\mathsf{S}$; they are
then translated into  (purely) probabilistic automata
$\mathcal{A}_\mathsf{P}$
and
$\mathcal{A}_\mathsf{S}$
by a game semantics-based tool \APEX~\cite{kiefer13algorithmicprobabilistic}. By establishing $L(\mathcal{A}_\mathsf{P})=L(\mathcal{A}_\mathsf{S})$, the protocol is shown to exhibit the same behaviors as the
specification---hence is verified. 
The grades protocol aims to calculate the summation of grades of students without
revealing each individual's grade.
The protocol has two parameters $G$ and $S$. 
Here, $G$ is the number of grades (i.e.\  the set of grades is $\{0,1,\ldots,G-1\}$) and $S$ is the number of students.

\begin{table}
\small
\begin{center}
\begin{tabular}{rr|rr|rr|r|r|r|r}
\multicolumn{2}{l|}{param.} & \multicolumn{2}{l|}{$\mathcal{A}_{\mathsf{P}}$} & \multicolumn{2}{l|}{$\mathcal{A}_{\mathsf{S}}$} & & \multicolumn{1}{l|}{direction,} & \multicolumn{1}{l|}{time} & \multicolumn{1}{l}{space}  \\
\multicolumn{1}{r}{$G$} & \multicolumn{1}{r|}{$S$} &
	 \multicolumn{1}{r}{\#st.} & \multicolumn{1}{r|}{\#tr.} &
		 \multicolumn{1}{r}{\#st.} & \multicolumn{1}{r|}{\#tr.} &
			 \multicolumn{1}{l|}{$|\Sigma|$} &
 \multicolumn{1}{l|}{fwd./bwd.}   & \multicolumn{1}{l|}{(sec)} &(GB)  \\ \hline\hline
2 & 8 & 578&1522&130 & 642& 11 &$\mathcal{A}_{\sf P}\!\!\fwd\!\!\mathcal{A}_{\sf S}$& 1.02 & 0.30 \\ 
  &   &    &    &    &    &    &        $\mathcal{A}_\mathsf{P}
\!\!\sqsupseteq_{\bf B}\!\!
\mathcal{A}_\mathsf{S}
$& 1.00 & 0.30 \\ \hline
2 &10 &1102&2982&202 &1202& 13 &$\mathcal{A}_{\sf P}\!\!\fwd\!\!\mathcal{A}_{\sf S}$& 5.37 & 1.04 \\ 
  &   &    &    &    &    &    &        $\mathcal{A}_\mathsf{P}
\!\!\sqsupseteq_{\bf B}\!\!
\mathcal{A}_\mathsf{S}
$& 5.30 & 1.05 \\ \hline
2 &12 &1874&5162&290 &2018& 15 &$\mathcal{A}_{\sf P}\!\!\fwd\!\!\mathcal{A}_{\sf S}$& 22.21&2.91 \\ 
  &   &    &    &    &    &    &        $\mathcal{A}_\mathsf{P}
\!\!\sqsupseteq_{\bf B}\!\!
\mathcal{A}_\mathsf{S}
$& 21.99&2.94 \\ \hline
2 &14 &2942&8206&394 &3138& 17 &$\mathcal{A}_{\sf P}\!\!\fwd\!\!\mathcal{A}_{\sf S}$& 76.71 & 6.86 \\ 
  &   &    &    &    &    &    &        $\mathcal{A}_\mathsf{P}
\!\!\sqsupseteq_{\bf B}\!\!
\mathcal{A}_\mathsf{S}
$& 76.03 & 6.94 \\ \hline
2 &16 &4354&12258&514 &4610& 19 &$\mathcal{A}_{\sf P}\!\!\fwd\!\!\mathcal{A}_{\sf S}$& 232.67&13.01 \\ 
  &   &    &    &    &    &    &        $\mathcal{A}_\mathsf{P}
\!\!\sqsupseteq_{\bf B}\!\!
\mathcal{A}_\mathsf{S}
$& 231.33&13.20 \\ \hline
3 & 8 &1923&7107&243 &2163& 20 &$\mathcal{A}_{\sf P}\!\!\fwd\!\!\mathcal{A}_{\sf S}$& 25.51&3.33 \\ 
  &   &    &    &    &    &    &        $\mathcal{A}_\mathsf{P}
\!\!\sqsupseteq_{\bf B}\!\!
\mathcal{A}_\mathsf{S}
$& 25.35&3.36 \\ \hline
3 & 10 &3803&14323&383 &4183& 24 &$\mathcal{A}_{\sf P}\!\!\fwd\!\!\mathcal{A}_{\sf S}$& 167.68&12.21 \\ 
  &   &    &    &    &    &    &        $\mathcal{A}_\mathsf{P}
\!\!\sqsupseteq_{\bf B}\!\!
\mathcal{A}_\mathsf{S}
$& 166.86&12.27 \\ \hline
4 & 6 &1636&7468&196 &1924& 23 &$\mathcal{A}_{\sf P}\!\!\fwd\!\!\mathcal{A}_{\sf S}$& 17.58&2.61 \\ 
  &   &    &    &    &    &    &        $\mathcal{A}_\mathsf{P}
\!\!\sqsupseteq_{\bf B}\!\!
\mathcal{A}_\mathsf{S}
$& 17.34&2.64 \\ \hline
4 & 8 &4052&19076&356 &4580& 29 &$\mathcal{A}_{\sf P}\!\!\fwd\!\!\mathcal{A}_{\sf S}$& 210.09&13.13 \\ 
  &   &    &    &    &    &    &        $\mathcal{A}_\mathsf{P}
\!\!\sqsupseteq_{\bf B}\!\!
\mathcal{A}_\mathsf{S}
$& 212.58&13.12 \\
\end{tabular}

\end{center}
\caption{Results for the grades protocol~\cite{kiefer11languageequivalence}}
\label{table:resultsForGradesProtocol}
\end{table}
In our experiment we proved
 $L(\mathcal{A}_\mathsf{P})=L(\mathcal{A}_\mathsf{S})$ by establishing
two-way language inclusion ($\sqsubseteq$ and $\sqsupseteq$). The results are shown in Table.~\ref{table:resultsForGradesProtocol}.
For all the choices of parameters $G$ and $S$, our program \ptsim was
 able to establish, without applying \ptPE:
$
\mathcal{A}_\mathsf{P}
\sqsubseteq_{\bf F}
\mathcal{A}_\mathsf{S}
$ (but not $\sqsubseteq_{\bf B}$) for the $\sqsubseteq$ direction; and
$
\mathcal{A}_\mathsf{P}
\sqsupseteq_{\bf B}
\mathcal{A}_\mathsf{S}
$
(but not $\sqsupseteq_{\bf F}$) for the $\sqsupseteq$ direction.
In the table, \#st.\ and \#tr.\ denote the numbers of states and
 transitions, respectively, and $|\Sigma|$ is the size of the
 alphabet. All these numbers are determined by \APEX.
The table indicates that space  is a bigger problem for
our approach than time.  In~\cite{kiefer11languageequivalence} four algorithms for
checking language \emph{equivalence} between $\spt$-weighted automata
are implemented and compared: two are
deterministic~\cite{tzeng92apolynomial,doyen08equivalenceof} and the other two
are randomized~\cite{kiefer11languageequivalence}. These algorithms can
process bigger problem instances (e.g.\ $G=2,S=100$ in ca.\ 10 sec) and, in
comparison, the results in Table~\ref{table:resultsForGradesProtocol}
are far from impressive.  Note however that our algorithm is for
 language \emph{inclusion}---an undecidable problem, unlike
language \emph{equivalence} that is in $\mathbf{P}$, see~Section~\ref{subsec:SPTcomparisonWithOtherSim}---and hence is
more general.

\begin{table}
\begin{center}
\small
\begin{tabular}{rrr|rr|rr|r|r|r|r|r}
\multicolumn{3}{l|}{param.} & \multicolumn{2}{l|}{$\mathcal{A}_{\sf P}$} & \multicolumn{2}{l|}{$\mathcal{A}_{\sf S}$} & & \multicolumn{1}{l|}{direction} & \multicolumn{1}{l|}{time} & \multicolumn{1}{l|}{space} & $d$ \\
\multicolumn{1}{r}{$n$} & \multicolumn{1}{r}{$c$} &
	 \multicolumn{1}{r|}{$p_f$} & \multicolumn{1}{r}{\#st.} &
		 \multicolumn{1}{r|}{\#tr.} & \multicolumn{1}{r}{\#st.} &
			 \multicolumn{1}{r|}{\#tr.} &
			     \multicolumn{1}{r|}{$|\!\Sigma\!|$} & 
 \multicolumn{1}{l|}{fwd./bwd.} & \multicolumn{1}{r|}{(sec)}  &(GB)&  \\ \hline\hline
 5 & 1&$\frac{9}{10}$& 7&  44& 7&  56& 18&$\mathcal{A}_{\sf P}\!\!\fwd\!\!\mathcal{A}_{\sf S}$& 0.10& 0.02&2\\ 
  &  &              &  &    &  &    &   &$\mathcal{A}_{\sf P}\!\!\bwd\!\!\mathcal{A}_{\sf S}$&  0.04&  0.01&2\\ \hline
7 & 1&$\frac{3}{ 4}$& 9&  88& 9& 118& 26&$\mathcal{A}_{\sf P}\!\!\fwd\!\!\mathcal{A}_{\sf S}$&  0.59&  0.14&2\\ 
  &  &              &  &    &  &    &   &$\mathcal{A}_{\sf P}\!\!\bwd\!\!\mathcal{A}_{\sf S}$&  0.05&  0.01&2\\ \hline
10& 2&$\frac{4}{ 5}$&12& 224&12& 280& 54&$\mathcal{A}_{\sf P}\!\!\fwd\!\!\mathcal{A}_{\sf S}$&42.63& 3.86&2\\ 
  &  &              &  &    &  &    &   &$\mathcal{A}_{\sf P}\!\!\bwd\!\!\mathcal{A}_{\sf S}$&  0.07& 0.01&2\\ \hline
20& 6&$\frac{4}{ 5}$&22&1514&22&1696&238&$\mathcal{A}_{\sf P}\!\!\fwd\!\!\mathcal{A}_{\sf S}$&   T/O&     &  \\ 
  &  &              &  &    &  &    &   &$\mathcal{A}_{\sf P}\!\!\bwd\!\!\mathcal{A}_{\sf S}$&  0.84&  0.20&2\\ \hline
30& 10&$\frac{4}{ 5}$&32&4732&32&5112&550&$\mathcal{A}_{\sf P}\!\!\fwd\!\!\mathcal{A}_{\sf S}$&   S/O&     &  \\ 
  &  &              &  &    &  &    &   &$\mathcal{A}_{\sf P}\!\!\bwd\!\!\mathcal{A}_{\sf S}$& 6.17& 1.44&2\\ \hline
40& 14&$\frac{4}{ 5}$&42&10742&42&11392&990&$\mathcal{A}_{\sf P}\!\!\fwd\!\!\mathcal{A}_{\sf S}$&   S/O&     &  \\ 
  &  &              &  &    &  &    &   &$\mathcal{A}_{\sf P}\!\!\bwd\!\!\mathcal{A}_{\sf S}$& 30.70& 6.07&2\\ \hline
50& 17&$\frac{4}{ 5}$&52&20504&52&21560&1494&$\mathcal{A}_{\sf P}\!\!\fwd\!\!\mathcal{A}_{\sf S}$&   S/O&     &  \\ 
  &  &              &  &    &  &    &   &$\mathcal{A}_{\sf P}\!\!\bwd\!\!\mathcal{A}_{\sf S}$& 102.89& 13.64&2\\ 
%
\end{tabular}
\end{center}
\caption{Results for the Crowds protocol}
\label{table:resultsForCrowdsProtocol}
\end{table}
\subsubsection{Crowds Protocol}
Our second experiment calls for checking language \emph{inclusion},
making the algorithms studied in~\cite{kiefer11languageequivalence}
unapplicable. We verified some instances of the \emph{Crowds
protocol}~\cite{reiter98crowdsanonymity} against a quantitative
anonymity specification called \emph{probable
innocence}~\cite{chatzikokolakis2006probableinnocence}.  We used a
general  trace-based verification
method in~\cite{hasuo10probabilisticanonymity} for probable innocence: language
inclusion $L(\mathcal{A}_{\sf P})\sqsubseteq L(\mathcal{A}_{\sf S})$,
from the model $\mathcal{A}_{\sf P}$ of a protocol in question to 
$\mathcal{A}_{\sf P}$'s suitable modification  $\mathcal{A}_{\sf S}$, guarantees probable
innocence.

The Crowds protocol has parameters $n$, $c$ and $p_{f}$. In fact, for
this specific protocol, a sufficient
condition for probable innocence is known~\cite{reiter98crowdsanonymity}  (namely 
 $n\ge \frac{p_f}{p_f - 1/2}(c+1)$); we used  parameters that satisfy
 this condition. We implemented a small program that takes a choice of $n,c,p_{f}$ and
 generates an automaton $\mathcal{A}_{\sf P}$; it is then passed to another
 program that generates $\mathcal{A}_{\sf S}$.

The results are  in Table.~\ref{table:resultsForCrowdsProtocol}.
For each problem instance we tried both $\fwd$ and $\bwd$. The last
column shows the final value of the parameter $d$ for \ptPE---i.e.\ how many times
partial execution (Section~\ref{sec:partExec}) was applied. 

The entry 
``S/O'' designates that \ptPE was killed because of stack overflow
caused by an oversized automaton.
 ``T/O'' means that alternate application of  \ptsim and \ptPE did not
 terminate within a time limit (one hour).



We observe that backward simulation matrices were much faster to be found than forward
 ones.
 This seems to result from the shapes of the automata for this
 specific problem; after all it is an advantage of our forward and backward
 approach that we can try two different directions and use the faster one.
Space consumption seems again serious.


\section{Simulation Matrices for $\smp$-Weighted
 Automata}\label{sec:maxplusWeightedAutomata}
In this section
we discuss $\smp$-weighted
automata, in which weights are understood as (best-case) profit or
(worst-case) cost (see Example~\ref{example:semirings}). Such automata
are  studied in~\cite{chatterjee10quantitativelanguages} 
 (called \emph{$\mathsf{Sum}$-automata} there). In fact we observe that their
notion of simulation---formulated in 
game-theoretic terms and hence called \emph{G-simulation}
here---coincides with the notion of forward simulation matrix. 
This observation---that is presented in 
Section~\ref{subsec:gameSimAsFwdSimMat}---follows from the 
game-theoretic characterization
in~\cite{akian12tropicalpolyhedra} of linear inequalities in $\smp$.
In~Section~\ref{subsec:expResForMaxPlus}
our
implementation is presented.


\subsection{G-Simulation by Forward Simulation Matrices
}\label{subsec:gameSimAsFwdSimMat}
In this section we restrict to finite-state
automata. In this case we can dispose of the weight $\infty$, and have
$[-\infty,\infty)$ as the domain of weights (see
Example~\ref{example:semirings}). 

%
What we shall call \emph{G-simulation} is  introduced in~\cite{chatterjee10quantitativelanguages}.
G-simulation is originally introduced 
to witness inclusion between weighted languages over \emph{infinite-length words}
$\Sigma^{\omega}\to[-\infty,\infty)$.
It is easy to adapt the definition to the current setting of \emph{finite-length words} $\Sigma^{*}\to[-\infty,\infty)$;
the adaptation is concerned with termination $\checkmark$.
%
\begin{mydefinition}
[
$\sqsubseteq_{\bf G}$]
\label{def:simgame}
Let
 $\mathcal{A}=(\mcr{Q}{A},\Sigma,\mcr{M}{A},\mcr{\alpha}{A},\mcr{\beta}{A})$
 and
 $\mathcal{B}=(\mcr{Q}{B},\Sigma,\mcr{M}{B},\mcr{\alpha}{B},\mcr{\beta}{B})$
 be finite-state $\smp$-weighted automata.
 A \emph{finite simulation game} from $\mathcal{A}$ to $\mathcal{B}$ is
 played by two players called \emph{Challenger} and \emph{Simulator}:
 a \emph{strategy for Challenger} is a pair
 $\bigl(\,\rho_1:1\to\mcr{Q}{A},\,\tau_1:(\mcr{Q}{A}\times\mcr{Q}{B})\times(\Sigma\times\mcr{Q}{A}\times\mcr{Q}{B})^*\to
 1+\Sigma\times\mcr{Q}{A}\,\bigr)$ of functions; and
 a \emph{strategy for Simulator} is a pair
 $\bigl(\,\rho_2:\mcr{Q}{A}\to\mcr{Q}{B},\,\tau_2:(\mcr{Q}{A}\times\mcr{Q}{B})\times(\Sigma\times\mcr{Q}{A}\times\mcr{Q}{B})^*\times\Sigma\times\mcr{Q}{A}\to \mcr{Q}{B}\,\bigr)$.




A pair $(p_0 a_1\dotsc a_n p_n,q_0 a_1\dotsc a_n q_n)$ of runs on
 $\mathcal{A}$ and $\mathcal{B}$ is called the \emph{outcome of
 strategies $(\rho_1,\tau_1)$ and $(\rho_2,\tau_2)$} if:

\begin{itemize}
 \item $ \rho_1(\bullet)=p_0$ and $\rho_2(p_0)=q_0$ where $\bullet$ is the unique element of the domain of $\rho_1$.
\item $\tau_1\bigl((p_0,q_0)(a_1,p_1,q_1)\dotsc(a_i,p_i,q_i)\bigr) = ( a_{i+1},p_{i+1})$ for each $i\in [0,n-1]$. 
\item $\tau_2\bigl((p_0,q_0)(a_1,p_1,q_1)\dotsc(a_i,p_i,q_i),(a_{i+1},p_{i+1})\bigr)=q_{i+1}$ for each $i\in [0,n-1]$. 
 \item $\tau_1((p_0,q_0)(a_1,p_1,q_1)\dotsc(a_n,p_n,q_n))=\checkmark$.
\end{itemize}
A strategy $(\rho_1,\tau_1)$ for Challenger is \emph{winning} if for any
strategy $(\rho_2,\tau_2)$ for Simulator, their outcome $(r_1,r_2)$ exists and it
satisfies $L(\mathcal{A})(r_1)>L(\mathcal{B})(r_2)$. Here the weight
$L(\mathcal{A})(r)$ of a run $r$ is defined in the obvious way, 
exploiting the structure of the semiring $\smp$.

Finally, we write $\mathcal{A}\sqsubseteq_{\bf G}\mathcal{B}$ if there
 is no winning strategy for Challenger.
\end{mydefinition}






\begin{mytheorem}\label{thm:KleisliSimIsGSim}
Let $\mathcal{A}$ and $\mathcal{B}$ be finite-state $\smp$-weighted automata.
 Assume that $\mathcal{A}$ has no trap states, that is, every
 state has a path to $\checkmark$ whose weight is not $-\infty$.
Then, 
$\mathcal{A}\fwd\mathcal{B}$
if and only if $\mathcal{A}\sqsubseteq_{\bf G}\mathcal{B}$.
\end{mytheorem}
The extra assumption  can be easily
enforced by eliminating trap states through backward reachability
check. This does not change the (finite) weighted language.

The proof of Theorem~\ref{thm:KleisliSimIsGSim} is sketched as follows.
We first reduce G-simulation (between the original automata $
\mathcal{A}, \mathcal{B}$ on finite words) to
G-simulation 
$\sqsubseteq_{\bf G}^{\sf Limavg}$ between \emph{{\sf Limavg} automata}, the original setting
in~\cite{chatterjee10quantitativelanguages} with infinite words. 
This reduction (that is the first equivalence below)
allows us to exploit the characterization
in~\cite{chatterjee10quantitativelanguages} of G-simulation in terms of
a \emph{mean
payoff game}, yielding the second equivalence below.
\begin{align}
 \mathcal{A}
\sqsubseteq_{\bf G}
\mathcal{B}
& \;\Longleftrightarrow\;
 \mathcal{A}^{\sf Limavg}
\sqsubseteq_{\bf G}^{\sf Limavg}
\mathcal{B}^{\sf Limavg}
\;\Longleftrightarrow\;
\text{Max wins in $\mathcal{G}_{\mathcal{A}^{\sf Limavg},\mathcal{B}^{\sf Limavg}}^{\bf G}$}\enspace.
\label{eq:GSimToLimAvg}
\end{align}
Conversely, starting from
\begin{math}
   \mathcal{A}
 \fwd
 \mathcal{B}
\end{math}, we use the fundamental result in~\cite{akian12tropicalpolyhedra}
 that characterizes feasibility of inequalities in $\smp$ in terms of
 mean payoff games.
\begin{align}
  \mathcal{A}
\fwd
\mathcal{B}
&\;\Longleftrightarrow\;
 \text{a certain linear inequality
is feasible}
\;\Longleftrightarrow\;
\text{Max wins in $\mathcal{G}_{\mathcal{A},\mathcal{B}}^{\bf F}$}\enspace.
\label{eq:FSimToMeanPayoffGame}
\end{align}
Finally, we observe that the mean payoff games
$\mathcal{G}_{\mathcal{A}^{\sf Limavg},\mathcal{B}^{\sf Limavg}}^{\bf G}$
in~(\ref{eq:GSimToLimAvg})
and $\mathcal{G}_{\mathcal{A},\mathcal{B}}^{\bf F}$ in~(\ref{eq:FSimToMeanPayoffGame})
are in fact the same,
establishing $\mathcal{A}\fwd\mathcal{B}$ if and only if $\mathcal{A}\sqsubseteq_{\bf G}\mathcal{B}$.

In what follows we introduce necessary definitions and lemmas,
eventually leading to the proof of Theorem~\ref{thm:KleisliSimIsGSim}.

\begin{mydefinition}[{\sf Limavg} automaton]
 A \emph{{\sf Limavg} automaton} $\mathcal{C}=(Q,\Sigma,M,q_0)$ consists of a finite state space $Q$, a finite alphabet $\Sigma$, 
transition matrices $M:\Sigma\to [-\infty,\infty)^{Q\times Q}$, and the initial state $q_0\in Q$.

For an infinite word $w= a_0 a_1\dotsc\in\Sigma^{\omega}$, the automaton $\mathcal{C}$ assigns a value 
$\lang(\mathcal{C})(w)$ that is calculated by $\lang(\mathcal{C})(w)=\mathrm{sup}_{q_0q_1\dotsc \in Q^{\infty}}\mathrm{liminf}_{N\to\infty}\frac{1}{N}\sum_{i=0}^{N}M( a_i)_{q_i,q_{i+1}}$.
\end{mydefinition}

\begin{mylemma}\label{lem:finiteGametoLimavg}
Given an $\smp$-weighted automaton $\mathcal{C}=(Q,\Sigma,M,\alpha,\beta)$, 
we define a {\sf Limavg} automaton $\mathcal{C}^{\sf Limavg}=(Q^{\sf Limavg},\Sigma^{\sf Limavg},M^{\sf Limavg},\star)$ by 
\begin{align*}
Q^{\sf Limavg}&=Q+\{\star\} \;, \\
\Sigma^{\sf Limavg}&=\Sigma+\{\mbox{$?$}\}+\{\mbox{$!$}\} \;\text{, and} \\
M^{\sf Limavg}(a)_{x,y}&=\begin{cases}
M(a)_{x,y} & (a\in\Sigma, x,y\neq\star) \\
\alpha_y & (a=\mbox{$?$}, x=\star) \\
\beta_x & (a=\mbox{$!$}, y=\star) \\
-\infty & (\text{otherwise}) \;.
\end{cases}
\end{align*}
Let
 $\mathcal{A}=(\mcr{Q}{A},\Sigma,\mcr{M}{A},\mcr{\alpha}{A},\mcr{\beta}{A})$
 and
 $\mathcal{B}=(\mcr{Q}{B},\Sigma,\mcr{M}{B},\mcr{\alpha}{B},\mcr{\beta}{B})$
be
 $\smp$-weighted automata; assume further that
 $\mathcal{A}$ has no trap states. Then we have
\begin{equation}
  \mathcal{A}
\sqsubseteq_{\bf G}
\mathcal{B}
 \;\Longleftrightarrow\;
 \mathcal{A}^{\sf Limavg}
\sqsubseteq_{\bf G}^{\sf Limavg}
\mathcal{B}^{\sf Limavg}\enspace.
\tag*{\qed}
\end{equation}
\end{mylemma}
Intuitively, the automaton $\mathcal{C}^{\sf Limavg}$ is obtained from
$\mathcal{C}$ by
connecting the initial and final states of $\mathcal{C}$.
More concretely, 
we add a new state $\star$ that represents both the initial and final state in $\mathcal{C}$,
add transitions from $\star$ to states in $\mathcal{C}$ according to the initial vector $\alpha$,
and add transitions from states in $\mathcal{C}$ to $\star$ according to
the final vector $\beta$. 

The proof of the lemma is technical and deferred to
Appendix~\ref{appendix:lemfiniteGametoLimavg}. 
The basic idea  is as follows. A winning strategy for the
(finite) game for 
$  \mathcal{A}
\sqsubseteq_{\bf G}
\mathcal{B}
$ yields that for the other (infinite) game, by repeating the
strategy. The other direction is similar, except that trap states 
call for special care.

We shall now describe the second equivalence
in~(\ref{eq:GSimToLimAvg}). 
The notion of mean payoff game is from~\cite{ehrenfeucht79positionalstrategies}.
\begin{mydefinition}[mean payoff game]
A \emph{weighted bipartite graph} $\mathcal{G}=(Q_{\Min},Q_{\Max},q_I,E,\gamma)$ consists of
a set
  $Q_{\Min}$  of states for $\Min$, a set  $Q_{\Max}$ of states for $\Max$,
the initial state $q_I\in Q_{\Min}$, a set  $E=Q_{\Min}\times
 Q_{\Max}+Q_{\Max}\times Q_{\Min}$ of edges, and a weight function $\gamma:E\to\mathbb{R}$.

A \emph{mean payoff game} is a game played by two players $\Min$ and
 $\Max$ on a  weighted bipartite graph $\mathcal{G}$.
A \emph{strategy for $\Min$} is a function $\tau_{\Min}:(Q_{\Min}\times Q_{\Max})^*\times Q_{\Min}\to Q_{\Max}$ 
and a \emph{strategy for $\Max$} is a function $\tau_{\Max}:(Q_{\Min}\times Q_{\Max})^+\to Q_{\Min}$.
An infinite run $p_0q_0p_1q_1\dotsc$ on $\mathcal{G}$ (here  $p_{i}\in Q_{\Min}$ and $q_{i}\in Q_{\Max}$)
is the \emph{outcome} of strategies $\tau_{\Min}$ and $\tau_{\Max}$ if $p_0=q_I$ and
for any $i\geq 1$, $\tau_{\Min}(p_0q_0\dotsc q_{i-1} p_i)=q_i$ and $\tau_{\Min}(p_0q_0\dotsc  p_iq_i)=p_{i+1}$.
A strategy $\tau_{\Max}$ for $\Max$ is \emph{winning} if for any strategy $\tau_{\Max}$ for $\Min$, 
their outcome $r_0r_1r_2r_3\dotsc$ satisfies
$\mathrm{liminf}_{N\to\infty}\frac{1}{N}\sum_{i=0}^{N}\bigl(\gamma(r_i,r_{i+1})\bigr)\geq 0$.
%
%
\end{mydefinition}

\begin{mylemma}[\cite{chatterjee10quantitativelanguages}]\label{lem:limavgtoMeanPayoff}
Let  $\mathcal{E}=(Q_{\mathcal{E}},\Sigma,M_{\mathcal{E}},q^0_{\mathcal{E}})$ and 
$\mathcal{F}=(Q_{\mathcal{F}},\Sigma,M_{\mathcal{F}},q^0_{\mathcal{F}})$ be {\sf Limavg} automata. 
We have $\mathcal{E}\sqsubseteq_{\bf G}\mathcal{F}$ if and only if Max wins in 
the mean payoff game $\mathcal{G}_{\mathcal{E},\mathcal{F}}$, where
the game  $\mathcal{G}_{\mathcal{E},\mathcal{F}}=(Q_{\Min},Q_{\Max},q_I,E,\gamma)$ is defined by 
\begin{align*}
&Q_{\Min} = Q_{\mathcal{E}}\times Q_{\mathcal{F}}\enspace, 
\qquad
Q_{\Max} = Q_{\mathcal{E}}\times Q_{\mathcal{F}}\times\Sigma\enspace, 
\qquad
q_I = (q^0_{\mathcal{E}},q^0_{\mathcal{F}})\enspace, 
\\
&E = \bigl\{((p,q),(p',q,a))\,\bigl|\bigr.\,
 M_{\mathcal{E}}(a)_{p,p'}\neq-\infty
 \bigr\}+\bigl\{((p,q,a),(p,q'))\,\bigl|\bigr.\,
 M_{\mathcal{F}}(a)_{q,q'}\neq-\infty \bigr\}\enspace, 
\\
&\gamma((p,q),(p',q,a)) = -M_{\mathcal{E}}(a)_{p,p'}
\quad\text{and}\quad
\gamma((p,q,a),(p,q')) = M_{\mathcal{F}}(a)_{q,q'}\enspace.
\end{align*}

\vspace{-3em}
\qed
\end{mylemma}

We turn to the equivalences in~(\ref{eq:FSimToMeanPayoffGame}). 
The following proposition is interesting for its own sake,
characterizing $\fwd$ for $\smp$-weighted automata
 in terms of mean payoff games. We crucially rely on a result in~\cite{akian12tropicalpolyhedra}.

\begin{myproposition}\label{prop:fwdSimtoGame}
For a pair of $\smp$-weighted automata $\mathcal{A}$ and $\mathcal{B}$,
there exists a mean payoff game
 $\mathcal{G}_{\mathcal{A},\mathcal{B}}^{\bf F}$ such that:
\begin{math}
   \mathcal{A}
\fwd
\mathcal{B}
\end{math}
if and only if Max wins in 
 $\mathcal{G}_{\mathcal{A},\mathcal{B}}^{\bf F}$.
\end{myproposition}

\begin{myproof}
Let $\mathcal{A} = (\mcr{Q}{A}, \Sigma, \mcr{M}{A}, \mcr{\alpha}{A}, \mcr{\beta}{A})$ and $\mathcal{B} = (\mcr{Q}{B}, \Sigma, \mcr{M}{B}, \mcr{\alpha}{B}, \mcr{\beta}{B})$.
By Definition~\ref{def:simMat}, a forward simulation matrix from $\mathcal{A}$ to $\mathcal{B}$ is a matrix $X\in\smp^{\mcr{Q}{B}\times\mcr{Q}{A}}$ that satisfies 
\begin{equation}
\mcr{\alpha}{A} \leq \mcr{\alpha}{B} X \; \land \;  \forall  a \in \Sigma. \, X\mcr{M}{A}( a) \leq \mcr{M}{B}( a)X \; \land \; X\mcr{\beta}{A} \leq \mcr{\beta}{B} \,.\label{eq:fwdSimtoGame1}
\end{equation}
The result in~\cite{akian12tropicalpolyhedra}  reduces: 
\begin{itemize}
 \item 
 existence of a
 nontrivial
 (i.e.\ not $-\infty$) solution of a
 linear inequality
 $A\mathbf{x}\le B\mathbf{x}$, where $A,B$ are matrices over $\smp$ and
       $\mathbf{x}$ is a column vector of variables, to
 \item a mean payoff game. 
\end{itemize}
We therefore need to transform~(\ref{eq:fwdSimtoGame1}) into
 the format $A\mathbf{x}\le B\mathbf{x}$. In particular, 
$\alpha_{\mathcal{A}}$ and $\beta_{\mathcal{B}}$ on both ends
 of~(\ref{eq:fwdSimtoGame1}) should be taken care of.

We shall prove that: there exists a matrix $X$ that satisfies (\ref{eq:fwdSimtoGame1}), if and only if,
there exist $x_{\star,\star}\in\smp$ 
and $X'\in\smp^{\mcr{Q}{B}\times\mcr{Q}{A}}$ 
that satisfy $x_{\star,\star}\neq-\infty$ and
\begin{equation}
x_{\star,\star}\mcr{\alpha}{A} \leq \mcr{\alpha}{B} X' \; \land \;  \forall  a \in \Sigma. \, X'\mcr{M}{A}( a) \leq \mcr{M}{B}( a)X' \; \land \; X'\mcr{\beta}{A} \leq x_{\star,\star} \mcr{\beta}{B}\,.
\label{eq:fwdSimtoGame2}
\end{equation}
Note here that $x_{\star,\star}\mcr{\alpha}{A}$ denotes the vector
$\mcr{\alpha}{A}$ multiplied by the scalar $x_{\star,\star}$. Here
 ``multiplication'' is by the semiring multiplication of $\smp$, that
 is,
addition of real numbers.


Indeed,
if $X$ satisfying (\ref{eq:fwdSimtoGame1}) exists, then $x_{\star,\star}=0$ and $X'=X$ satisfy (\ref{eq:fwdSimtoGame2}).
Conversely, if $x_{\star,\star}\in\smp$ (where
 $x_{\star,\star}\neq-\infty$) and
 $X'\in\smp^{\mcr{Q}{B}\times\mcr{Q}{A}}$ satisfy
 (\ref{eq:fwdSimtoGame2}), then
$X\in\smp^{\mcr{Q}{B}\times\mcr{Q}{A}}$ defined by
 $X_{q,p}=X'_{q,p}-x_{\star,\star}$ satisfies (\ref{eq:fwdSimtoGame1}). 
Here $-x_{\star,\star}$ denotes subtraction of the real number
 $x_{\star,\star}$. It is well-defined and constitutes the inverse of
 (semiring-)multiplication by $x_{\star,\star}$, since 
 $x_{\star,\star}\neq-\infty$.
%
%

It is straightforward to translate~(\ref{eq:fwdSimtoGame2}) into the
 format  $A\mathbf{x}\le B\mathbf{x}$. Then applying the
 result~\cite{akian12tropicalpolyhedra} yields the following mean payoff
 game, Max's winning in which is equivalent to the feasibility
 of~(\ref{eq:fwdSimtoGame2}), hence to that of~(\ref{eq:fwdSimtoGame1}).

The game is played on a graph
$\mathcal{G}_{\mathcal{A},\mathcal{B}}^{\bf F}=(Q^{\bf F}_{\Min},Q^{\bf F}_{\Max},q^{\bf F}_I,E^{\bf F},\gamma^{\bf F})$,
where
\begin{align*}
Q^{\bf F}_{\Min} &= \{x_{\star,\star}\}+\{x_{q,p}\mid q\in\mcr{Q}{B}, p\in\mcr{Q}{A}\}\,, &\\
 Q^{\bf F}_{\Max} &= \mcr{Q}{A} + \Sigma\times\mcr{Q}{B}\times\mcr{Q}{A} + \mcr{Q}{B}\,, &\\
q^{\bf F}_I &= x_{\star,\star} &\\
E^{\bf F} &= E^{\bf F}_1+E^{\bf F}_2 
\text{ where  }E^{\bf F}_1\subseteq Q^{\bf F}_{\Min}\times Q^{\bf F}_{\Max}, E^{\bf F}_2\subseteq Q^{\bf F}_{\Max}\times Q^{\bf F}_{\Min}\text{ and}  & \\
&E^{\bf F}_1 = \{(x_{\star,\star},p)\mid (\mcr{\alpha}{A})_p\neq-\infty\}\}+\{(x_{q,p},( a,q,p'))\mid \mcr{M}{A}( a)_{p,p'}\neq-\infty\}\\
& \mspace{30mu}+\{(x_{q,p},q)\mid (\mcr{\beta}{A})_p\neq-\infty\} &\\
&E^{\bf F}_2 = \{(p,x_{q,p})\mid (\mcr{\alpha}{B})_q\neq-\infty\}+\{(( a,q,p),x_{q',p})\mid \mcr{M}{B}( a)_{q,q'}\neq-\infty\} \\
& \mspace{30mu}+\{(q,x_{\star,\star})\mid (\mcr{\beta}{B})_q\neq-\infty\}\,,&
\end{align*}
\vspace*{-2em}
\begin{align*}
\gamma^{\bf F}(x_{\star,\star},p) &= -(\mcr{\alpha}{A})_p, & \gamma^{\bf F}(x_{q,p},( a,q,p'))&=-\mcr{M}{A}( a)_{p,p'}, & \gamma^{\bf F}(x_{q,p},q)&=-(\mcr{\beta}{A})_p \\
\gamma^{\bf F}(p,x_{q,p}) &= (\mcr{\alpha}{B})_q, & \gamma^{\bf F}((a,q,p),x_{q',p})&=\mcr{M}{B}( a)_{q,q'}, & \gamma^{\bf F}(q,x_{\star,\star})&=(\mcr{\beta}{B})_q \,.
\end{align*}
This concludes the proof.
\qed
\end{myproof}

Finally, we bridge the rightmost conditions
in~(\ref{eq:GSimToLimAvg}--\ref{eq:FSimToMeanPayoffGame}) and prove Theorem~\ref{thm:KleisliSimIsGSim}.

\newproof{proofKlisIG}{Proof of Theorem~\ref{thm:KleisliSimIsGSim}}
\begin{proofKlisIG}
Let $\mathcal{A} = (\mcr{Q}{A}, \Sigma, \mcr{M}{A}, \mcr{\alpha}{A}, \mcr{\beta}{A})$ and $\mathcal{B} = (\mcr{Q}{B}, \Sigma, \mcr{M}{B}, \mcr{\alpha}{B}, \mcr{\beta}{B})$.

By Lemma~\ref{lem:finiteGametoLimavg} and Lemma~\ref{lem:limavgtoMeanPayoff}, 
$\mathcal{A}\sqsubseteq_{\bf G}\mathcal{B}$ is equivalent to existence of a winning strategy for $\Max$ in 
the mean payoff game played on a graph $\mathcal{G}_{\mathcal{A}^{\sf
 Limavg},\mathcal{B}^{\sf Limavg}}^{\bf G}=(Q^{\bf G}_{\Min},Q^{\bf
 G}_{\Max},q^{\bf G}_I,E^{\bf G},\gamma^{\bf G})$. It is 
defined by 
\begin{align*}
Q^{\bf G}_{\Min} &= (\{\star\}+\mcr{Q}{A})\times(\{\star\}+\mcr{Q}{B}) \,,&\\
Q^{\bf G}_{\Max} &= (\{\star\}+\mcr{Q}{A})\times(\{\star\}+\mcr{Q}{B})\times(\{\mbox{$?$}\}+\{\mbox{$!$}\}+\Sigma) \,,&\\
q^{\bf G}_I &= (\star,\star) \,,&\\
E &= E^{\bf G}_1+E^{\bf G}_2 \text{ s.t.  }E^{\bf G}_1\subseteq Q^{\bf G}_{\Min}\times Q^{\bf G}_{\Max}, E^{\bf G}_2\subseteq Q^{\bf G}_{\Max}\times Q^{\bf G}_{\Min}\text{ and}  & \\
&E^{\bf G}_1 = \{((\star,\star),(p,\star,\mbox{$?$}))\mid (\mcr{\alpha}{A})_p\neq-\infty\}\}+\{((p,q),(p',q, a))\mid \mcr{M}{A}( a)_{p,p'}\neq-\infty\}\\
& \mspace{30mu}+\{((p,q),(\star,q,\mbox{$!$}))\mid (\mcr{\beta}{A})_p\neq-\infty\} &\\
&E^{\bf G}_2 = \{((p,\star,\mbox{$?$}),(p,q))\mid (\mcr{\alpha}{B})_q\neq-\infty\}+\{((p,q, a),(p,q'))\mid \mcr{M}{B}( a)_{q,q'}\neq-\infty\} \\
& \mspace{30mu}+\{((\star,q, \mbox{$!$}),(\star,\star))\mid (\mcr{\beta}{B})_q\neq-\infty\} \,,&
\end{align*}
\vspace*{-2em}
\begin{align*}
\gamma^{\bf G}((\star,\star),(p,\star,\mbox{$?$})) &\!=\! -(\mcr{\alpha}{A})_p, & \mspace{-14mu}\gamma^{\bf G}((p,q),(p',q, a))&\!=\!-\mcr{M}{A}( a)_{p,p'}, & \mspace{-16mu}\gamma^{\bf G}((p,q),(\star,q,\mbox{$!$}))&\!=\!-(\mcr{\beta}{A})_p \\
\gamma^{\bf G}((p,\star,\mbox{$?$}),(p,q)) &\!=\! (\mcr{\alpha}{B})_q, & \mspace{-14mu}\gamma^{\bf G}((p,q, a),(p,q'))&\!=\!\mcr{M}{B}( a)_{q,q'}, & \mspace{-16mu}\gamma^{\bf G}((\star,q, \mbox{$!$}),(\star,\star))&\!=\!(\mcr{\beta}{B})_q \,.
\end{align*}
It is not hard to see that the last graph $\mathcal{G}_{\mathcal{A}^{\sf
 Limavg},\mathcal{B}^{\sf Limavg}}^{\bf G}$ is equivalent to 
$\mathcal{G}_{\mathcal{A},\mathcal{B}}^{\bf F}$ in
 Proposition~\ref{prop:fwdSimtoGame}: the former has extra states but
 they are all unreachable.
%
\qed
\end{proofKlisIG}

\begin{myremark}[complexity]
 The decision problem of mean payoff games is known to be in 
 ${\bf NP} \cap {\bf
 co}\text{-}{\bf NP}$~\cite{zwick96thecomplexity}; it has  a pseudo polynomial-time
 algorithm, too~\cite{zwick96thecomplexity}.
 By~\cite{akian12tropicalpolyhedra} this problem is equivalent to the
 feasibility problem of linear inequalities in $\smp$. For the latter
 problem, the
 algorithm  proposed in~\cite{butkovic06astrongly} (for solving linear
 equalities) can be utilized; this algorithm is
 shown in~\cite{bezem08exponentialbehaviour} to be
  superpolynomial. These results give upper bounds for the complexity of
 $\fwd$ and  $\sqsubseteq_{\bf G}$, 
by Theorem~\ref{thm:KleisliSimIsGSim} and the
 subsequent lemmas.




 Similarly to $\spt$-weighted automata, language inclusion between
 $\smp$-weighted automata is known to be undecidable~\cite{krob92theequality}.
 We note that, by Theorem~\ref{thm:KleisliSimIsGSim}, applying FPE or BPE (Section~\ref{sec:partExec})
 increases the likelihood of $\sqsubseteq_{\bf G}$ (in the sense of
 Theorem~\ref{thm:adequacyOfFPEBPE}).
 We additionally note that, by exploiting symmetry of forward and backward simulation
 matrices (Remark~\ref{rem:oppositeAutom}), we could define ``backward
 G-simulation'' as a variation of Definition~\ref{def:simgame}.
\end{myremark}

\subsection{Algorithms}\label{subsec:algoMaxPlus}
We can give, much like in Section~\ref{subsec:algoSpt}, an algorithm $\mplangfwdalgo$ that tries to establish language
inclusion between two $\smp$-weighted automata. The algorithms
\mpfpealgo\ and \mpbpealgo\ that apply FPE and BPE, respectively, are
the same as in Section~\ref{subsec:algoSpt}. The only difference is in
the algorithm \mpsimfwdalgo\ that
  searches for a forward simulation matrix  between $\smp$-weighted
  automata.

\begin{myalgorithm}[\mpsimfwdalgo]
\noindent{\bf Input}: 
A pair of $\smp$-weighted automata $\mathcal{A}$ and $\mathcal{B}$.

\noindent{\bf Output}: 
A forward simulation matrix $X$ from $\mathcal{A}$ to $\mathcal{B}$ if such $X$ exists, and
``$\mathsf{No}$" otherwise.
%

\noindent{\bf Procedure}:
This algorithm first transforms the constraints (\ref{eq:simMatFwd}) in Definition~\ref{def:simMat} into an
       inequality $A\mathbf{x}\le B\mathbf{x}$, which in turn is made into a
       linear \emph{equality} $A'\mathbf{x'}=B'\mathbf{x'}$ by adding slack
       variables that absorb the difference between the left hand side
 and the right hand side (note that this trick depends on the choice of
 the semiring $\smp$).       
        The last equality is solved by the algorithm
in~\cite{butkovic06astrongly}.
If a solution is found by the algorithm in~\cite{butkovic06astrongly}, then the found solution is output.
Otherwise, ``$\mathsf{No}$'' is output.
%
\end{myalgorithm}
The algorithm in~\cite{butkovic06astrongly} is known not to be  a polynomial-time algorithm~\cite{bezem08exponentialbehaviour}.


Similarly we can obtain an algorithm $\mplangbwdalgo$ that tries to
establish language inclusion by searching for a \emph{backward}
simulation matrix. It uses an algorithm $\mpsimbwdalgo$ that searches
for a backward simulation matrix---the latter is similar to
$\mpsimfwdalgo$ and relies on the algorithm in~\cite{butkovic06astrongly}.

\subsection{Implementation, Experiments and Discussions}\label{subsec:expResForMaxPlus}
We implemented two programs: \mpsim and \mpPE. Both of them are available in~\cite{implPtMpTree}.

\begin{itemize}
 \item 
%
The program \mpsim executes \mpsimfwdalgo or \mpsimbwdalgo in the previous section.
Since we did not find  actual
software  available, we implemented (in C++) the algorithm
in~\cite{butkovic06astrongly} 
(for solving $\smp$-linear equalities)
as part of the program \mpsim.


 %
 \item 
 \mpPE 
 executes \mpfpealgo or \mpbpealgo in the previous section. It
 is as in Section~\ref{sec:plustimesWeightedAutomata}.
 It simply uses the whole state space as   the parameter $P$.
\end{itemize}


 Experiments were done on 
a MacBook Pro laptop with a Core i5 processor (2.6 GHz, 2 cores) and 16 GB RAM. 
There we faced a difficulty
of finding a benchmark example: although small examples are not hard to
come up with by human efforts, we could not find a good example that has
parameters (like $G,S$ in Table~\ref{table:resultsForGradesProtocol})
and allows for experiments with problem instances of a varying size.

We therefore ran \mpsim for:
\begin{itemize}
 \item the problem if $\mathcal{A}\sqsubseteq_{\bf F}\mathcal{A}$ for
       randomly generated $\mathcal{A}$, and
 \item the problem if $\mathcal{A}\sqsubseteq_{\bf F}\mathcal{B}$ for
       randomly generated $\mathcal{A,B}$,
\end{itemize}
and measured time and memory consumption. Although the answers are known
by construction (positive for the former, and almost surely negative for the
latter), actual calculation via linear inequality constraints  gives
us an idea about resource consumption of our simulation-based method
when it is applied to real-world problems.

\begin{figure}
\begin{minipage}{0.5\hsize}
\begin{flushleft}
\includegraphics[width=6.5cm]{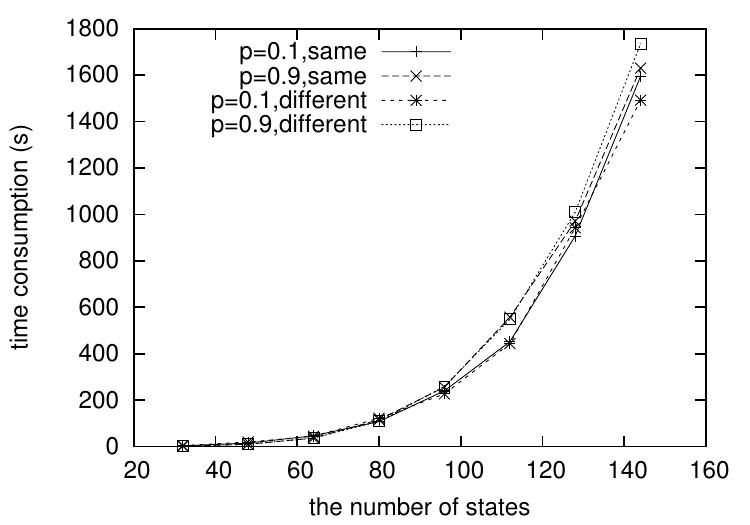}
\end{flushleft}
\end{minipage}
\begin{minipage}{0.5\hsize}
\begin{flushright}
\includegraphics[width=6.5cm]{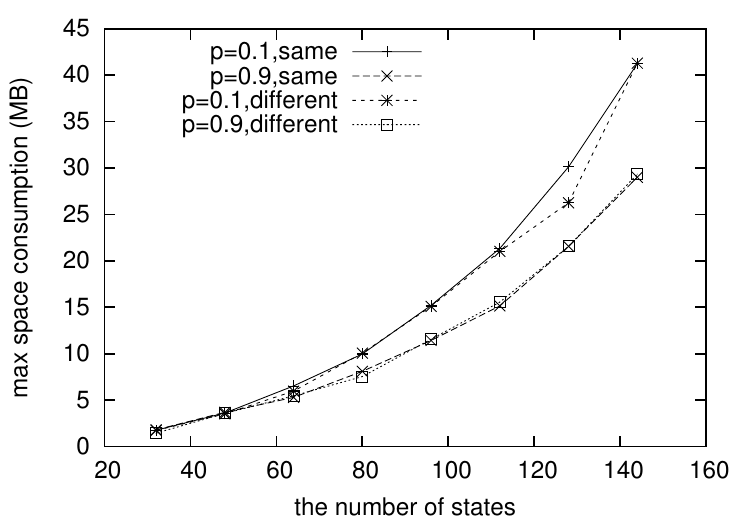}
\end{flushright}
\end{minipage}
\vspace*{0.0cm}
\caption{Time and max space consumption for \mpsim}
\label{fig:simtimespace}
\end{figure}

The outcome is as shown in Figure~\ref{fig:simtimespace}. The parameter
$p$ is the probability with which an $a$-transition exists given a source
state, a target state, and a character $a\in \Sigma$. Its weight is
chosen from $\{0,1,\dotsc, 16\}$ subject to the uniform distribution.
``Same'' means checking $\mathcal{A}\sqsubseteq_{\bf F}\mathcal{A}$ 
and ``difference'' means checking  $\mathcal{A}\sqsubseteq_{\bf
F}\mathcal{B}$ 
(see above). The two problem settings resulted in comparable performance.

We observe that space consumption is not so big a problem as in  the
$\spt$ case (Section~\ref{subsec:sptImpl}).  Somehow unexpectedly, there is
no big performance gap between the sparse case ($p=0.1$) and the dense
case ($p=0.9$); in fact the sparse case consumes slightly more memory.
Consumption of both time and space
grows  faster than linearly, which poses a question about the
scalability of our approach. That said, our current implementation
of the algorithm in~\cite{butkovic06astrongly} leaves a lot of room for
further optimization:  one possibility is use of dynamic programming (DP).
After all,
it is  an advantage of our  approach that a simulation problem is
reduced to linear inequality constraints, a subject of extensive
research  efforts (cf.~Section~\ref{subsec:SPTcomparisonWithOtherSim} and~Section~\ref{subsec:gameSimAsFwdSimMat}).



\section{Matrix Simulation for Polynomial
 Functors}\label{sec:matSimPolyF}
In the previous sections, we concentrated our attention on 
matrix simulations for weighted (word) automata.
They are special cases of $(T,F)$-systems where $T$ is a multiset monad over some semiring and $F=1+\Sigma\times(\place)$.
However, according to the general theory developed in \cite{hasuo07generictrace,hasuo06genericforward},
Kleisli simulation can be defined and its soundness can be proved for more general $(T,F)$-systems.

In this section, we generalize the functor $F$ from
$1+\Sigma\times(\place)$ (that we have been using) to an arbitrary polynomial functor:
\begin{equation}\label{eq:polynomialF}
F \;::=\; (\place) \;|\; \Sigma \;|\; F_1\times F_2 \;|\; \coprod_{i\in I}F_i\,.
\end{equation}
Such a generalized $(T,F)$-system---where $T$ is again  a multiset monad---represents a system called a 
\emph{weighted tree automaton},
whose concrete (not coalgebraic) theory can be found in~\cite[Chapter 9]{droste09handbookof}, for example.
Here, the choice of $F$ determines the shape of trees to which the automaton
assigns a weight.

This section is organized as follows.
In Section~\ref{subsec:partExecCat} we
build on~\cite{hasuo06genericforward} and introduce the  notion of
forward partial execution (FPE) on the coalgebraic level of abstraction. 
We also prove its correctness (soundness and adequacy); the
overall coalgebraic theory (i.e.\ the one
in~\cite{hasuo06genericforward} augmented with FPE)
generalizes the one in Section~\ref{sec:partExec} for weighted (word) automata.
The abstract theory thus obtained is applied 
in Section~\ref{subsec:theoryTree} to weighted tree
automata---i.e.\ $(T,F)$-systems with $T$ being a multiset monad and $F$
being a polynomial functor. Much like for word automata, Kleisli
simulations for tree automata are represented by matrices, subject
however to \emph{nonlinear} inequality constraints.
Finally
in Section~\ref{subsec:implTree} our proof-of-concept implementation is presented.
 


\subsection{Forward Partial Execution, Categorically}\label{subsec:partExecCat}



\begin{mydefinition}[FPE, categorically]\label{def:catFPE}
\emph{Forward partial execution} (FPE) for $(T,F)$-systems is 
a transformation that
 takes: a $(T,F)$-system $\mathcal{X}=(X,\,s:\{\bullet\}\kto X,\,c:X\kto FX)$ and
 a parameter $X_1\subseteq X$ as input; and returns a $(T,F)$-system $\mathcal{X}_{\FPE,X_1}$.

 The outcome $\mathcal{X}_{\FPE,X_1}$ is defined as follows.
Let $c_{1}$ and $c_{2}$ be the domain restrictions of $c$ to $X_{1}$
 and $X_{2}$, respectively, via the coprojections $\kappa_{i}:X_{i}\kto
 X$. That is explicitly:
\begin{displaymath}
 c_1\;=\;c\odot\kappa_1
 \;:\; X_{1}\kto FX
 \quad\text{and}\quad
c_2\;=\;c\odot\kappa_2
 \;:\; X_{2}\kto FX\enspace,
\end{displaymath}
where $\odot$
 denotes composition of Kleisli arrows (Definition~\ref{def:KlArrow}).
We define  $X_{2}=X\setminus X_{1}$ (hence $X=X_{1}+X_{2}$);
the system  $\mathcal{X}_{\FPE,X_1}$ is now given by
 \begin{displaymath}
\mathcal{X}_{\FPE,X_1}
\;=\;
\left(\,
\begin{array}{l}
  F(X)+X_2\,,
 \,
 \\
 \xymatrix@1@C+1em{
 {\{\bullet\}}
   \kar[r]^-{s}    
 &
 {X_{1}+X_{2}}
   \kar[r]^-{c_{1}+\id}
 &
 {F(X)+X_{2}}
 }
 \,,
 \\
 \xymatrix@1@C+1em{
 {F(X)+X_{2}}
   \kar[r]^-{[\id,c_{2}]}    
 &
 {F(X_{1}+X_{2})}
   \kar[r]^-{\overline{F}(c_{1}+\id)}
 &
 {F\bigl(F(X)+X_{2}\bigr)}
 }
\end{array}
\,\right)\,.
\end{displaymath}
Here $\overline{F}\colon \Kl(T)\to\Kl(T)$ is the canonical lifting of
$F:{\bf Sets}\to{\bf Sets}$ (see~\cite{jacobs12introductionto} for a concrete definition).

\end{mydefinition}

For the last categorical generalization of FPE, we shall establish 
its correctness---soundness, adequacy and monotonicity---much like in
Section~\ref{subsec:correctnessFPEBPEWord}.


\begin{mytheorem}[soundness of categorical
 FPE]\label{thm:catsoundnessFPE}
Let $X_{1}$ and $Y_{1}$ be  arbitrary subsets of the state spaces of
 $\mathcal{X}$ and  $\mathcal{Y}$, respectively.
 Each of the following implies $\tr(\mathcal{X})\sqsubseteq
 \tr(\mathcal{Y})$.
\begin{enumerate}
 \item $\mathcal{X}_{\FPE,X_1}\fwd\mathcal{Y}$ 
 \item $\mathcal{X}\bwd\mathcal{Y}_{\FPE,Y_1}$  \qed
\end{enumerate}
\end{mytheorem}


\begin{mytheorem}[adequacy of categorical FPE]\label{thm:catadequacyOfFPE}
Let $X_{1}$ and $Y_{1}$ be  arbitrary subsets of the state spaces of
 $\mathcal{X}$ and  $\mathcal{Y}$, respectively.
We have:
\begin{enumerate}
 \item
      $\mathcal{X}\fwd\mathcal{Y}\;\Rightarrow\;\mathcal{X}_{\FPE,X_1}\fwd\mathcal{Y}$ 
 \item 
$\mathcal{X}\bwd\mathcal{Y}\;\Rightarrow\;\mathcal{X}\bwd\mathcal{Y}_{\FPE,Y_1}$
 \qed
\end{enumerate}
\end{mytheorem}




 The last two theorems are immediate consequences of the following
 lemma. The last is a categorical generalization of Lemma~\ref{lem:peproperty}.

\begin{mylemma}\label{lem:catpeproperty}
For each subset $X_1$ of the state space of
 $\mathcal{X}$, we have:
\begin{enumerate}
 \item $\mathcal{X}\bwd\mathcal{X}_{\FPE,X_1}$
 \item $\mathcal{X}_{\FPE,X_1}\fwd\mathcal{X}$
\end{enumerate}
\end{mylemma}

\begin{myproof}
1. We define $g:X_1+X_2\kto F(X)+X_2$ by $g=c_1+\id$. Then we have
\begin{flalign*}
\overline{F}(g)\odot c &= \overline{F}(c_1+\id)\odot [c_1,c_2] \\
&= \overline{F}(c_1+\id)\odot [\id,c_2]\odot(c_1+\id) \\
&= \bigl(\overline{F}(c_1+\id)\odot [\id,c_2]\bigr)\odot g\enspace,\\
g\odot s &= (c_1+\id)\odot s \;.
\end{flalign*}
Note here that $\overline{F}(c_1+\id)\odot [\id,c_2]$ is 
the dynamics of the system $\mathcal{X}_{\FPE,X_1}$. The above
 equalities witness
 that $g$ indeed satisfies the inequalities  required in the
 definition of backward Kleisli simulation, a generalization of
 Definition~\ref{def:ksim} that is found
 in~\cite{hasuo06genericforward}. 
Hence $g$ is a backward simulation from $\mathcal{X}$ to $\mathcal{X}_{\FPE}$.
\auxproof{\[
\begin{xy}
0;/r0.8mm/:
(0,8)*{X_1} = "ldx1",
(4,4)*{+} = "ld+",
(8,0)*{X_2} = "ldx2",
(-8,44)*{F} = "luF",
(-5,44)*{\Biggl(} = "lulp",
(0,48)*{X_1} = "lux1",
(4,44)*{+} = "lu+",
(8,40)*{X_2} = "lux2",
(13,44)*{\Biggr)} = "lulp",
(40,8)*{F(X)} = "rdx1",
(47,4)*{+} = "rd+",
(52,0)*{X_2} = "rdx2",
(35,26)*{F} = "rF",
(38,26)*{\Biggl(} = "rlp",
(42,30)*{X_1} = "rx1",
(47,26)*{+} = "r+",
(52,22)*{X_2} = "rx2",
(56,26)*{\Biggr)} = "rlp",
(31,44)*{F} = "ruF",
(34,44)*{\Biggl(} = "rulp",
(40,48)*{F(X)} = "rux1",
(47,44)*{+} = "ru+",
(52,40)*{X_2} = "rux2",
(57,44)*{\Biggr)} = "rulp",
\kar ^{c_1} "ldx1"; (3,36)
\kar _{c_2} "ldx2"; (5,36)
\kar ^{c_1} "ldx1";"rdx1"
\kar ^{\id} "ldx2";"rdx2"
\kar ^{\id} "rdx1"; (46,19)
\kar _{c_2} "rdx2"; (48,19)
\kar ^{c_1} "lux1";"rux1"
\kar ^(.3){\id} "lux2";"rux2"
\kar ^{c_1} "rx1";(42,46)
\kar ^{\id} "rx2";"rux2"
\end{xy}
\]
}%

The item 2.\ is proved similarly: the same $g$ that we used in the proof of
 the item 1.\ is
shown to be a forward simulation from $\mathcal{X}_{\FPE}$ to $\mathcal{X}$.
\qed
\end{myproof}

Finally, we present a monotonicity result. It generalizes Proposition~\ref{prop:monotonicityOfFPEBPE}.

\begin{myproposition}[monotonicity of categorical FPE]\label{prop:catmonotonicityOfFPE}
Assume $X_{1}\subseteq X'_{1}$ and $X_{2}\subseteq X'_{2}$. We have:
\begin{enumerate}
 \item
      $\mathcal{X}_{\FPE,X_{1}}\fwd\mathcal{Y}\;\Rightarrow\;\mathcal{X}_{\FPE,X'_{1}}\fwd\mathcal{Y}$
 \item
      $\mathcal{X}\bwd\mathcal{Y}_{\FPE,Y_{2}}\;\Rightarrow\;\mathcal{X}\bwd\mathcal{Y}_{\FPE,Y'_{2}}$ 
 \qed
\end{enumerate}
\end{myproposition}



Categorical formalization of BPE  is still open---it seems that BPE 
in Section~\ref{sec:partExec} exists somewhat
coincidentally, for the specific functor $F=1+\Sigma\times (\place)$ for which an opposite
automaton is canonically defined (cf.\ 
Remark~\ref{rem:oppositeAutom}).



\subsection{Matrix Simulations for Weighted Tree Automata}\label{subsec:theoryTree}
Here we exploit the general theory we have just obtained (by
augmenting~\cite{hasuo06genericforward} with FPE). We shall apply it to a class
of systems that is more general than what we
have been dealing with in the previous sections (namely weighted (word) automata). Specifically, we use the
same monads for $T$ but allow arbitrary polynomial functors for $F$.
Such systems are naturally identified with \emph{weighted tree
automata}, where a finite-depth tree, instead of a finite word, gets a
weight assigned.


We first define the notion of tree.
\begin{mydefinition}
A \emph{ranked alphabet} is a family $\Sigma=(\Sigma_n)_{n\in\mathbb{N}}$ of countable sets that are indexed by natural numbers called \emph{arities}.

The set  $\text{Tree}(\Sigma)$  of (finite-depth) \emph{trees} over a ranked alphabet $\Sigma$
 is defined
 in the obvious way. Concretely,  $\text{Tree}(\Sigma)$ is  the smallest
 set such that: for each $a\in\Sigma_n$,
 $t_0,t_1,\dotsc,t_{n-1}\in \Tree(\Sigma)$ implies $a(t_0,t_1,\dotsc,t_{n-1})\in\Tree(\Sigma)$.
\end{mydefinition}
We introduce weighted tree automata, firstly in concrete terms.
\begin{mydefinition}[$\mathcal{S}$-weighted tree automaton, weighted tree language]\label{def:SWeightedTreeAutom}
Let $\mathcal{S}=(S,+_{\mathcal{S}},0_{\mathcal{S}},\times_{\mathcal{S}},
1_{\mathcal{S}},\sqsubseteq)$ be a commutative cppo-semiring.  
An \emph{$\mathcal{S}$-weighted tree automaton} is a quadruple $\mathcal{A}=(Q,\Sigma,M,\alpha)$
consisting of a countable state space $Q$, a ranked alphabet $\Sigma=(\Sigma_n)_{n\in\mathbb{N}}$,
transition matrices $M(a)\in \mathcal{S}^{Q\times Q^n}$ for each $a\in\Sigma_n$, and the initial row vector $\alpha\in\mathcal{S}^Q$.

An $\mathcal{S}$-weighted tree automaton $\mathcal{A}=(Q,\Sigma,M,\alpha)$ yields a \emph{weighted tree language} 
$L(\mathcal{A}):\text{Tree}(\Sigma)\to S$. 
It is defined by $L(\mathcal{A})(t)=\alpha\cdot \Phi(t)$---the 
 product of a row vector $\alpha$ and a column vector $\Phi(t)$---where
$\Phi:\Tree(\Sigma)\to\mathcal{S}^Q$ is defined as follows,
by induction on the depth of trees.
\[
\Phi\bigl(a(t_0,t_1,\dotsc,t_{n-1})\bigr)\;=\;M(a)\bigl(\,\Phi(t_0)\otimes\Phi(t_1)\otimes\dots\otimes\Phi(t_{n-1})\,\bigr)
\qquad
\text{for each $a\in\Sigma_n$.}
\] 
\end{mydefinition}
The final column vector $\beta$ in Definition~\ref{def:SWeightedAutom} do not appear here; transition matrices $M(a)$ 
for $a\in\Sigma_0$ play the corresponding role.

The way weighted tree automata in the above operate can be understood 
in two different modes---a \emph{top-down} one and a \emph{bottom-up} one. 
In the top-down mode, we regard the automaton as one that reads the input tree from the root towards the leaves with its input 
head being split in its course; while in the bottom-up mode, the automaton reads the input tree in the opposite direction with its input head merged in its course.

It is known that these two modes define different languages for \emph{deterministic} tree automata~\cite{comonDGLJLTT07treeautomata}.
However, 
for nondeterministic tree automata as well as weighted ones, they define the same languages (of finite-depth trees).

Language inclusion between two $\mathcal{S}$-weighted tree automata is
defined similarly to the case with $\mathcal{S}$-weighted automata.
\begin{mydefinition}[language inclusion]\label{def:treelangincl}
Let $\mathcal{A}$ and $\mathcal{B}$ be $\mathcal{S}$-weighted tree automata.
We say the language of $\mathcal{A}$ is \emph{included} in the language of $\mathcal{B}$ and 
write $L(\mathcal{A})\sqsubseteq L(\mathcal{B})$ if, for all $t\in\Tree(\Sigma)$, $L(\mathcal{A})(t)\sqsubseteq L(\mathcal{B})(t)$.
\end{mydefinition}

Similarly to $\mathcal{S}$-weighted (word) automata (see Proposition~\ref{prop:SemiringWeightedAutomAsCoalg}), 
$\mathcal{S}$-weighted tree automata are instances of $(T,F)$-systems.

\begin{mydefinition}[the functor $F_{\Sigma}$]
 It is standard that a ranked alphabet $\Sigma$ gives rise to a
 polynomial functor. It is given as follows and is denoted by $F_{\Sigma}$.
\begin{displaymath}
 F_{\Sigma}
 \;=\;
 \coprod_{n\in\mathbb{N}}\Sigma_n\times(\place)^n
 \quad:\;\Sets\longrightarrow\Sets\enspace.
\end{displaymath}
\end{mydefinition}
\begin{myproposition}[weighted tree automata as $(T,F)$-systems]\label{prop:SemiringWeightedTreeAutomAsCoalg}
Let $\mathcal{S}$ be a commutative cppo-semiring. 
An $\mathcal{S}$-weighted tree automaton $\mathcal{A}=(Q, (\Sigma_n)_{n\in\mathbb{N}}, M, \alpha)$
gives rise to an $\bigl(\mss,F_{\Sigma})$-system
 $\mathcal{X_{A}}=(Q,s_{\mathcal{A}},c_{\mathcal{A}})$ defined as follows. 
The function $s_{\mathcal{A}}\colon \{\bullet\}\to \mss Q$ is given by $s_{\mathcal{A}}(\bullet)(x)=\alpha_x$; and 
$c_{\mathcal{A}}\colon Q\to \mss(F_{\Sigma}Q)$  is given by
\[
c_{\mathcal{A}}(x)\bigl(a,(y_0,y_1,\dotsc,y_{n-1})\bigr)=M(a)_{x,(y_0,y_1,\dotsc,y_{n-1})}
\]
 where $a\in\Sigma_n$. \qed
\end{myproposition}

The last identification 
allows us to apply the general results
in~\cite{hasuo06genericforward,hasuo07generictrace} to 
 weighted tree automata. One of the results characterizes
 \emph{coalgebraic trace semantics} by a final coalgebra in the Kleisli
 category $\Kl(\mss)$; it is easy to see that, for weighted tree automata, 
coalgebraic trace semantics is nothing but the weighted tree language
concretely defined in Definition~\ref{def:SWeightedTreeAutom}.

The notions of forward and backward Kleisli simulation is defined
in~\cite{hasuo06genericforward} in categorical terms; and a
(categorical) proof of their soundness against coalgebraic trace
semantics is presented. Much like in the previous sections, we shall now 
characterize Kleisli simulations for weighted tree automata by matrices;
their soundness then follows from the above mentioned categorical proof.

The following lemma is a generalization of
Lemma~\ref{lem:actionOfFOnArrowsByMatrices}. It introduces the matrix
representation of the action of the functor $F_{\Sigma}$ on Kleisli arrows.

\begin{mylemma}\label{lem:actionOfFOnArrowsByMatricesTree}
Let $f:A\kto B$ be a Kleisli arrow in $\Kl(\mss)$ and $M_f$ be its matrix representation (see Definition~\ref{def:matrixReprOfKlArr}). 
Then the matrix representation 
$M_{\overline{F_{\Sigma}}f}$ 
of the arrow $\overline{F_{\Sigma}}f$ 
is given by
\begin{equation}\label{eq:matrixPolynomFunc}
\bigoplus_{n\in\mathbb{N}}\bigr(I_{\Sigma_n}\otimes M_f^{\otimes
 n}\bigr)
\quad\in\;\mathcal{S}^{(F_{\Sigma}A)\times (F_{\Sigma}B)}\,. 
\end{equation}
Here, for each $n\in\mathbb{N}$ and $X_n\in\mathcal{S}^{A_n\times B_n}$,
$\bigl(\bigoplus_{n\in\mathbb{N}}X_n\bigr) \in \mathcal{S}^{(\coprod_{n\in\mathbb{N}}A_n)\times(\coprod_{n\in\mathbb{N}}B_n)}$
is defined by 
\begin{equation*}
\Bigl(\bigoplus_{n\in\mathbb{N}} X_n\Bigr)_{x,y}=\begin{cases} (X_n)_{x,y} & (x\in A_n, y\in B_n) \\ 0 & (\text{\rm otherwise}) \end{cases}
\end{equation*}
that is a generalization of the binary operation $\oplus$. 
The matrix $X^{\otimes n}$ is defined by
\begin{equation*}
X^{\otimes n}=\underbrace{X\otimes X\otimes\dots\otimes X}_{n}\enspace.
 \tag*{\qed}
\end{equation*}
\end{mylemma}
We note that the matrix in~(\ref{eq:matrixPolynomFunc}) is equivalently
 expressed as
\begin{math}
\bigoplus_{n\in\mathbb{N}}\bigr(
\underbrace{M_f^{\otimes n}
\oplus
M_f^{\otimes n}
\oplus
\cdots
\oplus
M_f^{\otimes n}
}_{|\Sigma_{n}|}\bigr)
\end{math}.

In what follows, Definition~\ref{def:simMatTree} and  Theorem~\ref{thm:soundnessOfMatSimTree} 
are parallel to Definition~\ref{def:simMat} and Corollary~\ref{cor:soundnessOfMatSim}, respectively.

\begin{mydefinition}[simulation matrix]\label{def:simMatTree}
Let $\mathcal{A}=(\mcr{Q}{A},\Sigma,\mcr{M}{A},\mcr{\alpha}{A})$ and $\mathcal{B}=(\mcr{Q}{B},\Sigma,\mcr{M}{B},\mcr{\alpha}{B})$ be
$\mathcal{S}$-weighted tree automata.
 \begin{itemize}
  \item A matrix $X \in S^{\mcr{Q}{B}\times\mcr{Q}{A}}$ is a
	\emph{forward simulation matrix} from $\mathcal{A}$  to
	$\mathcal{B}$ if
	\begin{displaymath}
	          \mcr{\alpha}{A} \sqsubseteq \mcr{\alpha}{B} X\enspace,
	 \quad \text{and}\quad
            X\cdot\mcr{M}{A}(a) \sqsubseteq \mcr{M}{B}(a)\cdot (X^{\otimes n})
	 \quad\text{for any $n\in\mathbb{N}$ and  $a \in \Sigma_n$.}
	\end{displaymath}
  \item A matrix $X \in S^{\mcr{Q}{A}\times\mcr{Q}{B}}$ is  a
	\emph{backward simulation matrix} from $\mathcal{A}$  to
	$\mathcal{B}$ if
	\begin{displaymath}
	 \alpha_{\mathcal{A}}X \sqsubseteq \alpha_{\mathcal{B}}
	 \enspace,
	 \quad  \text{and}\quad
	 M_{\mathcal{A}}(a)\cdot (X^{\otimes n}) \sqsubseteq X\cdot
	 M_{\mathcal{B}}(a)
	 \quad\text{for any $n\in\mathbb{N}$ and  $a \in \Sigma_n$.}
	\end{displaymath}
 \end{itemize}
\end{mydefinition}

Similarly to the case of semiring weighted automata, we write $\mathcal{A}\fwd\mathcal{B}$ and $\mathcal{A}\bwd\mathcal{B}$
if there exists a forward and backward matrix simulation from $\mathcal{A}$ to $\mathcal{B}$, respectively.

\begin{mytheorem}[soundness]\label{thm:soundnessOfMatSimTree}
Let $\mathcal{A}$ and $\mathcal{B}$ be $\mathcal{S}$-weighted tree automata.
Existence of a forward or backward simulation matrix from $\mathcal{A}$ to $\mathcal{B}$---i.e.\ 
 $\mathcal{A}\fwd\mathcal{B}$ or  $\mathcal{A}\bwd\mathcal{B}$---witnesses 
language inclusion $\lang(\mathcal{A})\sqsubseteq\lang(\mathcal{B})$.
\end{mytheorem}
\begin{myproof}
Simulation matrices in Definition~\ref{def:simMatTree} coincide with
 Kleisli simulations in the general theory
 of~\cite{hasuo06genericforward}. The latter is sound with respect to 
 coalgebraic trace
 semantics~\cite{hasuo06genericforward,hasuo07generictrace}; and the
 last coincides with the weighted tree language in
 Definition~\ref{def:SWeightedTreeAutom}. \qed
\end{myproof}

We note that, differently from matrix simulations for semiring weighted automata (Definition~\ref{def:simMat}),
the inequalities in Definition~\ref{def:simMatTree} are not necessarily
linear. For example,
\begin{displaymath}\scriptsize
\begin{aligned}
 X^{\otimes 2} 
 \;=\; \begin{pmatrix}
  & \vdots &  \\
 \cdots & \begin{xy}(0,0)*+[F]{x_{i,j}X}\end{xy} & \cdots \\
 & \vdots & 
 \end{pmatrix}\enspace.
\end{aligned}
\end{displaymath}
This nonlinearity poses an algorithmic  challenge: many known algorithms
for feasibility of inequalities are restricted to linear ones. See
Section~\ref{subsec:implTree} for further discussions.


In the remainder of this section, we present a concrete definition of
forward partial execution for weighted tree automata. It is an instance of Definition~\ref{def:catFPE}.
Its soundness, adequacy and monotonicity follow from the general results in Section~\ref{subsec:partExecCat}.

\begin{mydefinition}[FPE for weighted tree automata]
\emph{Forward partial execution (FPE)} is transformation of a weighted tree automata such that:
given an  $\mathcal{S}$-weighted tree automaton $\mathcal{A}=(Q, \Sigma,
 M, \alpha)$ where $\Sigma=(\Sigma_n)_n$, 
and a parameter $P\subseteq Q$,
 the resulting automaton $\mathcal{A}_{\FPE,
 P}=(Q',\Sigma,M',\alpha')$ is as follows. It has a state space
\begin{align}\label{eq:FPEStateSpaceTree}
 Q' &\;=\; \bigl\{(a,(y_0,y_1,\dotsc,y_{n-1}))\,\bigl|\bigr.\,  a\in\Sigma_n, \exists x\in P.\, M(a)_{x,(y_0,y_1,\dotsc,y_{n-1})}\neq 0_{\mathcal{S}}\bigr\} + (Q \setminus P)\enspace,
\end{align}
that replaces each state $x\in P$ with its one-step behaviors $(a,(y_0,y_1,\dotsc,y_{n-1}))$ as new states. 
As for the transition matrices $M'$,
 \begin{align*}
  & M'(a)_{(a,(x_0,\dotsc, x_{n-1})),\bigl((a_0,(y^0_0,\dotsc, y^0_{m_0-1})),\dotsc,(a_{n-1},(y^{n-1}_0,\dotsc, y^{n-1}_{m_{n-1}-1}))\bigr)}=\prod_{i=0}^{n-1}M(a_i)_{x_i,(y^i_0,\dotsc,y^i_{m_i-1})} \\
& M'(a)_{(a,(x_0,x_1,\dotsc,x_{n-1})),(x_0,x_1,\dotsc,x_{n-1})}=1_{\mathcal{S}} \\
& M'(a)_{x,\bigl((a_0,(y^0_0,\dotsc, y^0_{m_0-1})),\dotsc,(a_{n-1},(y^{n-1}_0,\dotsc, y^{n-1}_{m_{n-1}-1}))\bigr)} \\
& \qquad=\Bigl(M(a)\cdot\Bigl(\bigotimes_{i=0}^{n-1} M(a_i)\Bigr)\Bigr)_{x,\bigl((y^0_0,\dotsc, y^0_{m_0-1}),\dotsc,(y^{n-1}_0,\dotsc, y^{n-1}_{m_{n-1}-1})\bigr)} \\
 & M'(a)_{x,y}=M(a)_{x,y}  
 \end{align*}
 where $a\in\Sigma_n$, $a_i\in\Sigma_{m_i}$, and $x, x_i, y^i_j\in Q$. For all the other cases we define $M'(a)_{u,v}=0_{\mathcal{S}}$.
As for the initial vector $\alpha'$, the definition is shown below.
 \begin{displaymath}
 \alpha'_{(a,(x_0,\dotsc,x_{n-1}))}\;=\;(\alpha M(a))_{(x_0,\dotsc,x_{n-1})}\enspace, \qquad \alpha'_{x}\;=\;\alpha_{x}
 \end{displaymath}
%
%
\end{mydefinition}

We do not yet have a good definition 
of backward partial execution for weighted tree automata, probably for the reason
that we argued at the end of Section~\ref{subsec:partExecCat}.


\subsection{Algorithms}\label{subsec:algoTree}
The algorithm \pttreesimfwdalgo\ that searches for a forward simulation matrix between $\spt$-weighted tree automata is as follows.

\begin{myalgorithm}[\pttreesimfwdalgo]
 \noindent{\bf Input}: 
 A pair of $\spt$-weighted tree automata $\mathcal{A}$, $\mathcal{B}$.

 \noindent{\bf Output}: 
 A forward simulation matrix $X$ from $\mathcal{A}$ to $\mathcal{B}$ if such $X$ exists,
 and
 ``$\mathsf{No}$" otherwise.
 %

 \noindent{\bf Procedure}:
 For the input automata $\mathcal{A}$ and $\mathcal{B}$, 
 this algorithm first combines the constraints in Definition~\ref{def:simMatTree}
 into a system of (possibly nonlinear) polynomial inequalities,
 and then solves them.
 If a solution is found, then the algorithm outputs the solution and
 it outputs ``$\mathsf{No}$'' otherwise.
\end{myalgorithm}

Let us discuss the size of the system of inequalities to be solved. 
Assume that our goal is to establish $\mathcal{A}\fwd\mathcal{B}$.
Let $n$ be the number of states of $\mathcal{A}$ and
$m$ be that of $\mathcal{B}$. (In case our goal is
$\mathcal{A}\bwd\mathcal{B}$ we swap $n$ and $m$.) 
 Let $d$ be the
 maximum
 arity in the ranked alphabet $\Sigma$. 
Then the number of inequality
constraints
is at most $\sum_{k\in\mathbb{N}}|\Sigma_{k}|\cdot m\cdot n^{k}$;
and the degree of
each
polynomial inequality
constraint is at most $d$.

The algorithm \pttreesimbwdalgo\ that searches for a backward simulation matrix is similar.

\subsection{Implementation, Experiments and Discussions}\label{subsec:implTree}
%
A program \pttsim, 
in OCaml, 
implements the algorithms \pttreesimfwdalgo\ and \pttreesimbwdalgo\ in the previous section. The program is available in~\cite{implPtMpTree}.
In our implementation polynomial inequality constraints are solved  with the \texttt{FindInstance} function of
\emph{Mathematica}~\cite{mathematica}. 


Experiments were done on a MacBook Pro laptop with a Core i5 processor
(2.6 GHz, 2 cores) and 16 GB RAM. 
In our experiments we let the program  \pttsim\ try to establish
$\mathcal{A}\fwd\mathcal{A}$ (or $\mathcal{A}\bwd\mathcal{A}$) for a
randomly generated tree automaton $\mathcal{A}$. 

Our current
implementation turned out to be far from scalable: for the maximum
arity $d=2,3$ and an automaton $\mathcal{A}$ with several states and
transitions, the program barely manages to
establish the goal; it becomes hopeless for bigger problem instances.
This is in a sense as we expected: use of a general purpose algorithm
like \texttt{FindInstance} in
\emph{Mathematica} would never be a performance advantage; in fact, 
\texttt{FindInstance} tends to abort after tens of seconds, consuming
a few MBs of memory, for reasons that we cannot know. An obvious 
alternative is to use  special purpose algorithms---like the ones that are
known in the field of convex optimization. This is left as future work.


\section{Related Work}\label{sec:relatedWork}
Throughout this paper, we followed the framework
in~\cite{hasuo07generictrace} that uses Kleisli categories to
coalgebraically capture finite trace semantics of weighted automata and
weighted tree automata.  In~\cite{Jacobs0S15}, a different
approach to coalgebraic finite trace semantics is introduced where,
instead of Kleisli categories, \emph{Eilenberg-Moore categories} are
used.  Another coalgebraic approach to weighted tree automata---that is
based on Stone-type dualities---is found
in~\cite{klinR15coalgebraictrace}. 
Very roughly the \emph{Stone-duality
approach} in~\cite{klinR15coalgebraictrace} is similar to the
\emph{Kleisli approach} in the current paper, in that a single trace
(that is a word or a tree) is thought of as a logical formula.  The
\emph{Eilenberg-Moore approach} in~\cite{Jacobs0S15} comes
in quite a different flavor---it depends crucially on the
\emph{generalized determinization} construction. The latter is categorically
formulated for the first time in~\cite{Bartels04} and studied further on 
e.g.\ in~\cite{AdamekBHKMS12}.

 One clear advantage of the Kleisli approach (and hence of the
 Stone-duality approach,  possibly) is that it generalizes smoothly to
 the theory of \emph{(possibly) infinite} traces and simulations, such
 as in $\omega$-automata as opposed to ordinary automata. This advantage
 is witnessed in our recent work~\cite{urabeH15coalgebraicinfinite}, where
 forward and backward Kleisli simulations are shown to be sound (under
 certain minor conditions) with respect to the inclusion of infinite
 trace semantics. This soundness works both for the nondeterministic and
 probabilistic settings---the latter calls for measurable structures in 
 the setting of infinite traces---suggesting the versatility of the
 Kleisli approach. 

 Such generalization to (possibly) infinite traces seems unlikely for
 the Eilenberg-Moore approach to coalgebraic (finite) traces: the
 Eilenberg-Moore approach is essentially via determinization; and, as is
 well-known, for B\"{u}chi automata nondeterminism strictly increases
 expressive power. That being said, focusing on finite traces, it seems
 possible to exploit the \emph{algebraic} and \emph{syntactic} nature of
 the Eilenberg-Moore approach in the problem domain of
 language inclusion.\footnote{This is based on a comment by one of the
 reviewers, for which we are grateful.}  More specifically, the
 Eilenberg-Moore approach (and generalized determinization) has a close
 tie to the theory of \emph{generalized regular expressions} that is
 recently studied extensively (see
 e.g.~\cite{BonsangueRS09lics,bonsangueMS13soundand} and very
 recent~\cite{KozenMP015}).  This resulted in a sound and complete
 axiomatization of language \emph{equivalence} for automata weighted with a
 certain class of semiring~\cite{bonsangueMS13soundand} (see also
 Remark~\ref{rem:literatureForLanguageEquivalence}); it is conceivable
 that a generalization of this deductive framework to language \emph{inclusion},
 by replacing equalities with inequalities, is possible. 
 Such an ``Eilenberg-Moore approach'' to language inclusion would form a
 \emph{deductive} alternative to our current approach where we use
 matrices and continuous optimization (in the form of LP). Another
 potential advantage of the Eilenberg-Moore approach is that it does not
 need the $\omega$-cpo structures that we need for the Kleisli approach,
 and hence can  accommodate some semirings that our framework does not.

(Bi)simulation notions and (weighted) language equivalence/inclusion for quantitative systems have been an active
research topic in the field of formal verification and concurrency. Some related work in this direction has
already discussed
 in the earlier sections,
 including~\cite{jonsson91specificationand,chatterjee10quantitativelanguages,BaierHK04,kiefer11languageequivalence,blondel03undecidableproblems}. See
 also Remark~\ref{rem:literatureForLanguageEquivalence}.
Moreover, works
like~\cite{hughes04simulationsin,hasuo06genericforward,hasuo10probabilisticanonymity,Sokolova05}
take a categorical/coalgebraic approach.

Use of matrices as witnesses of quantitative language equivalence/inclusion is in
fact not uncommon. The rest of this section is devoted to the discussion
of such works, and their comparison to the current work. Overall, the
current work is distinguished in the following aspects.
\begin{itemize}
 \item 
  The
 categorical backend of Kleisli simulation that allows clean theoretical
       developments. The latter include: the duality between forward and
       backward simulation; a general soundness proof; and
 generalization to tree automata (Section~\ref{sec:matSimPolyF}).
 \item Our simulations witness language \emph{inclusion}, a problem 
       that is harder than language \emph{equivalence} (see Section~\ref{subsec:SPTcomparisonWithOtherSim}).
 \item Forward/backward partial execution (Section~\ref{sec:partExec})
       that enhances  effectivity of the approach by matrix
       simulations.
 \item Actual implementation of the algorithms and experiments.
\end{itemize}



In \cite{beal05equivalenceZautomata}, a notion called \emph{conjugacy}
between semiring weighted (word) automata is introduced. It is an
equivalence notion---it is a special case of $\fwd$ in the current work,
with the
inequalities
in Definition~\ref{def:simMat} replaced with equalities.
The notion of conjugacy comes with ``completeness'': assuming that the
weight semiring is so-called a  \emph{division ring}, two automata are equivalent if and only if
they are connected by some finite chain of conjugacies.


The notion of \emph{simulation}
in~\cite{esik10simulationequivalence} is  essentially the same as
conjugacy in~\cite{beal05equivalenceZautomata} (it is therefore an
equivalence notion unlike the name).
A simulation  in~\cite{esik10simulationequivalence} witnesses language
equivalence.
In~\cite{esik10simulationequivalence}  a semiring $\mathcal{S}$ is
called \emph{proper} when: two $\mathcal{S}$-weighted automata are
language equivalent if and only if they are connected by a finite chain 
of simulations. The authors go on to study proper semirings: they
present a necessary condition for a semiring to be proper, and an
example  that is not proper (namely $\smp$).

The results in~\cite{esik10simulationequivalence} have been extended to
weighted \emph{tree} automata in~\cite{esik11categorysimulations}. Their
simulation is a special case of ours (Definition~\ref{def:simMatTree})
where inequalities are replaced with equalities; soundness with respect
to tree language equivalence is proved; and completeness of a combined
simulation
(with an intermediate automaton, much like in our $\sqsubseteq_{\bf BF}$
and $\sqsubseteq_{\bf FB}$) is shown, under some assumptions.


Unlike the work discussed in the above, the
works~\cite{ciric12bisimulationfuzzy,ciric12computationgreatest,damljanovic14bisimulationsweighted}
study simulations given by matrices in the context of \emph{fuzzy
automata}. Here simulation is an oriented notion and witnesses language
inclusion (instead of language equivalence); its definition is essentially the
same
as ours (Definition~\ref{def:simMat}).
A principal difference
between~\cite{ciric12bisimulationfuzzy} and the current work is in the
domain of weights: in~\cite{ciric12bisimulationfuzzy} it is a structure
called \emph{residuated lattice}.

Algorithmic aspects of the simulation notion
in~\cite{ciric12bisimulationfuzzy} is pursued
in~\cite{ciric12computationgreatest}, where
an algorithm for computing the greatest simulation is presented.
Their algorithm works for a general residuated lattice, unlike ours
where linear inequalities are solved in semiring-specific manners.


These results
in~\cite{ciric12bisimulationfuzzy,ciric12computationgreatest} are
adapted in~\cite{damljanovic14bisimulationsweighted} to automata
weighted in a \emph{semiring} (instead of a residuated
lattice)---although some assumptions are imposed on a semiring and this makes
the semiring $\spt$ unqualified.

\section{Conclusions and Future Work}\label{sec:concl}
We introduced simulation matrices for weighted automata. While they
are instances of (categorical) Kleisli simulations, their concrete
presentation by matrices and linear inequalities yields concrete
algorithms for 
simulation-based quantitative verification. Generalization to weighted
tree automata follows immediately from the categorical theory behind,
too, although linearity is lost in general.

There are some directions in which the current matrix-based simulation
framework can be further generalized.  Our idea of $\spt$-weighted
automata was that they are probabilistic systems; when we wish to
accommodate uncountable state spaces (for which discrete probabilities
are hardly meaningful), we would need suitable measure theoretic
machinery. In the context of the current work of traces and simulations,
this will involve replacing $\Sets$ with $\mathbf{Meas}$ (the category
of measurable sets and measurable functions), and 
matrix multiplication with Lebesgue integration. 
Trace semantics for
probabilistic automata in 
$\mathbf{Meas}$ has been studied e.g.\
in~\cite{Cirstea11,kerstan13coalgebraictrace}. 

In fact, use of $\mathbf{Meas}$ as a base category becomes necessary
if we consider \emph{infinite} trace semantics---i.e.\ a language
of accepted infinite words---even if a state space is countable (i.e.\
discrete). This is simply because the set $\Sigma^{\omega}$ of all
infinite words is not countable. We are currently working on the
soundness
of matrix simulations against languages of infinite words; details 
will be presented in another venue.

As we mentioned in Section~\ref{sec:relatedWork}, 
there are other approaches~\cite{Jacobs0S15,
klinR15coalgebraictrace} to coalgebraic trace semantics than the approach
that
 we followed in this paper.
In Section~\ref{sec:relatedWork} we briefly discussed about the
relationship to them and potential research directions. Working out
their
details 
is future work.



Further generalization of the current theory will be concerned with acceptance
conditions that are unique to infinite words. An example is the
B\"{u}chi acceptance condition, for which a simulation notion (for
the nondeterministic setting) has been studied in~\cite{EtessamiWS05}.
A coalgebraic generalization of the simulation in~\cite{EtessamiWS05} and derivation of definitions of simulations
for systems like probabilistic B\"uchi automata~\cite{baierG05recognizingomega} are interesting future work.

Finally, further
optimization of our implementation is obvious future work.
Existing techniques for optimizing (bi)simulation (e.g.\ up-to techniques~\cite{sangiorgi05beyondbisimulation} whose coalgebraic characterization is studied in~\cite{bonchiPPR14coinductionupto}) are expected to be useful for this purpose.

\section*{Acknowledgments} Thanks are due to 
Andrzej Murawski,
Shota Nakagawa and the
anonymous referees (for the conference version and the current extended version) for useful discussions and
comments;
and to Bj\"{o}rn Wachter for the user support of the tool {\APEX}. The authors are
supported by:
Grants-in-Aid No.\ 24680001 \& 15KT0012, JSPS; and
Support Program for Researchers Traveling Abroad (No.\ 2014.1.2.099), International Information Science Foundation, Japan.




\newpage
\appendix
\section{Omitted Proofs}\label{appendix:omittedProofs}
\subsection{Proof of Lem.~\ref{lem:finiteGametoLimavg}}
\label{appendix:lemfiniteGametoLimavg}
\begin{myproof}
We prove the contraposition of each direction.

Assume that $\mathcal{A} \mathrel{\cancel{\sqsubseteq}_{\bf G}}\mathcal{B}$, 
i.e.\ there exists a winning strategy $(\rho_1,\tau_1)$ for Challenger in a finite simulation game played on $\mathcal{A}$ and $\mathcal{B}$.
Then a strategy that repeats $\rho_1$ and $\tau_1$ is a winning strategy for Challenger in an infinite simulation game played on  
$\mathcal{A}^{\sf Limavg}$ and $\mathcal{B}^{\sf Limavg}$.
Hence  $\mathcal{A}^{\sf Limavg}\mathrel{\cancel{\sqsubseteq}_{\bf G}^{\sf Limavg}}\mathcal{B}^{\sf Limavg}$.

Conversely, assume $\mathcal{A}^{\sf Limavg}\mathrel{\cancel{\sqsubseteq}_{\bf G}^{\sf Limavg}}\mathcal{B}^{\sf Limavg}$.
Then there exists a winning strategy $\tau^{\infty}_1$ for Challenger in an infinite simulation game played on $\mathcal{A}^{\sf Limavg}$ and $\mathcal{B}^{\sf Limavg}$.
By~\cite{ehrenfeucht79positionalstrategies}, without loss of generality, 
we may assume that $\tau^{\infty}_1$ is a positional strategy; i.e.\ the
 value of $\tau^{\infty}_1((p_0,q_0)(a_1,p_1,q_1)\dotsc(a_i,p_i,q_i))$ only depends on
 $p_i$ and $q_i$. 

One might hope to use $\tau^{\infty}_1$ itself as Challenger's winning
 strategy
for the finite game for 
$  \mathcal{A}
\sqsubseteq_{\bf G}
\mathcal{B}
$. This does not work in general, since a resulting run may not come
 back to $\star$. Modification of $\tau^{\infty}_1$ to force to visit 
$\star$ may make the strategy less advantageous. We shall show that such 
modification is nevertheless feasible, by finding an upper bound $r$ for
 ``the disadvantage that results from visiting $\star$.'' In the
 infinite game for 
$ \mathcal{A}^{\sf Limavg}
\sqsubseteq_{\bf G}^{\sf Limavg}
\mathcal{B}^{\sf Limavg}$, a winning strategy can always additionally
 ``save'' the advantage $r$; and it will be  spent to visit $\star$.


To each $x\in Q_{\mathcal{A}}$ we shall choose and assign an 
\emph{exit
 path}
$\pi_{x}$. Specifically,
by the assumption that $\mathcal{A}$ has no trap states, each state
$x\in Q_{\mathcal{A}}$ has a finite path $\pi_x=xb_1u_1b_2\dotsc
 b_mu_m$ in $\mathcal{A}$  that ``reaches the final state,'' that is,
the path satisfies
\begin{displaymath}
 M_{\mathcal{A}}(b_1)_{x,u_1}+M_{\mathcal{A}}(b_2)_{u_1,u_2}+\dots +M_{\mathcal{A}}(b_n)_{u_{m-1},u_{m}}+(\beta_{\mathcal{A}})_{u_{m}}\;\neq\;-\infty\enspace.
\end{displaymath}
Within the path $\pi_{x}$,  the advantages that Simulator can
 make are bounded: this follows from the assumptions that
no weight in $\mathcal{B}$ is $\infty$, and that $\mathcal{B}$ is finite
 state
(so that there are only finitely many choices of an initial state).
Since there are only finitely many $x\in Q_{\mathcal{A}}$, we can take a
 global upper bound $r\in \mathbb{R}$. To summarize, the real number $r$
 is chosen so that:
 for each $x\in Q_{\mathcal{A}}$, for the choice of an exit path $\pi_{x}
 = x\,b_{x,1}\,u_{x,1}\,b_{x,2}\,\dotsc\,
 b_{x,m(x)}\,u_{x,m(x)}$ as described 
 above, and for each path  $\pi=y\,b_{x,1}\,y_1\,b_{x,2}\,\dotsc\, b_{x,m(x)}\,y_{m(x)} $  in
 $\mathcal{B}$ on the same word $b_{x,1}\,b_{x,2}\,\dotsc\, b_{x,m(x)}$, we have
\begin{equation}\label{eq:saving}
 \begin{aligned}
M_{\mathcal{A}}(b_{x,1})_{x,u_{x,1}}+M_{\mathcal{A}}(b_{x,2})_{u_{x,1},u_{x,2}}+\dots
  +M_{\mathcal{A}}(b_{x,m(x)})_{u_{x,m(x)-1},u_{x,m(x)}}+(\beta_{\mathcal{A}})_{u_{x,m(x)}}
  + r
\\ >\;
M_{\mathcal{B}}(b_{x,1})_{y,y_1}+M_{\mathcal{B}}(b_{x,2})_{y_1,y_2}+\dots +M_{\mathcal{B}}(b_{x,m(x)})_{y_{m(x)-1},y_{m(x)}}+(\beta_{\mathcal{B}})_{y_{m(x)}}\,.
\end{aligned}
\end{equation}


Using this $r\in\mathbb{R}$ (``an upper bound for the cost of visiting $\star$''), we shall construct a strategy $\bigl(\rho'_1:1\to\mcr{Q}{A},\tau'_1:(\mcr{Q}{A}\times\mcr{Q}{B})\times(\Sigma\times\mcr{Q}{A}\times\mcr{Q}{B})^*\to
 1+\Sigma\times\mcr{Q}{A}\,\bigr)$ for Challenger in the finite
 simulation game
 for 
\begin{math}
   \mathcal{A}
\sqsubseteq_{\bf G}
\mathcal{B}
\end{math}
as follows.
 %
 %
 For a pair of runs $R=(p_0 a_1\dotsc a_i p_i,q_0 a_1\dotsc a_i q_i)$ on $\mathcal{A}$ and $\mathcal{B}$, 
 we define the accumulated weights $S^R_{\mathcal{A}}$ and $S^R_{\mathcal{B}}$ by 
\begin{displaymath}
 S^R_{\mathcal{A}}  =
 (\alpha_{\mathcal{A}})_{p_0}+M_{\mathcal{A}}(a_1)_{p_0,p_1}+\dots
 +M_{\mathcal{A}}(a_i)_{p_{i-1},p_{i}}
 \;\text{and}\;
S^R_{\mathcal{B}}  =
 (\alpha_{\mathcal{B}})_{q_0}+M_{\mathcal{B}}(a_1)_{q_0,q_1}+\dots
 +M_{\mathcal{B}}(a_i)_{q_{i-1},q_{i}}\enspace.
\end{displaymath}
While $S^R_{\mathcal{A}}\le S^R_{\mathcal{B}}+r$ (i.e.\ the saving is not
 enough for the cost of visiting $\star$), 
 $\rho'_1$ and $\tau'_1$ are 
defined as follows. They do essentially the same thing as
 $\tau^{\infty}_1$ does, except that $\tau'_{1}$ terminates when $\star$
 is visited.
\begin{equation}\label{eq:strategySavingMode}
\begin{aligned}
 &\rho'_1(\bullet)=\tau^{\infty}_1((\star,\star)) \\
 &\tau'_1\bigl((p_0,q_0)(a_1,p_1,q_1)\dotsc(a_i,p_i,q_i)\bigr)
 \\
 &=\begin{cases}
 \checkmark &  \bigl(\tau^{\infty}_1\bigl((p_0,q_0)(a_1,p_1,q_1)\dotsc(a_i,p_i,q_i)\bigr)=(\mbox{$!$},\star)\bigr) \\
 \tau^{\infty}_1\bigl((p_0,q_0)(a_1,p_1,q_1)\dotsc(a_i,p_i,q_i)\bigr) & (\text{otherwise}).
 \end{cases}
\end{aligned}
\end{equation}
Assume that $S^R_{\mathcal{A}}>S^R_{\mathcal{B}}+r$ is satisfied at a
 certain stage, say at the state $p_n\in Q_{\mathcal{A}}$ for the first time. 
After that, $\tau'_1$ is defined for each $1\le j\le m(p_{n})$ by:
\begin{equation}\label{eq:strategyExitMode}
\begin{aligned}
 &\tau'_1\bigl((p_0,q_0)(a_1,p_1,q_1)\dotsc(a_n,p_n,q_n)(b_{p_{n},1},u_{p_{n},1},y_1)\dotsc(b_{p_{n},j},u_{p_{n},j},y_j)\bigr)
 \\
 &\qquad=\begin{cases}
 (b_{p_{n},j+1},u_{p_{n},j+1}) & \bigl(j<m(p_{n})\bigr) \\ \checkmark & \bigl(j=m(p_{n})\bigr)
 \end{cases}
\end{aligned}
\end{equation}
where $\pi_{p_{n}}
 = p_{n}\,b_{p_{n},1}\,u_{p_{n},1}\,b_{p_{n},2}\,\dotsc\,
 b_{p_{n},m(p_{n})}\,u_{p_{n},m(p_{n})}$ is the exit path for $p_{n}$
 that we have fixed in the above. Here $\tau'_{1}$ is headed to the exit
 along $\pi_{p_{n}}$, no matter what  Simulator's move $y_{j}$ is.

It remains to show that the strategy $(\rho'_1,\tau'_1)$ of Challenger's
 is winning
in the finite game for 
$   \mathcal{A}
\sqsubseteq_{\bf G}
\mathcal{B}
$. 
In the case where the clause~(\ref{eq:strategyExitMode}) is never
 invoked, we must have that
 $\tau^{\infty}_1\bigl((p_0,q_0)(a_1,p_1,q_1)\dotsc(a_i,p_i,q_i)\bigr)=(\mbox{$!$},\star)$
 holds for some $i$. At such $i$ Challenger must have an accumulated weight
 that is strictly bigger than Simulator does: by the assumption that
 $\tau^{\infty}_1$ is positional, the strategy $\tau^{\infty}_1$ will
 just repeat what it has done, and it will not win unless it is winning
 so far. 

Finally, in the case where the clause~(\ref{eq:strategyExitMode}) is
 invoked, the advantage that Simulator makes between the time $n+1$ and 
$n+m(p_{n})$ is at most $r$ by the definition of $r$. This does not eat
 up the ``saving'' $r$. 

Therefore we have shown that $\mathcal{A}
 \mathrel{\cancel{\sqsubseteq}_{\bf G}}\mathcal{B}$. 
This concludes the proof. 
%
\qed
\end{myproof}

%

%
%
\section*{References}

\end{document}